\documentclass[universe,article,accept,moreauthors,pdftex]{Definitions/mdpi}

\firstpage{1}
\pubvolume{1}
\issuenum{1}
\articlenumber{0}
\pubyear{2024}
\copyrightyear{2024}
\externaleditor{Academic Editor: Firstname\\ Lastname}
\datereceived{28 March 2024}
\daterevised{22 April 2024} 
\dateaccepted{ }
\datepublished{ }
\hreflink{https://doi.org/} 

\usepackage{epsfig}
\usepackage{amssymb}
\usepackage[utf8]{inputenc}
\usepackage{graphicx}
\usepackage{dcolumn}
\usepackage{bm}
\newcommand{\be}{\begin{equation}}
\newcommand{\ee}{\end{equation}}
\newcommand{\bea}{\begin{eqnarray}}
\newcommand{\eea}{\end{eqnarray}}
\newcommand{\ba}{\begin{array}}
\newcommand{\ea}{\end{array}}
\newcommand{\bi}{\begin{itemize}}
\newcommand{\ei}{\end{itemize}}
\newcommand{\lan}{\langle}
\newcommand{\ran}{\rangle}

\Title{Theory of Majorana-Type Heavy Ion Double Charge Exchange Reactions by Pion--Nucleon Isotensor Interactions}
\TitleCitation{Theory of Majorana-Type Heavy Ion Double Charge Exchange Reactions by Pion--Nucleon Isotensor Interactions}

\Author{Horst Lenske
 $^{1,}$*
$^{,\dagger}$\orcidA{}, Jessica Bellone $^{2,\dagger}$, Maria Colonna $^{2,\dagger}$ and Danilo Gambacurta $^{2,\dagger}$}

\AuthorNames{Horst Lenske, Jessica Bellone, Maria Colonna and Danilo Gambacurta}

\AuthorCitation{Lenske, H.; Bellone, J.; Colonna, M.; Gambacurta, D.}

\address{%
$^{1}$ \quad Institut f\"ur Theoretische Physik, Justus--Liebig--Universit\"at Giessen, D-35392 Giessen, Germany
\\
$^{2}$ \quad Istituto
 Nazionale di Fisica Nucleare, Laboratori Nazionali del Sud, I-95123 Catania, Italy; bellone@lns.infn.it~(J.B.); colonna@lns.infn.it~(M.C.); gambacurta@lns.infn.it~(D.G.)
}

\corres{Correspondence: horst.lenske@physik.uni-giessen.de; Tel.: +49-641-9933361}

\firstnote{
 The NUMEN Collaboration, LNS Catania, I-95123 Catania, Italy.}

\date{\today}

\abstract{
The theory of heavy ion double charge exchange (DCE) reactions proceeding by effective rank-2 isotensor interactions is presented.
Virtual pion--nucleon charge exchange interactions are investigated as the source for induced isotensor interactions, giving rise to the Majorana DCE (MDCE) reaction mechanism. MDCE is of a generic character, proceeding through pairs of complementary ($\pi^\pm,\pi^\mp$) reactions in the projectile and target nucleus. The dynamics of the elementary
processes is discussed, where the excitation of pion--nucleon resonances are of central importance. Investigations of initial and final state ion--ion interactions show that these effects are acting as vertex renormalizations.
In closure approximation, well justified by the finite pion mass, the second-order transition matrix elements reduce to pion potentials and effective two-body isotensor DCE interactions, giving rise also to two-body correlations in either of the participating nuclei. Connections to neutrinoless Majorana double beta decay (MDBD) are elucidated at various levels of the dynamics, from the underlying fundamental electro-weak and QCD scales to the physical scales of nuclear MDBD and MDCE physics. It is pointed out that heavy ion MDCE reactions may also proceed by competing electro-weak charge exchange processes, leading to lepton MDCE by electrons, positrons, and neutrinos.}

\keyword{double charge exchange reactions; reaction theory; nuclear structure theory; double beta decay; induced interactions; nuclear matrix elements}

\begin{document}


\section{Introduction}\label{sec:intro}

Heavy ion double charge exchange (DCE) reactions are unique as a new tool for investigations of the rather unexplored sector of higher-order nuclear dynamics. DCE research is of generic interest for nuclear reaction and nuclear structure physics because of its large potential for high-precision investigations of nuclear modes, which otherwise are almost impossible to access. A~central topic of this article is to show that DCE physics is going significantly beyond the standard approach to peripheral heavy ion reactions as dominated by mean-field dynamics. DCE research is located at the intersection of nuclear and hadron physics, thus broadening the view on the dynamics of nuclear many-body~systems.

In a previous paper~\cite{Lenske:2024dsc}, the emergence of an effective isotensor interaction and the role of ion--ion elastic interactions in second-order double single charge exchange (DSCE) reactions were investigated. DSCE reactions proceed by acting twice with the nucleon--nucleon (NN) isovector T matrix, where each of the actions generates a single charge exchange (SCE) transition. It was shown that by proper transformations of the operator structures, defined by central spin--scalar, spin--vector and rank-2 spin tensor interactions, effective operators are obtained, acting as rank-2 isotensor operators intrinsically in each nucleus. In~addition, in~\cite{Lenske:2024dsc}, the role of initial state (ISI) and final state (FSI) ion--ion elastic interactions was investigated. The~DSCE investigations led to three significant and far-reaching~results:
\begin{itemize}
  \item ISI and FSI interactions lead to distortion coefficients, which act as quenching factors.  As~a result, the DSCE reaction amplitude and consequently the observed DSCE nuclear matrix elements are strongly suppressed  by orders of magnitudes compared to the results expected without ISI/FSI.
 \item The relative motion degree of freedom induces in DSCE reactions in each nucleus a correlation between the pair of SCE vertices, where the correlation length is determined by the kinematical conditions of the reaction.
 \item The pair of NN T matrices can be recast into a set of spin--scalar and spin--vector rank-2 isotensor interactions, acting in each nucleus as effective two-body interactions and forming together a four-body ion--ion interaction.
\end{itemize}

In this work, we investigate the competing \emph{Majorana DCE} (MDCE) scenario. While DSCE theory is the second-order extension of the conventional direct reaction single charge exchange (SCE) theory~\cite{Lenske:2018jav,Lenske:2019cex}, MDCE theory takes a completely different view by describing a heavy ion DCE reaction as a combination of pion--nucleon DCE reactions in projectile and target nucleus. A~consequence of such an approach is that the second-order aspects inherent to a DCE reaction are treated on the level of isovector pion--nucleon scattering, giving rise to dynamically created effective rank-2 isotensor interactions in the projectile and target~nucleus.

By definition, a~heavy ion DCE reaction relies finally on an interaction of rank-2 isotensor character. Hitherto, searches for such a kind of nuclear interaction of generic character have been unsuccessful. To date, the existence of neither elementary isotensor mesons~\cite{Dover:1984zq,Wu:2003wf,Anikin:2005ur} nor signatures of interactions of that kind in single, isolated nuclei~\cite{Leonardi:1976zz} could be confirmed with convincing certainty. Most likely, rank-2 isotensor interactions do not exist as an elementary mode of their own right. The~conditions, however, might change if two nuclei are in close contact as in a peripheral ion--ion collision. In~such a situation, an~effective isotensor interaction can be generated dynamically as a transient phenomenon. MDCE reactions proceed by virtual pion--nucleon double charge exchange scattering, involving sequences of $\pi^\pm,\pi^0$ and $\pi^0,\pi^\pm$ pion--nucleon SCE reactions. Their proper combination leads finally to virtual ($\pi^\pm,\pi^\mp$) pion--nucleon DCE reactions in the reacting nuclei. Under~nuclear structure aspects, a DCE reaction is determined by excitations of $n^2p^{-2}$ and $p^2n^{-2}$ two particle--two hole configurations in the interacting~nuclei.

In the past, pion beams were used extensively for DCE research on nuclei at the Los Alamos Meson Physics Facility (LAMPF) \cite{Haxton:1997rn}. LAMPF was shut down a long time ago, but the physics issues studied there have become of renewed interest for heavy ion DCE research. The~theoretical understanding achieved at that time for pion--nucleon isovector dynamics~\mbox{\cite{Johnson:1983sa,Johnson:1983pb,Greene:1984svl,Gilman:1985gt,Siciliano:1986kj,Gilman:1986rg}} and DCE nuclear structure theory~\cite{Auerbach:1987gn,Auerbach:1988ir,Auerbach:1993mm,Auerbach:1996tb,Auerbach:2018byu} are worth being rediscovered because they are of high value for research on the MDCE mechanism of heavy DCE~reactions.

\begin{figure}[H]
\includegraphics[width = 8.5cm]{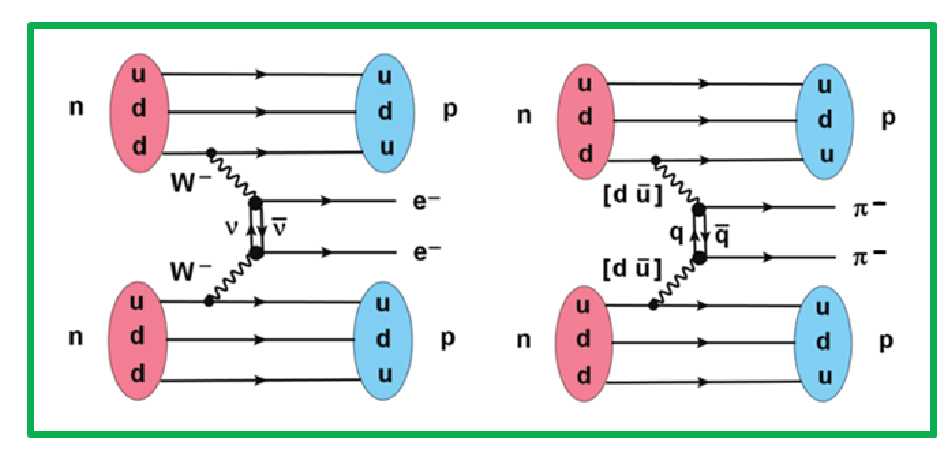}
\caption{Comparison of the elementary mechanisms underlying weak neutrinoless Majorana double beta decay (MDBD) (left), leading to the creation of a lepton pair on the mass shell~\cite{Tomoda:1990rs,Ejiri:2019ezh}, and~strong Majorana double charge exchange (MDCE) leading to the emission of a pair of virtual mesons off the mass shell. See text for further~discussions.}
\label{fig:MDCE_MDBD}

\end{figure}

In~\cite{Lenske:2024var,Lenske:2024dsc}, the similarity of DSCE and two--neutrino DBD was emphasized.
A special aspect of  the pionic MDCE scenario is the striking similarity to the heavily discussed neutrinoless \emph{Majorana DBD} (MDBD). That similarity is illustrated in Figure~\ref{fig:MDCE_MDBD} on the elementary level of virtual weak $W^\pm$ gauge bosons and highly virtual strong quark--antiquark $q\bar{q}$ modes, the~former materializing into a lepton pair on the mass shell, the~latter into a pair of mesons off the mass shell. MDBD is searched for as a possible signature for \emph{Beyond the Standard Model} (BSM) physics because MDBD relies on the still hypothetical Majorana neutrinos with the claimed property $\nu_M\equiv \bar{\nu}_M$, see e.g.,~\cite{Tomoda:1990rs,Ejiri:2019ezh}. MDBD would lead to the spontaneous creation of matter in the form of lepton pairs, thus violating lepton number conservation. As~will be seen, the~MDCE mechanism is described by graphs resembling those of neutrinoless DBD. However, MDCE is determined finally by strong hadronic interactions of a quite different range and strength. Spectroscopically, the same nuclear states as in MDBD are involved, and the transitions are induced by the same kind of isovector multipole operators, exciting spin--scalar and spin--vector~modes.

\vspace{-3pt}

In the forthcoming sections, we present a concise, unified picture of the physics of MDCE dynamics and the relation to neutrinoless Majorana double beta decay (MDBD). The~theoretical foundations and methods are discussed much beyond the level presented in previous publications~\cite{Lenske:2019cex,Cappuzzello:2022ton}.
The overall aspects, the~essential features, and~theoretical principles of MDCE reaction physics are presented in Section~\ref{sec:RMDCE}. As~mentioned before, the~MDCE reaction amplitude is formally given by one-step distorted wave matrix element. The~quenching of reaction yields caused by the strongly absorptive ion--ion optical potential discussed in~\cite{Lenske:2024dsc} for the DSCE amplitude is less pronounced but still a highly important effect of significant strength. Therefore, the~role of ISI and FSI is elucidated in Section~\ref{sec:ISIFSI}.
A different view on ISI and FSI as vertex renormalization is presented in Section~\ref{sec:View}, where we point to the formal similarity of ISI/FSI with the treatment of short-range correlations in nuclear structure calculations, especially also used in DBD theory.
The MDCE transition form factors and nuclear matrix elements are investigated in Section~\ref{sec:TNME}. There, we also address in some detail the essential features of the box diagram, introduce the closure approximation, which allows to define second-order pion potentials as effective two-body DCE interactions. Pion--nucleon scattering and the construction of the pion--nucleon T matrix, used to describe the excitation of $np^{-1}$ and $pn^{-1}$ states, are the subjects of Section~\ref{sec:TpiN}.
Illustrating numerical results are presented in Section~\ref{sec:Numerics}. The connections of DCE reactions to DBD are discussed in Section~\ref{sec:DCE-DBD}.
A summary and an outlook are found in Section~\ref{sec:SumOut}. Additional material on distortion amplitudes, details of the box diagram, the~pion--nucleon T matrix, and~more on the theoretical background of the pion potentials  is presented in several~appendices.

\section{Theory of Heavy Ion MDCE~Reactions}\label{sec:RMDCE}

\begin{figure}[H]

\includegraphics[width = 6.5cm]{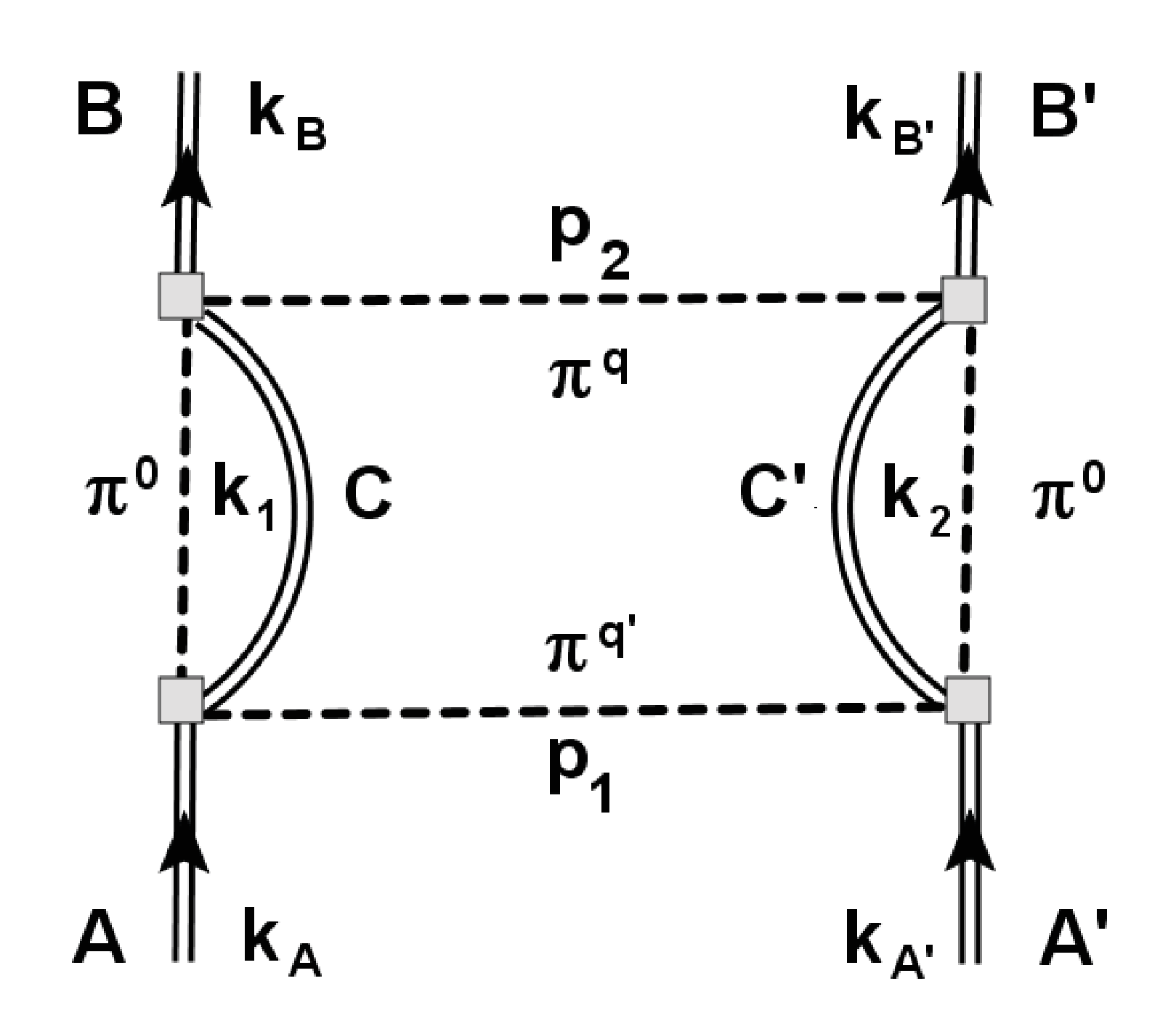}
\caption{\textls[-15]{The MDCE diagram for the reaction} $A(Z,N)+A'(Z',N')\to B(Z\pm 2,N\mp 2)+B'(Z'\mp 2,$ $N'\pm 2)$. The~isovector pion--nucleon T matrices are denoted by filled gray boxes. The~intermediate neutral pions induce a correlation between the SCE events, thus making MDCE a two-nucleon process.
The co-propagating core states are denoted by $C=C(Z\pm 1,N\mp 1)$ and $C'=C'(Z\mp 1,N\pm 1)$.  Charged pions
$\pi^q$ and $\pi^{q'}$, $q,q'=\pm 1$ are exchanged  with four-momenta $p_{1,2}$ between the nuclei.
The four momenta in the incident ($k_{A,A'}$), the~intermediate ($k_{1,2}$), and~the exit channel ($k_{B,B'}$) are~indicated.}
\label{fig:MDCE}

\end{figure}

The MDCE interaction process for a reaction $A(Z,N)+A'(Z',N')\to B(Z\pm 2,N\mp 2)$ $+~B'(Z'\mp 2,N'\pm 2)$ is illustrated graphically in Figure~\ref{fig:MDCE}. Formally, the~MDCE scenario is described in box diagrams, where the dynamical key elements are pion--nucleon isovector interactions. The~reaction is described by a first-order distorted wave (DW) reaction amplitude $M^{(1)}_{\alpha\beta}$.
The differential cross section for an unpolarized beam and target nuclei is defined as
\be\label{eq:dsigma}
d\sigma^{(1)}_{\alpha\beta}=\frac{m_\alpha m_\beta}{(2\pi\hbar^2)^2}\frac{k_\beta}{k_\alpha}\frac{1}{(2J_a+1)(2J_A+1)}
\sum_{M_a,M_A\in \alpha;M_b,M_B\in \beta}{\left|M^{(1)}_{\alpha\beta}(\mathbf{k}_\alpha,\mathbf{k}_\beta)\right|^2}d\Omega_{\alpha\beta},
\ee
The cross section is
averaged over the initial nuclear spin states ($J_{a,A},M_{a,A}$) and summed over the final nuclear spin states  ($J_{b,B},M_{b,B}$), respectively. Reduced masses in the incident and exit channels, respectively, are denoted by $m_{\alpha,\beta}$. $\mathbf{k}_\alpha$ and $\mathbf{k}_\beta$  are the (Lorentz-invariant) momenta in the incident and exit channels, respectively.

Formally, the~reaction amplitude has the structure of a first-order distorted wave matrix element:
\be\label{eq:Mab}
\mathcal{M}^{(1)}_{\alpha\beta}(\mathbf{k}_\alpha,\mathbf{k}_\beta)=
\lan \chi^{(-)}_\beta|
\lan B|\mathcal{T}_{\pi N}\mathcal{G}_{\pi C}\mathcal{T}_{\pi N}|A\ran
\mathcal{D}_{\pi^c\pi^{c'}}
\lan B'|\mathcal{T}_{\pi N}\mathcal{G}_{\pi' C'}\mathcal{T}_{\pi N}|A'\ran
|\chi^{(+)}_\alpha\ran.
\ee
The charged pions, described by the propagator $\mathcal{D}_{\pi^c\pi^{c'}}$, connect the nuclear transition matrix elements (TMEs). Since they describe the intranuclear DCE transitions, they are the key elements for spectroscopic investigations.
We introduce the pion and nucleon isospin operators $\mathbf{T}$ and $\bm{\tau}$ and rewrite the T matrices as
\be
\mathcal{T}_{\pi N}(i,i')=T_{\pi N}(i,i')\mathbf{T}(i')\cdot \bm{\tau}(i)
\ee

Since in a DCE transition only the ladder parts $T^{\pm}\tau^\mp$ are relevant and pion and nucleon operators commute, we rearrange the ladder operators to pion and nucleon rank-2 isotensor operators of complementary charge-lowering and charge-raising properties:
\bea
\mathcal{I}^{(\pi)}_{2\pm 2}(i',j')&=&T^{\pm}(i')T^{\pm}(j')\\
\mathcal{I}^{(N)}_{2\mp 2}(i,j)&=&\tau^{\mp}(i)\tau^{\mp}(j).
\eea

In r--space formulation, the MDCE amplitude is given by
\be\label{eq:Mab}
M^{(1)}_{\alpha\beta}(\mathbf{k}_\alpha,\mathbf{k}_\beta)=
\lan \chi^{(-)}_\beta|\mathcal{R}_{\alpha\beta}|\chi^{(+)}_\alpha\ran=
\int d^3r_\alpha\int d^3r_\beta
\chi^{(-)*}_\beta(\mathbf{r}_\beta)\mathcal{R}_{\alpha\beta}(\mathbf{r}_\alpha,\mathbf{r}_\beta)
\chi^{(+)}_\alpha(\mathbf{r}_\alpha).
\ee
The distorted waves $\chi^{(\pm)}_{\alpha,\beta}$ with asymptotically outgoing and incoming spherical waves, respectively, depend on the invariant channel momenta $\mathbf{k}_{\alpha,\beta}$ and the channel coordinates $\mathbf{r}_{\alpha,\beta}$, the~latter describing the relative distance between the initial nuclei $A,A'$ and the final nuclei $B,B'$, respectively. The~reaction is described in the ion--ion rest~frame.

The distorted waves are of central importance for the quantitative description of direct nuclear reactions like heavy ion SCE and DCE scattering. They account for diffractive and dispersive initial state and final state elastic ion--ion interactions. In~direct reaction (DR) theory, they are described globally by complex-valued optical model potentials including the long-range Coulomb potential and real and imaginary nuclear potentials of ranges which are defined by the sizes of the density distribution of the colliding nuclei. A~key role is played by the strong imaginary parts. They describe the absorption of the probability flux by the coupling to the multitude of non-elastic channels and as such are essential for a realistic description of the magnitudes of SCE and DCE cross sections and shapes of the related angular distributions; see ~\cite{Lenske:2021bpk}. As~stated in~\cite{Siciliano:1990yf}, a proper treatment of distortion effects was badly missed in the theory of pion--DCE, leaving many open questions for a realistic description of pion--DCE~data.

The spectroscopic content of the DCE reaction is contained in the transition kernel $\mathcal{R}_{\alpha\beta}$. An~instructive and successful approach is to use the momentum representation:
\be\label{eq:Rab}
\mathcal{R}_{\alpha\beta}(\mathbf{r}_\alpha,\mathbf{r}_\beta)=\int \frac{d^3p_1}{(2\pi)^3}\int \frac{d^3p_2}{(2\pi)^3}
e^{i(-\mathbf{p}_1\cdot \mathbf{r}_\alpha+\mathbf{p}_2\cdot \mathbf{r}_\beta)} \mathcal{F}_{\alpha\beta}(\mathbf{p}_1,\mathbf{p}_2).
\ee
As a remarkable first achievement, we have succeeded in  separating the nuclear and relative motion degrees of freedom. The~latter are represented by the plane waves depending on the relative ion--ion coordinates in the incident and the exit channels, denoted by $\mathbf{r}_{\alpha}$ and $\mathbf{r}_{\beta}$, respectively.
The MDCE transition form factor
\be\label{eq:Fab}
\mathcal{F}_{\alpha\beta}(\mathbf{p}_1,\mathbf{p}_2)=
\mathcal{W}_{AB}(\mathbf{p}_1,\mathbf{p}_2)D_{\pi^q\pi^{q'}}(p_1,p_2)\mathcal{W}_{A'B'}(\mathbf{p}_1,\mathbf{p}_2),
\ee
defined by the diagram of Figure~\ref{fig:MDCE} contains as key elements the nuclear transition matrix elements (TMEs)
\bea\label{eq:Wab}
\mathcal{W}_{AB}(\mathbf{p}_1,\mathbf{p}_2)&=&
\lan B|\mathcal{T}_{\pi N}(\mathbf{p}_2,\mathbf{k}|\bm{\sigma}_3)\mathcal{G}_{A}(k_1)
\mathcal{T}_{\pi N}(\mathbf{p}_1,\mathbf{k}|\bm{\sigma}_1)|A\ran ,\\
\mathcal{W}_{A'B'}(\mathbf{p}_1,\mathbf{p}_2)&=&
\lan B'|\mathcal{T}_{\pi N}(\mathbf{p}_2,\mathbf{k}'|\bm{\sigma}_4)\mathcal{G}_{A'}(k_2)
\mathcal{T}_{\pi N}(\mathbf{p}_1,\mathbf{k}'|\bm{\sigma}_2)|A'\ran .
\eea
The TMEs are of central interest for DCE research because they account for the spectroscopy of the reaction. For~example, the~transition $A\to B$ is induced by two consecutive actions of the pion--nucleon isovector T matrices $\mathcal{T}_{\pi N}$, each giving rise to a SCE transition, $A\to C$ and $C\to B$, respectively. The~vertices are connected by the Green's function $G_{A}(k_1)$, describing the s-channel propagation, i.e.,~in the direction of the left and right vertical branches of Figure~\ref{fig:MDCE}, of~the intermediate $\pi^0+C$ system. The~transition $A'\to B'$ follows the same rules. The~TMEs will be investigated in more detail in a later~section.

The DCE process is driven by the t-channel exchange of charged pions between the projectile and the target nucleus as~indicated by the lower and upper horizontal branches in Figure~\ref{fig:MDCE}. In~lowest order, the exchange is described by the
the symmetrized product propagator
\be
D_{\pi^q\pi^{q'}}(p_1,p_2)\approx \frac{1}{2}\left(D_{\pi^q}(p_1)D_{\pi^{q'}}(p_2)+D_{\pi^q}(p_2)D_{\pi^{q'}}(p_1)\right).
\ee
Possible pion--pion and pion--matter interactions are neglected.

As~discussed in Appendix \ref{app:BoxPW}, in the ion--ion rest frame, the four-momentum $p_{1}=(0,\mathbf{p}_{1})^T$, $p^2_{1}=-\mathbf{p}^2_{1}$ is purely space-like, while $p_2=(E_A-E_B,\mathbf{p}_2)^T$ includes the reaction Q value. For~ $|E_A-E_B|\ll m_\pi$, we may safely neglect the Q-value dependence and describe the exchange of both mesons by static pion propagators
\be
D_{\pi^q}(\mathbf{p})=-\frac{m_\pi}{\mathbf{p}^2+m^2_{\pi^q}} .
\ee

In the ion-ion rest frame and at the energies relevant for heavy ion MDCE reactions, the isovector pion--nucleon T-matrix $\mathcal{T}_{\pi N}$  is described adequately by the operator structure~\cite{Moorhouse:1969va,Johnson:1994na}
\be\label{eq:TpiN}
\mathcal{T}_{\pi N}(\mathbf{p},\mathbf{p}'|\bm{\sigma})=
\left[T_0(S_{\pi n})+\frac{1}{m^2_\pi}\left(T_1(s_{\pi n})\mathbf{p}\cdot \mathbf{p}'+
iT_2(s_{\pi N})\bm{\sigma}\cdot(\mathbf{p}\times \mathbf{p}')\right)\right]\mathbf{T}_\pi\cdot \bm{\tau}_N .
\ee
Nucleon spin degrees of freedom are involved via the spin operators $\bm{\sigma}$. The~form factors $T_{0,1,2}$ depend on the invariant pion--nucleon energy $s_{\pi N}=(k_\pi + k_N)^2$ . They are playing the role of energy-dependent coupling~constants.

\section{Initial State and Final State~Interactions}\label{sec:ISIFSI}
Before further investigating  MDCE interactions and form factors, we must understand first the contributions of ISI and FSI to the reaction process. In~momentum representation, the MDCE reaction amplitude attains an intriguing form:
\be\label{eq:Mab_momentum}
M^{(1)}_{\alpha\beta}(\mathbf{k}_\alpha,\mathbf{k}_\beta)=\int d^3p_1\int d^3p_2 D^{(-)*}_\beta(\mathbf{p}_2)\mathcal{F}_{\alpha\beta}(\mathbf{p}_1,\mathbf{p}_2)D^{(+)}_\alpha(\mathbf{p}_1).
\ee

The ion--ion ISI/FSI parts are contained in the distortion coefficients $D^{(\pm)}_{\alpha\beta}$.
The distortion coefficients are 3D Fourier transforms of the incoming and outgoing distorted waves
\bea\label{eq:Dpm}
D^{(+)}_\alpha(\mathbf{p}_1)&=&\int \frac{d^3r_\alpha}{(2\pi)^3}
e^{i\mathbf{p}_1\cdot \mathbf{r}_\alpha}\chi^{(+)}_\alpha(\mathbf{k}_\alpha,\mathbf{r}_\alpha),\\
D^{(-)*}_\beta(\mathbf{p}_2)&=&\int \frac{d^3r_\beta}{(2\pi)^3}
e^{i\mathbf{p}_2\cdot \mathbf{r}_\beta}\chi^{(-)*}_\beta(\mathbf{k}_\beta,\mathbf{r}_\beta).
\eea

In Appendix \ref{app:DW}, the properties of distorted wave, derived from an optical model wave equation, are investigated in detail. On~general theoretical grounds, two important results are obtained, namely, that formally
the distorted waves are factorizable into plane waves and residual amplitudes $h_{\alpha,\beta}$, which are determined essentially by the half off-shell optical model elastic scattering~amplitudes. As~the central result, the~distortion coefficients are derived in closed form.

Anticipating the results of Appendix \ref{app:DW}, we write
\bea\label{eq:chiab}
\chi^{(+)}_\alpha(\mathbf{k}_\alpha,\mathbf{r}_\alpha)&=&e^{i\mathbf{k}_\alpha\cdot \mathbf{r}_\alpha}\left(1-h_\alpha(\mathbf{r}_\alpha)\right),\\
\chi^{(-)*}_\beta(\mathbf{k}_\beta,\mathbf{r}_\beta) &=&e^{-i\mathbf{k}_\beta\cdot \mathbf{r}_\beta}\left(1-h_\beta(\mathbf{r}_\beta)\right).
\eea
In the second equation, the well-known relation $\chi^{(-)*}(\mathbf{k},\mathbf{r})=\chi^{(+)}(-\mathbf{k},\mathbf{r})$ is exploited; see~\cite{GoldbergerWatson:1964,Joachain:1975,Lenske:2021bpk}.

By defining the 3D Fourier transforms
\bea\label{eq:fafb}
f_{\alpha}(\mathbf{p})&=&\int \frac{d^3r}{(2\pi)^3}e^{i(\mathbf{p}-\mathbf{k}_\alpha)\cdot \mathbf{r}}h_{\alpha}(\mathbf{r}),\\
f_{\beta}(\mathbf{p})  &=&\int \frac{d^3r}{(2\pi)^3}e^{-i(\mathbf{p}-\mathbf{k}_\beta)\cdot \mathbf{r}}h_{\beta}(\mathbf{r}) ,
\eea
the distortion amplitudes become
\bea
D^{(+)}_\alpha(\mathbf{p}_1)&=&D^{(DW)}_\alpha(\mathbf{p}_1)-f_{\alpha}(\mathbf{p}_1)\\
D^{(-)*}_\beta(\mathbf{p}_2)&=&D^{(DW)}_\beta(\mathbf{p}_2)-f_{\beta}(\mathbf{p}_2).
\eea
For vanishing  elastic interactions, also the residual amplitudes vanish, and the distortion coefficients approach the plane distribution
\be\label{eq:DpmPW}
D^{(\pm )}_{\alpha,\beta}(\mathbf{p}_i)\mapsto D^{(PW)}_{\alpha,\beta}(\mathbf{p}_i)=\delta(\mathbf{p}_i- \mathbf{k}_{\alpha,\beta}).
\ee

For realistic optical potentials, accurately describing ion--ion elastic angular distributions and total reaction cross sections, the~residual amplitudes attain values of order unity $|f_{\alpha,\beta}|\to 1$, resulting in $|D^{(\pm)}_{\alpha,\beta}|\ll 1$. These results explain the pronounced quenching of the cross sections of heavy ion reactions by orders of magnitudes compared to the yields observed in reactions with particles not suffering from the strong absorption of the incoming probability~flux.

As implied by Equation~\eqref{eq:Mab_momentum}, the MDCE reaction amplitude is determined by the product of the initial and final state distortion coefficients. Together, they form the reaction~kernel
\be
\mathcal{K}_{\alpha\beta}(\mathbf{p}_1,\mathbf{p}_2)=D^{(-)*}_\beta(\mathbf{p}_2)D^{(+)}_\alpha(\mathbf{p}_1).
\ee

From Equation~\eqref{eq:DpmPW}, we find that the total kernel is a superposition of two kernels of the diagonal products of plane wave (PW) and DW distributions and two mixed PW/DW kernels. Combining the latter two into a single term, the~MDCE kernel becomes a sum of three distinct terms
\be\label{eq:Kab}
\mathcal{K}_{\alpha\beta}(\mathbf{p}_1,\mathbf{p}_2)=\sum_{n=0}^{2} K^{(n)}_{\alpha\beta}(\mathbf{p}_1,\mathbf{p}_2).
\ee
The product of plane wave coefficients defines the reaction kernel
\bea\label{eq:K0}
\mathcal{K}^{(0)}_{\alpha\beta}(\mathbf{p}_1,\mathbf{p}_2)&=&D^{(PW)}_{\alpha}(\mathbf{p}_1)D^{(PW)}_{\beta}(\mathbf{p}_2)\\
&=&\delta(\mathbf{p}_2-\mathbf{k}_{\beta})\delta(\mathbf{p}_1-\mathbf{k}_{\alpha}).
\eea

By exploiting the properties of the Dirac delta distributions, we find the on-shell relations
\be
\mathcal{K}^{(0)}_{\alpha\beta}(\mathbf{p}_1,\mathbf{p}_2)=
\delta(\frac{1}{2}(\mathbf{p}_1+\mathbf{p}_2)-\mathbf{P}_{\alpha\beta})\delta(\mathbf{p}_1-\mathbf{p}_2-\mathbf{q}_{\alpha\beta}).
\ee
where $\mathbf{P}_{\alpha\beta}=\frac{1}{2}(\mathbf{k}_\alpha+\mathbf{k}_\beta)$ and $\mathbf{q}_{\alpha\beta}=\mathbf{k}_\alpha-\mathbf{k}_\beta$ denote the average channel three momentum and the three-momentum transfer of the reaction, respectively.

Thus, in~the plane wave limit, the momenta $\mathbf{p}_{1,2}$ are fixed unambiguously by the (invariant) momenta in the initial and the final channels as derived in Appendix \ref{app:BoxPW}.

The ISI/FSI contributions are contained in the remaining two terms, which are determined by the amplitudes of Equation~\eqref{eq:fafb}.
Two types of ISI/FSI distortion kernels are~found:
\bea
\mathcal{K}^{(1)}_{\alpha\beta}(\mathbf{p}_1,\mathbf{p}_2)&=&
-\left(\delta(\mathbf{\mathbf{p}}_1-\mathbf{k}_\alpha)f_\beta(\mathbf{p}_2)
+\delta(\mathbf{\mathbf{p}}_2-\mathbf{k}_\beta)f_\alpha(\mathbf{p}_1) \right),\\
\mathcal{K}^{(2)}_{\alpha\beta}(\mathbf{p}_1,\mathbf{p}_2)&=&
f_\alpha(\mathbf{p}_1)f_\beta(\mathbf{p}_2) .
\eea

The kernel $\mathcal{K}^{(1)}_{\alpha\beta}$ describes the distortion effects exerted on the reaction by one of channels, while the other channel is in the PW mode, i.e.,~ISI and FSI act separately. $\mathcal{K}^{(s)}_{\alpha\beta}$ accounts for the combined action of ISI and FSI.
In the momentum space approach, the MDCE reaction amplitude is understood as a superposition of essentially three interfering contributions of different origin and structure but of comparable magnitude:
\bea\label{eq:MDCE_DW}
\mathcal{M}^{(1)}_{\alpha\beta}(\mathbf{k}_\alpha,\mathbf{k}_\beta)&=&\mathcal{F}_{\alpha\beta}(\mathbf{k}_\alpha,\mathbf{k}_\beta)\\
&-&\int d^3p\left(f_\beta(\mathbf{p})\mathcal{F}_{\alpha\beta}(\mathbf{k}_\alpha,\mathbf{p})+
                f_\alpha(\mathbf{p})\mathcal{F}_{\alpha\beta}(\mathbf{p},\mathbf{k}_\beta) \right)\nonumber\\
&+&\int d^3p_1\int d^3p_2f_\alpha(\mathbf{p}_1)
\mathcal{F}_{\alpha\beta}(\mathbf{p}_1,\mathbf{p}_2)f_\beta(\mathbf{p}_2)\nonumber.
\eea

The PW contribution reflects the bare nuclear transition matrix element before ISI/FSI renormalization. The~contributions in the second line introduce ISI in the initial channel while the exit channel is in PW mode, and FSI in the exit channel while the initial channel remains in PW mode. In~the term of the last line, ISI and FSI act in both channels~simultaneously.

\section{A Different View: ISI and FSI as Vertex~Renormalizations}\label{sec:View}
A standard problem of nuclear many-body theory is to incorporate interactions from outside of the model space into the operators acting between the states in the limited model space. Formally, the~projection techniques going back to Feshbach~\cite{FeshbachDeShalit:1974nuc} provide first insight into the problem of induced interactions. Over~the years, nuclear many-body theory has developed powerful techniques on how to incorporate induced interactions as consistently as possible into all parts of the theory. Examples are many-body shell model studies of double beta decay as, for example,~in~\cite{Coraggio:2017bqn,Coraggio:2019lsz,Coraggio:2020iht,Coraggio:2020hwx,Jokiniemi:2021qqv,Ejiri:2022zdg,Kostensalo:2022lzk} and under slightly different aspects also in~\mbox{\cite{Civitarese:2014rza,Gambacurta:2020dhb,Jokiniemi:2023bes}}, regarding even neutrino effective masses by induced interaction from the coupling to axions~\cite{Civitarese:2023cqt}. For~example, a~widely used approach, introduced into DBD theory by \v{S}imkovic~et~al., is the Jastrow method which implements short-range correlations into matrix elements by a function, acting repulsively at small~distances.

The considerations which led to Equation~\eqref{eq:MDCE_DW} are in fact following the same theoretical rationality as in nuclear structure theory, however, as~will be seen, in a complementary manner. In~order to recognize the relationship, we recall that the model space of MDCE reactions includes the incoming and outgoing channel configurations, where the incoming nuclei are assumed to be in their ground states and the outgoing nuclei are assumed to be again in their ground states or in a well-identified excited state. In~addition, the spectrum of intermediate SCE configurations will contribute. However, the~intermediate states are acting mainly as a reservoir of unresolved spectroscopic strength, being responsible in the first place for generating the effective two-body interactions for transitions from the incoming nuclei to the emerging ejectiles. Thus, the~explicitly treated model space contains only an extremely small subset of states of the total $\{A\}\otimes \{A'\}$ configuration space. In~nuclear reaction theory, the respective optical potentials account for the induced interaction as far as they affect elastic scattering. Hence, to~a large extent, ISI and FSI correspond to induced interactions from the vast background of non-elastic channels. As~known from nuclear many-body theory, once effective interactions are important in one sector, they also affect all other sectors of the theory. In~particular, transition operators have to be renormalized in accordance with the renormalization scheme. In~the above cited works, the proper implementation of renormalization into all parts of the theory is a topic of central~importance.

Reconsidering under these aspects the MDCE reaction amplitude, we arrive at the conclusion that in Equation~\eqref{eq:MDCE_DW}, the distortion amplitudes $f_{\alpha,\beta}$ are playing exactly that role, namely, to renormalize the SCE vertices in agreement and consistently with the induced ion--ion initial and final state interactions. This is performed in a systematic manner starting from the bare matrix element, represented by the PW amplitude, then renormalizing one of the vertices but retaining the second vertex as a bare vertex, and~finally renormalizing both vertices simultaneously. Hence, ISI and FSI account for the proper renormalization of the DCE--nuclear matrix element (NME) under the conditions of a heavy ion nuclear~reaction.

While in the nuclear structure context, renormalizations typically refer to short-range effects, ISI/FSI renormalization, however, accounts for scales defined by the ion--ion self-energies, subsumed in the respective optical potentials. A~decisive role is played by elastic scattering amplitude as discussed in Appendix \ref{app:DW}. The~most relevant observable, however, is the total reaction cross section as the measure for the amount of probability flux leaving the elastic channel. The~redirected flux is absorbed into channels ranging from transfer channels, which are dominated by mean-field dynamics, and~channels where the nuclei are excited inelastically by soft vibrational excitations and giant resonances, eventually leading to fission or fusion, to~hard central collisions, possibly upending in the complete fragmentation of the incoming nuclei. Thus, renormalization by optical model interaction is of a genuine character by covering a broad range of nuclear modes and interactions from the soft to the hard scale. That mechanism is not specific for first-order DW reactions as considered here. As~discussed in~\cite{Lenske:2024dsc}, a similar renormalization scheme is also present in the second-order reactions double single charge exchange (DSCE) reaction. In~DSCE reactions, the matrix elements, however, are renormalized by second-order distortion~amplitudes.

\section{Transition Form Factors and Nuclear Matrix~Elements}\label{sec:TNME}
\unskip

\subsection{The MDCE Box~Diagram}
Diagrams of the topology of the MDCE graph in Figure~\ref{fig:MDCE} may be rare in nuclear physics. However, such planar box diagrams are encountered frequently in other fields of physics, from~electro-weak theory, e.g.,~\cite{Fleischer:2006ht}, to~QCD and hadron phenomenology, e.g.,~\cite{Nierste:2009wg}. The~physics behind the diagram of Figure~\ref{fig:MDCE} has, however, several peculiarities: we have to treat a process which extends over two complex nuclei in a state of relative motion, the~particles involved are of a complex many-body structure, and, as~an additional challenge, we have to account for strong initial and final state~interactions.

An important aspect of ISI and FSI is that the incoming and outgoing waves become (stationary) wave packets, as~is emphasized in Section~\ref{sec:ISIFSI} and in Appendix \ref{app:DW}. The~momentum distributions are centered at the respective physical on-shell momentum. The width and shape of the distribution are governed by the properties of the elastic ion--ion self-energies, described by optical potentials. The~depth of the imaginary potential plays a special role due to controlling the amount of flux absorption into other reaction channels in a \emph{never-come-back} manner. The~wave packet properties of the distorted waves induce a certain amount of \emph{off-shellness}, described by the ISI and FSI distortion coefficients. As~a result, the~ISI/FSI momentum distributions will be imprinted on the MDCE form factors, reaction amplitudes, and~cross~sections.

Following Appendix \ref{app:BoxPW}, the reaction will be described in the rest frame of the colliding nuclei.
In that frame, the~incoming and outgoing ions carry, under on-shell conditions, asymptotically the four-momenta $k_{A,A'}=(E_{A,A'},\mp \mathbf{k}_\alpha)^T$ and
$k_{B,B'}=(E_{B,B'},\mp \mathbf{k}_\beta)^T$, with~the on-shell energies $E_{A,A'}=\sqrt{M^2_{A,A'}+k^2_\alpha}$ and $E_{B,B'}=\sqrt{M^2_{B,B'}+k^2_\alpha}$, respectively. At~the mass shell, the~charged pions are described by four-momenta $p_{1,2}$, which in the rest frame are $p_{1}=(0,\mathbf{k}_{\alpha})^T$, $p_{2}=(E_A-E_B,\mathbf{k}_{\beta})^T$. Hence,  $p^2_{1}=-\mathbf{k}^2_{\alpha}\leq 0$ is a purely space-like four-vector. This is also the case for $p^2_{2}=Q^2_{\alpha\beta}-\mathbf{k}^2_{\beta}$, provided that the reaction Q-value $Q_{\alpha\beta}=E_A-E_B$ is not of an extraordinary large~value.

ISI and FSI introduce off-shell three-momentum distributions for the three-momenta $p_{1,2}$ appearing as variables in the MDCE reaction amplitude in Equation~\eqref{eq:Mab_momentum}. Their origin and the properties are discussed in Section~\ref{sec:ISIFSI}. A~significant consequence for the box diagram is that it has be to evaluated for whole set of momenta allowed by ISI/FSI. Since the uncertainty in momentum is a purely virtual effect as~further elucidated in Appendix \ref{app:BoxDW}, the~nuclear four-momenta are to be evaluated for $\mathbf{p}_{1,2}$  $k_{A,A'}=(E_{A,A'}(\mathbf{p}_1),\mp \mathbf{p}_1)^T$, $k_{B,B'}=(E_{B,B'}(\mathbf{p}_2),\pm \mathbf{p}_2)^T$   but still obey mass-shell conditions $k^2_{A,A'}=M^2_{A,A'}$ and $k^2_{B,B'}=M^2_{B,B'}$, respectively.
Hence,
scanning through the ISI/FSI-induced momentum distributions, the nuclear four-momenta are kept on the mass shell. The~
charged pions are described by four-momenta  $p_{1}=(0,\mathbf{p}_{1})^T$ and $p_{2}=(Q_{\alpha\beta},\mathbf{p}_{2})^T$. While $p_1$ remains space-like also in the off-shell region, $p_2$ is time-like for  $Q^2_{\alpha\beta}\geq \mathbf{p}^2_2$ and changes back to being space-like for larger values of $|\mathbf{p}_2|$.

Obviously, the~role of $p_1$ and $p_2$ may be exchanged. Thus, the~box diagram displayed in Figure~\ref{fig:MDCE} is of a generic character. When evaluating the transition form factor, this particular symmetry is taken into account by a multiplicity factor~2.

\subsection{Pion--Nucleus and Pion--Nucleon Kinematics and~Interactions}
The intermediate $\pi^0 + C$ and $\pi^0+C'$ systems are populated with the time-like four-momenta $k_1=p_1+k_A=(E_A(\mathbf{p}_1),\mathbf{0})^T)^T$ and $k_2=p_1-k_A=(-E_{A'}(\mathbf{p}_1),\mathbf{0})^T$, respectively.
Thus, $s_{\gamma,\gamma'}(\mathbf{p}_1)=E^2_{A,A'}(\mathbf{p}_1)$ are the energies available for the intermediate systems. The~corresponding on-shell relative momenta are
\be
k^2_{\gamma}=\frac{1}{4s_{\gamma}}\left((s_{\gamma}-(M^*_C+m_\pi)^2)(s_{\gamma}-(M^*_C-m_\pi)^2) \right)\nonumber
\ee
\be
k^2_{\gamma'}=\frac{1}{4s_{\gamma'}}\left((s_{\gamma'}-(M^*_{C'}+m_\pi)^2)(s_{\gamma'}-(M^*_{C'}-m_\pi)^2)\right)\nonumber,
\ee
respectively.
At the on-shell points, we find $k_1=k_\pi+k_C$ leading to $\mathbf{k}_\pi+\mathbf{k}_C=\mathbf{0}$ and accordingly $k_2=k'_\pi+k_C'$ with  $\mathbf{k}'_\pi+\mathbf{k}_{C'}=\mathbf{0}$.

As it is evident from Equation~\eqref{eq:TpiN}, the~longitudinal and the transversal operators depend on the three-momenta of the incoming and outgoing pions. For~the $A\to C$ vertex, these are $\mathbf{p}_1$ and $\mathbf{k}_\pi$, while at the $C\to B$ vertex, these are $\mathbf{k}_\pi$ and $\mathbf{p}_2$. The~$A'\to C'$ vertex is determined by $-\mathbf{p}_1$ and $-\mathbf{k}'_\pi$, and at the $C'\to B'$ vertex, these are $-\mathbf{k}_\pi$ and $-\mathbf{p}_2$. Since the momenta occur always in binomials, the~minus signs are irrelevant. The~deeper reason for this ambiguity is the symmetry of the box diagram under the exchange $\{k_1,p_1\} \leftrightarrow \{-k_2,-p_2\}$.

The strength of the form factors $T_{0,1,2}$, acting as effective coupling constants, are determined by the energy available in the pion--nucleon systems. In~Appendix \ref{app:T012}, this is accomplished by the \emph{mean energy approach} which allows to derive the pion--nucleon (pseudo) kinematics from the intermediate channels, containing explicitly a pion in the s channel. The~proper energy per nucleon $\sqrt{s_{\pi N}}=\sqrt{s_{\gamma}}/A2$ is used to define the invariant relative momentum and the energies in the pion--nucleon system. By~Equation~\eqref{eq:Tlab} in Appendix~\ref{app:BoxPW}, we obtain the equivalent energy in the laboratory frame, which is used in some of the figures shown~below.

\subsection{The Intermediate~Propagator}
The intermediate channels deserve closer considerations because of their internal structure given by a $\pi^0$ and a SCE-excited nucleus. As~an example, we investigate the $\pi^0+C$ systems. The~intermediate propagator is expanded into $|\gamma=[\pi^0\otimes C]$ configurations:
\be
\mathcal{G}_{A}(k_1)=\sum_{\gamma}\int\frac{d^3k}{(2\pi)^3}|\gamma, \phi^{(+)}_\mathbf{k}\ran g^{(+)}_\gamma(\mathbf{k}|k_1)\lan \widetilde{\phi}^{(+)}_\mathbf{k},\gamma|,
\ee
where the relative motion of the  pion--nucleus system is described by the wave functions $\phi^{(+)}_\mathbf{k}$ and the dual state  $\widetilde{\phi}^{(+)}_\mathbf{k}$, obeying
$\lan \widetilde{\phi}^{(+)}_\mathbf{q}|\phi^{(+)}_\mathbf{k}\ran =(2\pi)^3\delta(\mathbf{k}-\mathbf{q})$.

In the intermediate channels, the~neutral pions will interact with the co-propagating nuclei by their own version of optical potentials. Pion optical potentials are discussed and applied widely in the literature, e.g.,~\cite{Ericson:1966fm,Oset:1981ih,Doring:2007qi,Bender:2009cj,Lukyanov:2016qrs}.
For our purpose, we neglect those interaction and replace
the pion--nucleus wave functions by plane waves, $\lan \mathbf{r}|\varphi^{(+)}_\gamma(\mathbf{k})\ran=\varphi^{(+)}_\gamma(\mathbf{k},\mathbf{r})\approx e^{i\mathbf{k}\cdot \mathbf{r}}$ and correspondingly
$\widetilde{\varphi}^{(+)}_\gamma(\mathbf{k},\mathbf{r})\approx e^{-i\mathbf{k}\cdot \mathbf{r}}$.

The on-shell energy for the $\pi +C$ system $s_{\pi C}=k^2_1=M^2_A+\mathbf{p}^2_1$ is defined by $k_1$. The~related invariant on-shell three-momentum is $k^2_\gamma=(s_{\pi C}-(m_\pi + M^*_C)^2)(s_{\pi C}-(m_\pi - M^*_C)^2)/(4s_{\pi C})$. The~reduced \emph{retarded} channel propagator becomes
\be\label{eq:gc_red}
g^{(+)}_\gamma(\mathbf{k}|\mathbf{p}_1)=
\frac{1}{E_{A}(\mathbf{p}_1)-E_\pi(\mathbf{k})-E_C(\mathbf{k})+i\eta}+
\frac{1}{E_{A}(\mathbf{p }_1)+E_\pi(\mathbf{k})+E_C(\mathbf{k})+i\eta},
\ee
and we note that in the ion--ion rest frame, the $k_1$ dependence is in fact a dependence on the three-momentum $\mathbf{p}^2_1$.

The Cauchy formula allows to decompose the propagator into a principal value part $\mathcal{P}$ and a pole term:
\be\label{eq:gc_Cauchy}
g^{(+)}_\gamma(\mathbf{k}|\mathbf{p}_1)=2E_A(\mathbf{p}_1) \frac{\mathcal{P}}{E^2_{A}(\mathbf{p}_1)-(E_\pi(\mathbf{k})+E_C(\mathbf{k}))^2}
-i\pi\mu_{\pi C}(k_\gamma)\frac{1}{k_\gamma  }\delta(k-k_\gamma)
\ee
where $\mu_{\pi C}(k_\gamma)=E_\pi(k_\gamma)E_C(k_\gamma)/(E_\pi(k_\gamma)+E_C(k_\gamma))\sim E_\pi(k_\gamma)$ is the reduced energy of the $\pi +C $ system at the pole position $k=k_\gamma$.

The delta distribution of the pole part contributes only if $k_\gamma$ is real valued, i.e.,~$k^2_\gamma >0$ is positive. From~the definition of $k_\gamma$, we find that the latter condition is fulfilled if the external momentum obeys $\mathbf{p}^2_1> (m_\pi +M^*_C)^2-M^2_A\sim 2M_A(m_\pi+\varepsilon_C)$, where $\varepsilon_C= M^*_C-M_A$ is the excitation energy of the SCE daughter nucleus $C$. The~three-momentum $|\mathbf{p}|_1$ must be large enough to compensate for the pion rest mass appearing in the intermediate channel and the nuclear excitation energies. This constraint establishes an important difference from on-shell pion--DCE reactions,  which obviously contain an incoming charged pion on the mass shell. In~MDCE reactions, however, the~charged pions are in purely virtual states, thus not contributing with their rest mass to the energy balance of the reaction. That missing energy---plus the excitation energy contained in $M^*_C$---must be compensated for solely by the momenta exchanged between the ions.
The same rues apply to the reaction $A'\to B'$ under the proper conditions and constraints belonging to the $A'$ system.

\subsection{The Nuclear Transition Matrix~Elements}
In channel representation, the~MDCE transition form factor is given by
\be\label{eq:WAB}
\mathcal{W}_{AB}(\mathbf{p}_1,\mathbf{p}_2)= -\sum_C \int \frac{d^3k}{(2\pi)^3}
\mathcal{M}_{BC}(\mathbf{p}_2,\mathbf{k})g^{(+)}_{\gamma}(\mathbf{k}|\mathbf{p}_1)\mathcal{M}_{CA}(\mathbf{p}_1,\mathbf{k}).
\ee
The summation extends over the ground state and the excited states of the $C(Z\pm 1,N\mp 1)$ SCE daughter nucleus, underlining again that in a DCE reaction, the charge number partition is changed, but the nucleon number partition is~conserved.

The two charge-converting processes are described by SCE-type nuclear matrix elements,
\bea\label{eq:MBCMCA}
\mathcal{M}_{CA}(\mathbf{p}_1,\mathbf{k})&=&
\lan C|e^{i(\mathbf{p}_1-\mathbf{k})\cdot \mathbf{r}_1}\mathcal{T}_{\pi N}(\mathbf{k},\mathbf{p}_1|\bm{\sigma}_1)|A\ran ,\\
\mathcal{M}_{BC}(\mathbf{p}_2,\mathbf{k})&=&
\lan B|e^{-i(\mathbf{p}_2-\mathbf{k})\cdot \mathbf{r}_2} \mathcal{T}_{\pi N}(\mathbf{k},\mathbf{p}_2|\bm{\sigma}_3)|C\ran .
\eea

According to Equation~\eqref{eq:TpiN}, these matrix elements are given by a superposition of three terms. We denote the isospin wave functions of the pions by their charge states, $\{\pi^0,\pi^\pm\}$ and introduce the isospin matrix elements $\mathcal{I}(0,\pm)=\lan\pi^0|T_\mp|\pi^\pm\ran=\mathcal{I}^\dag(\mp,0)=\sqrt{2}$. With~the spin--scalar ($S=0$) and the spin--vector ($S=1$) nuclear matrix elements
\be
M^{(0)}_{CA}(\mathbf{p}_1,\mathbf{k})=
\lan C|e^{i(\mathbf{p}_1-\mathbf{k})\cdot \mathbf{r}_1}\tau^{\pm}|A\ran\quad ; \quad
\mathbf{M}^{(1)}_{CA}(\mathbf{p}_1,\mathbf{k})=
\lan C|e^{i(\mathbf{p}_1-\mathbf{k})\cdot \mathbf{r}_1}\bm{\sigma}\tau^{\pm}|A\ran
\ee
and considering that the form factors depend on the invariant pion--nucleon energy $s_{\pi N}=(p_\pi+p_N)^2$,
we find
\bea
&&\mathcal{M}_{CA}(\mathbf{p}_1,\mathbf{k})=\mathcal{I}(0,\pm)\\
&\times&\bigg(
\left(T_0(s_{\pi N})+
T_1(s_{\pi N})\frac{\mathbf{k}\cdot\mathbf{p}_1}{m^2_\pi}\right)M^{(0)}_{CA}(\mathbf{p}_1,\mathbf{k})
+T_2(s_{\pi N})\frac{\mathbf{k}\times\mathbf{p}_1}{m^2_\pi}
\cdot \mathbf{M}^{(1)}_{CA}(\mathbf{p}_1,\mathbf{k})\nonumber
\bigg).
\eea

Correspondingly, the~second SCE matrix element is
\bea
&&\mathcal{M}_{BC}(\mathbf{p}_2,\mathbf{k})=\mathcal{I}(\pm,0)\\
&\times&\bigg(
\left(T_0(s_{\pi N})+
T_1(s_{\pi N})\frac{\mathbf{p}_2\cdot \mathbf{k}}{m^2_\pi}\right)M^{(0)}_{BC}(\mathbf{p}_2,\mathbf{k})
+T_2(s_{\pi N})\frac{\mathbf{p}_2\times\mathbf{k}}{m^2_\pi}
\cdot \mathbf{M}^{(1)}_{BC}(\mathbf{p}_2,\mathbf{k})\nonumber
\bigg).
\eea
The matrix elements $\mathcal{M}_{C'A'}$ and $\mathcal{M}_{B'C'}$ are defined~accordingly. As mentioned before, pion--nucleus elastic interactions are neglected.

\subsection{Pion Mass as a Scale Separator and Closure~Approximation}\label{ssec:Closure}
The principal value part of the propagator, Equation~\eqref{eq:gc_Cauchy}, is worth considering in more detail. Obviously, the~theoretical and numerical efforts will be reduced drastically if the propagator is independent of the quantum numbers of the intermediate states $C$ and $C'$, respectively. Under~such conditions, the~summation over the spectrum of intermediate SCE configurations could be performed by exploiting the completeness relations for each multipolarity, thus applying closure. This  can be achieved in two~ways.

First, we can approach the problem as in~\cite{Lenske:2024dsc}, namely, we replace the excitation energies of $C$ and $C'$ by an auxiliary state-independent average excitation energy $\varepsilon_{C,C'}\mapsto \omega_{\gamma,\gamma'}$. As~a result, we obtain that only the ground state masses $M_{C,C'}$ in the SCE channels are left as channel indicators, which are uncritical because the $(Z\pm 1,N\mp 1)$ nuclei are unique and well defined. In~that approximation, the~propagator, for example,~in the $A$ system, becomes
\bea\label{eq:gc_bar}
&&g^{(+)}_\gamma(\mathbf{k}|\mathbf{p}_1)\approx \bar{g}^{(+)}_\gamma(\mathbf{k}|\mathbf{p}_1)=\\
&&\frac{1}{E_A(\mathbf{p}_1)-E^{(0)}_C(\mathbf{k})-E_\pi(\mathbf{k})-\omega_\gamma+i\eta}+
\frac{1}{E_A(\mathbf{p}_1)+E^{(0)}_C(\mathbf{k})+E_\pi(\mathbf{k})+\omega_\gamma+i\eta}\nonumber
\eea
where $E^{(0)}_C(\mathbf{k}=\sqrt{\mathbf{k}^2+M^2_C}$. A~meaningful criterion for the choice of the auxiliary energy is the pion energy, which serves as a scale separator. Thus, for~excitation energies less than the pion rest mass, we may safely replace $\varepsilon_{C,C'}$ by an average value $\omega_\gamma$. The~first neglected, next-to-leading-order terms are at least of the order $\mathcal{O}((\varepsilon_{C,C'}-\omega_{\gamma,\gamma'})/m_\pi)$. $\omega_\gamma$ and $\omega_{\gamma'}$ may be chosen separately in each nucleus and for each multipolarity $J^\pm$, which allows well-adopted adjustments to the spectral properties of the nuclei. Assuming that $\omega_\gamma$ is chosen as a global parameter which is not dependent on the multipolarity, we find the pion~potential
\be\label{eq:PionPot1}
\mathcal{U}_\pi(\mathbf{x}|\mathbf{p}_{1,2}\bm{\sigma}_{1,3})=
\int \frac{d^3k}{(2\pi)^3}
\mathcal{T}_{\pi N}(\mathbf{p}_2,\mathbf{k}|\bm{\sigma}_3)\bar{g}^{(+)}_\gamma(\mathbf{k}|\mathbf{p}_1)e^{i\mathbf{k}\cdot \mathbf{x}}
\mathcal{T}_{\pi N}(\mathbf{p}_1,\mathbf{k}|\bm{\sigma}_1).
\ee
where $\mathbf{x}=\mathbf{r}_1-\mathbf{r}_2$ is the distance between the two protons or neutrons, respectively, participating in the MDCE transition. The~integrals define, in fact, monadic and dyadic tensors, which is seen by expressing the momentum vectors in the basis of spherical unit vectors.  Details and the resulting formalism are discussed in Appendix \ref{app:PiPot}.

Second, we may use even a more drastic simplification. From~the energies involved, it is found that the nuclear energies may indeed be replaced in first approximation by the rest masses $E_A\sim M_A$, $E_C\sim M_C+\varepsilon_C$. Using in addition $M_A+M_C+\varepsilon_C\sim 2M_A$ and neglecting terms of order $E_\pi/(2M_A)$, we obtain
\be
g^{(+)}_\gamma(\mathbf{k}|\mathbf{p}_1)\approx -\frac{m_\pi-\varepsilon_C}{m^2_\pi+k^2-\varepsilon^2_C}\sim
\frac{m_\pi}{m^2_\pi+k^2}.
\ee
Thus, for~not-too-large momenta and moderate excitation energies, $\varepsilon_C\ll m_\pi$, in~leading order, the propagator becomes independent of all quantum numbers of the intermediate systems, which allows to evaluate the transition form factors in closure approximation. Under~those conditions, we obtain an effective isotensor two-body interaction of the second order in $T_{\pi N}$:
\be \label{eq:PionPot2}
\mathcal{U}_\pi(\mathbf{x}|\mathbf{p}_{1,2}\bm{\sigma}_{1,3})=-m_\pi\int \frac{d^3k}{(2\pi)^3}
\mathcal{T}_{\pi N}(\mathbf{p}_2,\mathbf{k}|\bm{\sigma}_3)\frac{e^{i\mathbf{k}\cdot \mathbf{x}}}{m^2_{\pi}+k^2}
\mathcal{T}_{\pi N}(\mathbf{p}_1,\mathbf{k}|\bm{\sigma}_1).
\ee
Equation~\eqref{eq:PionPot2} also shows that the s-channel $\pi^0$ exchange induces a dynamical short range correlation between two nucleons of the same kind, connecting a pair of  particle--hole SCE transitions, either of $np^{-1}$ or $pn^{-1}$ type. The~same scenario is found in the $A'$ system.

In the closure approximation of the second kind, the TME is obtained as
\be \label{eq:WAB_Clos}
\mathcal{W}_{AB}(\mathbf{p}_1,\mathbf{p}_2)=
\lan \mathcal{I}^{(\pi)}_{2,\mp 2}\ran\lan B|e^{-i\mathbf{p}_2\cdot \mathbf{r}_2}\mathcal{U}_\pi(\mathbf{x}|\mathbf{p}_{1,2}\bm{\sigma}_{1,3})e^{i\mathbf{p}_1\cdot \mathbf{r}_1}
\mathcal{I}^{(N)}_{2\pm 2}|A\ran ,
\ee
including the (expectation value of the) pionic and the nucleonic rank-2 isotensors
$\mathcal{I}^{(N)}_{2\pm 2}=\left[\bm{\tau}_1\otimes \bm{\tau}_2\right]_{2\pm 2}$ and
$\mathcal{I}^{(\pi)}_{2\mp 2}=\left[\mathbf{T}_1\otimes \mathbf{T}_2\right]_{2\mp 2}$, respectively.
In this form, it is recognized immediately that
the MDCE pion potentials are two-body operators enforcing complementary $n^{\mp 2}p^{\pm 2}$ transitions in the interacting nuclei, while conserving the total charge of the projectile--target~system.

\section{The Pion--Nucleon Partial Wave Amplitudes and the Isovector~T Matrix}\label{sec:TpiN}
\unskip
\subsection{Pion--Nucleon Interactions and Scattering~Amplitudes}
\begin{figure}[H]
\includegraphics[width = 6.5cm]{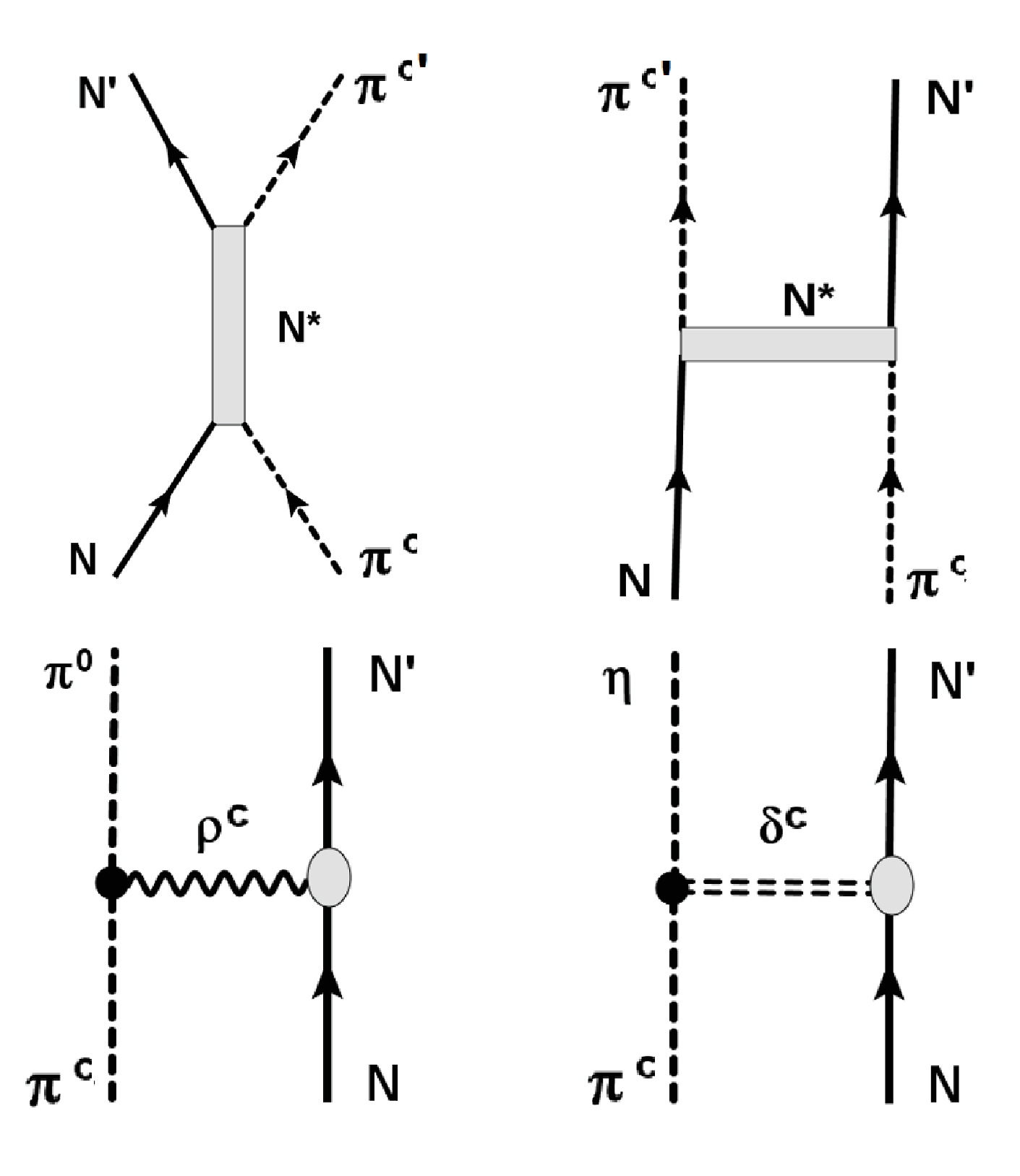}
\caption{Pion--nucleon isovector interactions either by the formation of $N^*$ resonances in elastic s-channel scattering (upper row, left) or by t-channel exchange (upper row, right) and t-channel vector--isovector $\rho$ meson (lower row, left) and scalar--isovector $\delta/a_0(980)$ meson exchange (lower row, right).}
\label{fig:piNSCE_st}

\end{figure}

As depicted in Figure~\ref{fig:piNSCE_st}, pion--nucleon scattering is determined by the formation of $N^*$ resonances in the s-channel and t-channel meson exchange. Meson exchange will contribute to all pion--nucleon partial waves as an omnipresent, smooth background contribution. The~formation of elastic resonances, however, is an interaction mode which depends critically on the partial wave.
The most prominent example of a $N^*$ state is the $\Delta(1232)$ resonance at centroid energy $M=1232$~MeV and with $\Gamma=120$~MeV. In~spectroscopic notation, $L_{2I2J}(M)$, with~orbital angular momentum $L=S,P,D\ldots$ in the $\pi N$ system, isospin $I=\frac{1}{2},\frac{3}{2}$, and~total angular momentum $J=\frac{1}{2},\frac{3}{2},\frac{5}{2}\ldots$, the~Delta resonance is denoted by $P_{33}(1232)$, hence indicating a P-wave resonance with stretched isospin and spin--orbital coupling. The~next higher resonance is the Roper resonance, $P_{11}(1440)$. At~higher energies, up to about 2.5~GeV S-, D-, and F-wave resonances have been confirmed as being listed and regularly updated by the Particle Data Group~\cite{PDG:2022}.
Meson--nucleon spectroscopy is an intensively studied field. Among~several other approaches, the Giessen coupled channels model was successfully used in the past to describe the photo production of mesons on the nucleon and meson--nucleon dynamics, see~\cite{Lenske:2018bgr} for an~overview.

\vspace{-6pt}

The diagrams of Figure~\ref{fig:piNSCE_st} together with a few other graphs and appropriately chosen form factors, see~\cite{Lenske:2018bgr}, define the bare pion--nucleon $\pi N$ interactions $\mathcal{V}_{\pi N}$. Since we are dealing with nuclear interactions of considerable strength and additional resonant enhancements, the~scattering series must be summed to all orders. This is achieved by the Lippmann--Schwinger integral equation~\cite{GoldbergerWatson:1964} for the T matrix, which in non-relativistic notation is:
\be
\mathcal{T}_{\pi N}(\mathbf{k},\mathbf{k}')=\mathcal{V}_{\pi N}(\mathbf{k},\mathbf{k}')+\int \frac{d^3q}{(2\pi)^3}\mathcal{V}_{\pi N}(\mathbf{k},\mathbf{q})G_{\pi N}(s|q)\mathrm{T}_{\pi N}(\mathbf{q},\mathbf{k}'),
\ee
to be solved numerically as a set of coupled integral equations. The~essence of the T-matrix formalism is to shift dynamics from wave functions to the interaction operator~\cite{GoldbergerWatson:1964,Joachain:1975} such that $\mathcal{T}_{\pi N}(\mathbf{k},\mathbf{k}')$ is defined as the plane wave matrix element of the correlated pion--nucleon scattering operator. Thus, in~matrix elements, $\mathcal{T}_{\pi N}(\mathbf{k},\mathbf{k}')$ always has to be combined with the incoming and outgoing pion--nucleon plane waves $e^{i\mathbf{k}'\cdot \mathbf{x}}$ and $e^{-i\mathbf{k}\cdot \mathbf{x}}$, respectively, where $\mathbf{x}=\mathbf{r}_N-\mathbf{r}_\pi$. It is worth mentioning that the earlier pion--DCE studies tried to describe  pion--nucleon interactions in the isovector channel in a perturbative approach, focusing on the $\Delta(1232)$ resonance. Such a reductive approach is not supported by our results. An~alternative approach utilizing an effective potential is presented in the next~section.

\begin{figure}[H]

\includegraphics[width = 10.5cm]{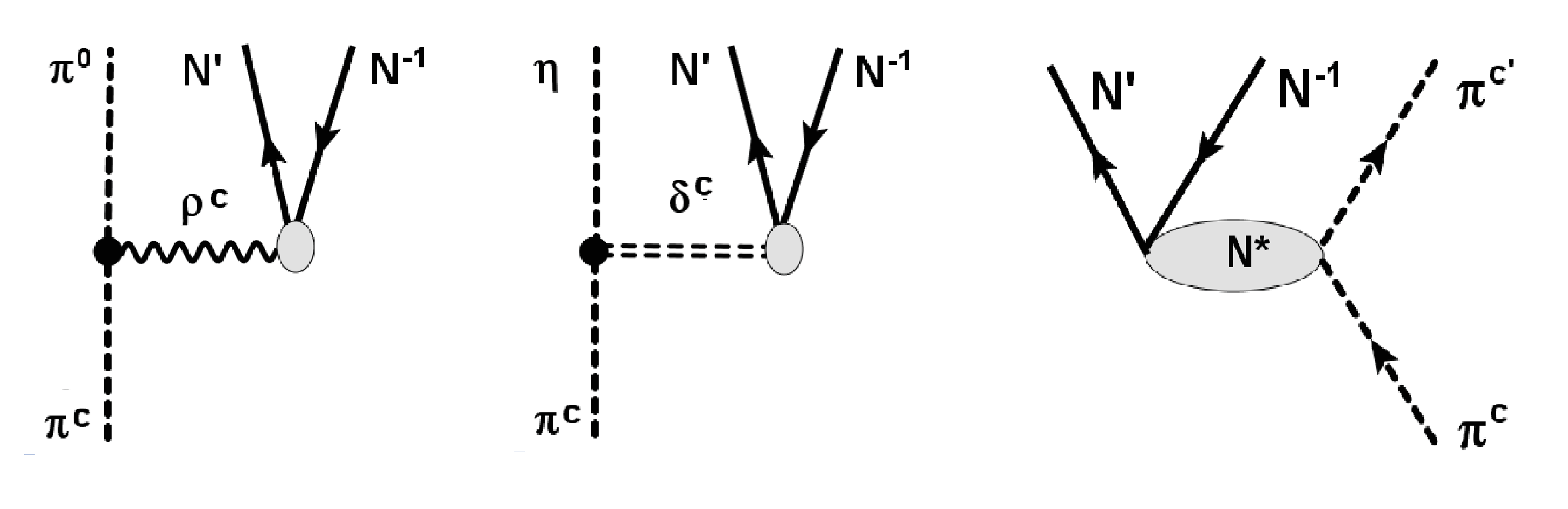}
\caption{Excitation
 of $np^{-1}$ or $pn^{-1}$ SCE particle--hole configurations by pion--nucleon isovector interactions through meson exchange (left and center) and formation and decay of $N^*$ resonances (right). The~$\pi^\pm\to \eta$ conversion via delta--meson exchange shown in the center indicates that the intermediate meson could also be an $\eta(540)$ meson. }
\label{fig:piNSCE_ph}
\end{figure}

For pionic SCE and DCE reactions, the focus is on the interactions of mesons with nucleons immersed in matter. As~a result, $N^*N^{-1}$ particle--hole configurations are excited as depicted in Figure~\ref{fig:piNSCE_ph}. In~a different context, such a scenario is discussed in detail in~\cite{Lenske:2023mis}. Since the $N^*$ particle state is unstable and finally decaying by strong interactions into $N'N^{-1}$ states under emission of a meson, in~our case, the~decay leads to an outgoing neutral or charged~pion.

While pion--nucleon scattering trivially proceeds on the basis of the definite partial of well-defined orbital and total angular momenta, $N^*N^{-1}$ dynamics is determined by the full pion--nucleon scattering amplitude, summed over partial waves. Hence, the~vertices of Figure \ref{fig:piNSCE_st} being active in $\pi N$ scattering and the ones of Figure~\ref{fig:piNSCE_ph} describing $\pi^{c'}N^*N^{-1}\pi^c$, $c,c'=0,\pm 1$ processes, are quite different as will be seen in the~following.

\vspace{-4pt}

\unskip

\subsection{Pion--Nucleon Potential Model for the Scattering~Amplitudes}
For the present purpose, a full-scale coupled channels calculation as in the Giessen model and comparable approaches is of little sense. Here, we are not interested in a detailed spectroscopic study of $N^*$ states and their excitation and decay by coupled meson--nucleon channels. Rather, our interest is specifically focused on the isovector pion--nucleon T matrix as an effective t-channel pion--$N'N^{-1}$ interaction. A~meaningful approach is to use an effective optical potential (OP) model, where the self-energies from coupled channels dynamics are treated by complex dispersive optical potentials. Pion--nucleon potential models have been used before with surprising success in reproducing the spectral distributions, see ~\cite{Coronis:1981vf,McLeod:1984cu}. We account for the opening of nucleon--multipion decay channels by partial wave-dependent pion--nucleon optical potentials (OPs). The~imaginary parts are modeled according to the opening of nucleon--multimeson decay channels, which finally are the observable configurations. In~our r-space approach, the~best results were obtained by using Wood--Saxon form factors with very small diffusivities which are well approximated by step functions. The~parameters were adjusted to the partial wave cross sections of full-scale coupled channels (CC) calculations, where the latter were fitted to the available meson--nucleon~data.

\section{Numerical~Studies}\label{sec:Numerics}
\unskip
\subsection{Pion--Nucleon Partial Wave Cross~Sections}
\begin{figure}[H]

\includegraphics[width = 13.5cm]{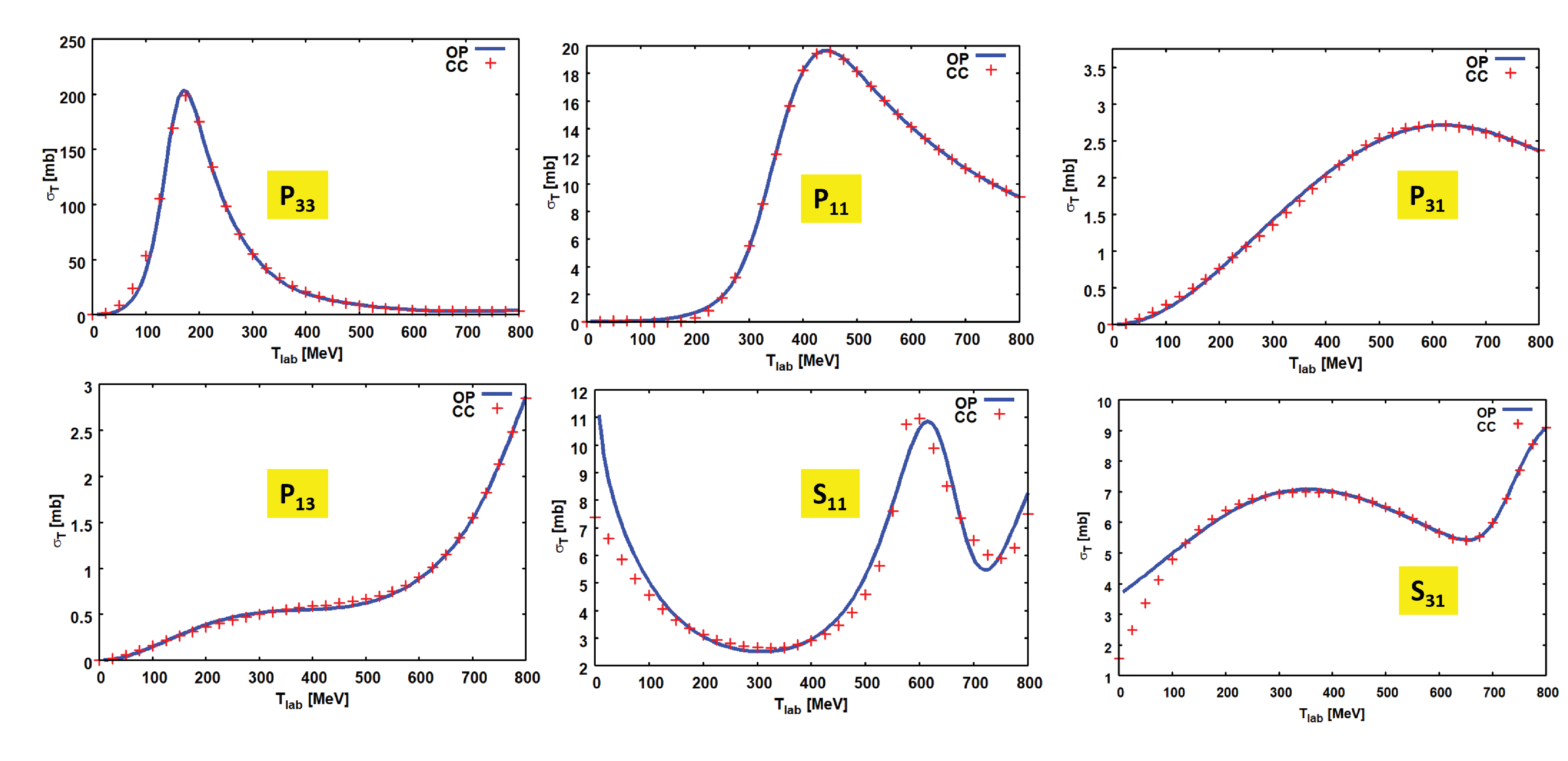}
\caption{Total $\pi^{-}+p$ partial wave cross sections for P waves and S waves. Cross sections obtained with the pion--nucleon optical potential model (OP) are compared to coupled channel results (CC). In~the upper row,
$P_{33}$ (left), $P_{11}$ (center),  and~$P_{31}$ (right) are shown, and in~the lower row, $P_{13}$, $S_{11}$, and~$S_{31}$
cross sections are displayed as functions of the pion energy in the laboratory frame.
The $P_{33}(1232)$ Delta resonance at $T_{lab}\sim 190$~MeV and the $P_{11}(1440)$ Roper  resonance  at $T_{lab}\sim 484$~MeV are well reproduced. The~low-energy tail of the $P_{13}(1710)$ resonance is visible at the end of the displayed $P_{13}$ cross section. The~$S_{11}(1520)$ resonance sticks out as a rather narrow structure at $T_{lab}\sim 620$~MeV.  Note the differences in scales. }
\label{fig:P33P11_sigT}

\end{figure}

Representative results illustrating the quality of the description for P- and S-wave total cross sections are shown in Figure~\ref{fig:P33P11_sigT}. The~reference data from explicit coupled channels calculations are surprisingly well described, especially in view of the simplicity of the potential approach.  In~detail, the~Delta and the Roper resonances are well reproduced as is the case for the $I=\frac{1}{2}$ and $I=\frac{3}{2}$ S-wave sector. In~the S-wave spectra, the highly disputed $S_{11}(1520)$ resonance  is most prominently visible as a rather narrow structure on a non-resonant background. Interestingly, the~$S_{11}(1520)$ peak is largely the result of interferences of a virtual s-channel state with the smooth t-channel background. A~long tome ago, the~same explanation was already obtained in coupled channels calculations~\cite{Shklyar:2012js}, and more recent studies have come to similar conclusions. The~present potentials model results may be taken as an interesting independent confirmation of the earlier CC results.
Overall, the~agreement of the present results with the CC-generated reference data is surprisingly good in view of the extremely simplified model. Larger deviations occur in the S-wave spectra. Close to the threshold, the~S-wave cross sections show some deficiencies, and deviations are seen in the $S_{31}$ channel also towards the highest considered energies. They are, however, of~minor importance for the present use in MDCE studies because $T_{\pi N}$ is dominated by P-wave~interactions.

\begin{figure}[H]

\includegraphics[width = 13.5cm]{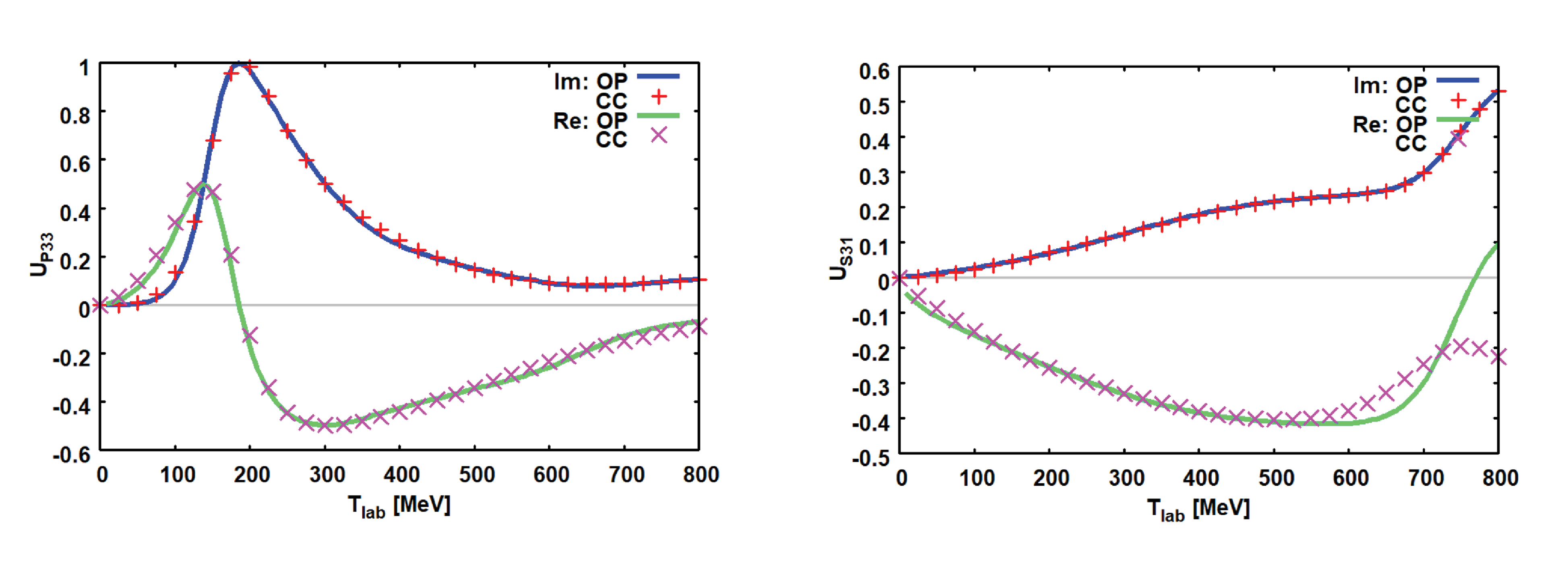}
\caption{$P_{33}$ (left)
 and $S_{31}$ (right) partial wave scattering amplitudes. Real and imaginary parts from the potential model (OP) are compared to coupled channels (CC) results. Note that the OP parameters are fitted to total cross sections, defined by the imaginary part of the scattering amplitude, \mbox{$\sigma_{tot}\sim Im(T)$}.  }
\label{fig:P33S31_amp}

\end{figure}

Partial wave total cross sections are defined by the imaginary parts of the scattering amplitudes,
$\sigma^{L_{2I2J}}_T\sim Im(T_{L_{2I2J}})$. Thus,
a first important test of the reliability of the model calculations is to compare real and imaginary parts of scattering amplitudes.
The agreement between OP and CC scattering amplitudes is very satisfying. An~example is shown in Figure~\ref{fig:P33S31_amp}, where the $P_{33}$ and the $S_{31}$ partial waves scattering amplitudes are compared to the corresponding CC~amplitudes.

The P-wave cross sections and scattering amplitudes are slightly better reproduced than the corresponding S-wave quantities. The~CC calculations show that the S-wave components, which are generally located at higher energies, are strongly affected by coupled channel dynamics. Physically, an~important source of CC effects are multimeson decay channels, either by direct $N^*\to n \pi$ decay, possibly passing through intermediate heavy mesons, or~sequentially by decay chains passing through lower lying resonances, e.g.,~$N^*\to \Delta(1232)+\pi\to N+2\pi$. Such details, of~course, have not been resolved in the present approach but are taken into account globally by the dispersive parts of the partial wave~potentials.

\vspace{-3pt}

\unskip

\subsection{Construction of the Pion--Nucleon~T Matrix}
In order to construct the pion--nucleon T matrix, Equation~\eqref{eq:TpiN}, we need to determine the three vertex form factors $T_{0,1,2}$. That goal is achieved by considering the partial wave structure of the T matrix and collecting terms of the proper multipolarity and dependencies on the nucleon spin. That task is well documented in the literature, e.g.,~\cite{Feshbach:2003} and reviewed briefly in Appendix \ref{app:T012}.

In the energy region of our interest, the~vertex form factors are obtained with sufficient accuracy by the two S-wave amplitudes $S_{11}$ and $S_{31}$ and the three P-wave contributions, $P_{11}$, $P_{31}$, and~$P_{33}$, respectively. Within~this basis, the~form factors are
\bea\label{eq:T012}
T_0(k_{\pi N})&=&\frac{1}{3}F(k_{\pi N})\left(U_{S11}(s_{\pi N})-U_{S31}(k_\gamma)\right) \\
T_1(k_{\pi N})&=&\frac{1}{3}F(k_{\pi N})\left(U_{P11}(k_{\pi N})+2U_{P13}(k_{\pi N})-U_{P31}(k_{\pi N})-2U_{P33}(k_{\pi N})\right) \\
T_2(k_{\pi N})&=&\frac{1}{3}F(k_{\pi N})\left(U_{P13}(k_{\pi N})-  U_{P11}(k_{\pi N})-U_{P33}(k_{\pi N})+U_{P31}(k_{\pi N})\right)
\eea
where $k_{\pi N}=k_{\pi N}(s_{\pi N})$ is the invariant pion--nucleon~three momentum.

The partial wave-scattering amplitudes are normalized to units of 1/MeV. By~means of the kinematical factor $F(k)=-4\pi/(2m_{\pi N}k)$, the T-matrix amplitudes are normalized to units of 1/MeV$^2$.
$m_{\pi N}$ is the pion--nucleon reduced mass and
$k=k(s_{\pi N})$ denotes the invariant relative pion--nucleon momentum. For~the numerical results displayed below, we follow, however, the~widely used practice to present the form factors as function of the pion kinetic energy in the laboratory frame, which is obtained by $T_{lab}=(s_{\pi N})-(m_\pi +M_N)^2)/(2M_N)$.

Although each of the (complex-valued) partial wave-scattering amplitudes varies considerably with energy as seen in Figure~\ref{fig:P33S31_amp}, their superpositions are much smoother functions
as Figure~\ref{fig:TP33} and Figure~\ref{fig:T012} confirm. By~multiplication with $(\hbar c)^3$, the units may be changed to MeVfm$^3$, which is a typical unit for volume integrals and momentum space form factors of~NN interactions.

\subsection{Extrapolation into the Subthreshold~Region}
\begin{figure}[H]

\includegraphics[width = 6.5cm]{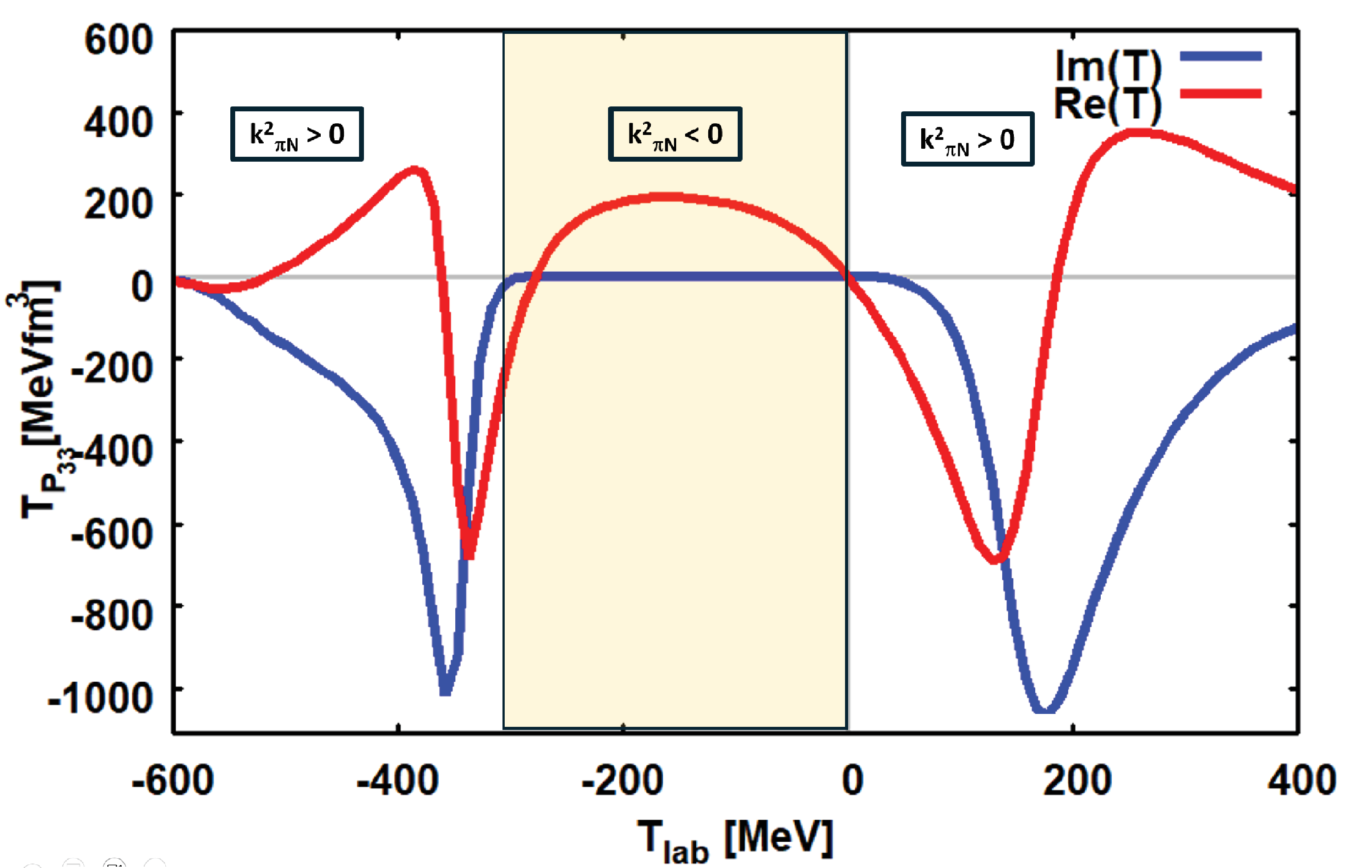}
\caption{The $P_{33}$ T-matrix in the kinematical regions relevant for MDCE reactions. Real and imaginary parts from the potential model (OP) are shown for energies above threshold, $s_{\pi N} >0$, $T_{Lab}>0$, and~the two subthreshold regions $s_{\pi N} <0$, $T_{Lab}<0$ and , $s_{\pi N} >0$, $T_{Lab}<0$.}\label{fig:TP33}

\end{figure}

The most important advantage of the OP approach for MDCE theory, however, is to have at hand a method which allows to extrapolate reliably and easily into the subthreshold region. As~illustrated in Figure~\ref{fig:TP33} for the $P_{33}$ partial wave, three different sheets are covered kinematically. The~sheets are distinguished by the values of the invariant relative
pion--nucleon momentum $k_{\pi}$:
\begin{itemize}
  \item In the physical region, $s_{\pi N}>(m_\pi+m_N)^2$ and the invariant momentum $k^2_{\pi} >0$ and $T_{Lab}>0$ are positive.
  \item In the interval $(m_\pi-m_N)^2<s_{\pi N}<(m_\pi+m_N)^2$, one finds $k^2_{\pi N} <0$ and $T_{Lab}<0$.
  \item If also $s_{\pi N}<(m_\pi-m_N)^2$, positive values of $k^2_{\pi N} >0$ are recovered but $T_{Lab}<0$ remains negative.
\end{itemize}
In Figure~\ref{fig:TP33}, it is seen that the T matrix changes in  a characteristic manner: real and imaginary parts are non-vanishing in the physical region while in the first subthreshold sheet, the imaginary parts vanish but~recover as soon as the second subthreshold sheet is~entered.

Because of the intrinsic momentum spread introduced by ISI and FSI, in~a heavy ion MDCE reaction, in principle, all three kinematical sheets will be visited while propagating through the intermediate s-channel pion--nucleon systems. In~other words, ISI and FSI lead effectively to a sampling over the distribution of MDCE box diagrams of different kinematical and dynamical~content.

\begin{figure}[H]

\includegraphics[width = 13.5cm]{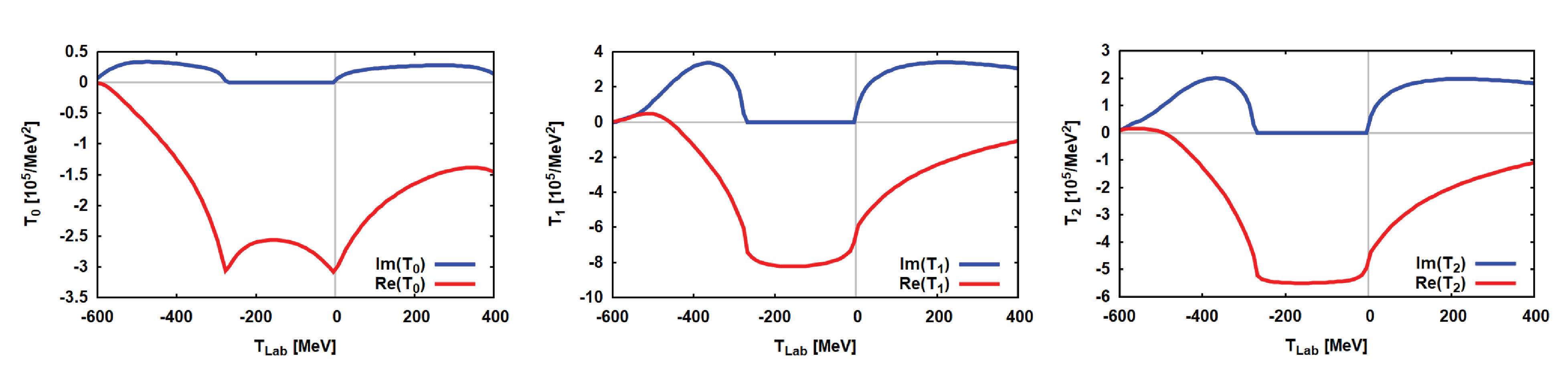}
\caption{The pion-nucleon
 vertex form factors $T_0$ (left), $T_1$ (center), and~$T_2$ (right) are shown as functions of the pion energy in the laboratory frame.  The~imaginary parts of $T_k$ vanish in the physically inaccessible region, where the invariant Mandelstam energy $(m_\pi -M_N)^2<s_{\pi N}<(m_\pi +M_N)^2$ as demanded by the analytic properties of the~T matrix. }
\label{fig:T012}

\end{figure}

The vertex form factors $T_{0,1,2}$ are shown as functions of the pion energy in the laboratory system  in
Figure~\ref{fig:T012}. When traversing the boundaries between the kinematical sheets, the amplitudes develop cusps. In~the $T_0$  amplitude, defined by the S-wave scattering amplitudes, the~cusps are most pronounced, while they are washed out in the P-wave amplitudes $T_{1,2}$.
A closer inspection shows that the P-wave amplitudes are, in magnitude, about a factor of 1.5 to 2 times larger than $T_0$. That difference will be enhanced further in matrix elements by the fact that in $T_{\pi N}$, the P-wave terms scale by $p^2$ for SCE transitions and even by $p^4$ in DCE transitions. Thus, already from these considerations, we expect a prevalence of the momentum-dependent P-wave terms in a DCE~reaction.

\vspace{-6pt}

\subsection{Form Factors of the Pion~Potentials}
Since the pion--nucleon T matrix, Equation~\eqref{eq:TpiN}, consists of three terms, the~pion potential $\mathcal{U}_\pi$, Equation~\eqref{eq:PionPot1} or Equation~\eqref{eq:PionPot2}, respectively, is in general, in either version, a superposition of nine terms $U_{ij}(\mathbf{x}|\mathbf{p}_1,\mathbf{p}_2)$, $i,j=0,1,2$, which depend on the three-momenta $\mathbf{p}_1$ and $\mathbf{p}_2$. Likewise, because~of $\mathbf{p}_1-\mathbf{p}_2=\mathbf{q}_{\alpha\beta}$, we may choose one of the momenta and the three-momentum transfer $\mathbf{q}_{\alpha\beta}$ of the reaction and momentum~variables.

A simplification is obtained for vanishing total momentum transfer $|q_{\alpha\beta}|=0$, which implies the collinearity of the momenta, $\mathbf{p}_1 || \mathbf{p}_2$ and $|\mathbf{p}_1|=|\mathbf{p}_2|=p$.  Then, the~number of elements reduces to six independent scalar form factors $U_{ij}(\mathbf{x}|\mathbf{p})$, $i\leq j=0,1,2$.  Under~these conditions, we find the diagonal potentials
\bea\label{eq:Uii}
U_{00}(\mathbf{x}|\mathbf{p})&=&U^2_0(k_{\pi N})\int \frac{d^3k}{(2\pi)^3}g^{(+)}_\gamma(\mathbf{k}|\mathbf{p}_1)e^{i\mathbf{k}\cdot \mathbf{x}}\\
U_{11}(\mathbf{x}|\mathbf{p})&=&U^2_1(k_{\pi N})\frac{p^2}{3m^4_\pi}\int\frac{d^3k}{(2\pi)^3}g^{(+)}_\gamma(\mathbf{k}|\mathbf{p}_1)k^2\cos^2(\theta)e^{i\mathbf{k}\cdot \mathbf{x}}\\
U_{22}(\mathbf{x}|\mathbf{p})&=&U^2_2(k_{\pi N})\frac{p^2}{3m^4_\pi}\int\frac{d^3k}{(2\pi)^3}g^{(+)}_\gamma(\mathbf{k}|\mathbf{p}_1)k^2\sin^2(\theta)e^{i\mathbf{k}\cdot \mathbf{x}}
\eea
and three non-diagonal potentials
\bea\label{eq:Uij}
U_{01}(\mathbf{x}|\mathbf{p})&=&2U^2_0(k_{\pi N})\int \frac{d^3k}{(2\pi)^3}g^{(+)}_\gamma(\mathbf{k}|\mathbf{p}_1)e^{i\mathbf{k}\cdot \mathbf{x}}\\
U_{02}(\mathbf{x}|\mathbf{p})&=&2U^2_1(k_{\pi N})\frac{p^2}{3m^2_\pi}\int \frac{d^3k}{(2\pi)^3}g^{(+)}_\gamma(\mathbf{k}|\mathbf{p}_1)k^2\cos^2(\theta)e^{i\mathbf{k}\cdot \mathbf{x}}\\
U_{12}(\mathbf{x}|\mathbf{p})&=&2U^2_2(k_{\pi N})\frac{p^2}{3m^4_\pi}
\int \frac{d^3k}{(2\pi)^3}g^{(+)}_\gamma(\mathbf{k}|\mathbf{p}_1)k^2\sin(\theta)\cos(\theta)e^{i\mathbf{k}\cdot \mathbf{x}}.
\eea

For simplicity, the~potentials are evaluated numerically for the special case that $\mathbf{x}$ and $\mathbf{p}$ are collinear as well, and~$\mathbf{x}||\mathbf{p}$ implies $\mathbf{k}\cdot \mathbf{x}=kx\cos{(\theta)}$.

By expressing the sine and cosine functions in terms of Legendre polynomials or Legendre functions, respectively, the~angle integrations can be performed in closed form. The~momentum integrals are regularized by dipole form factors with cut-off $\Lambda = 1000$~MeV/c. The~resulting k integrals, given by products of ordinary or spherical Bessel functions and Legendre functions of the second
kind, all combined with powers of $k$, have to be evaluated numerically.  The~full propagator, Equation~\eqref{eq:gc_red}, is used. Excitation energies, however, are neglected, which is justified in view of the rather weak dependence on energies well below the pion rest~mass.

\begin{figure}[H]

\includegraphics[width = 13.5cm]{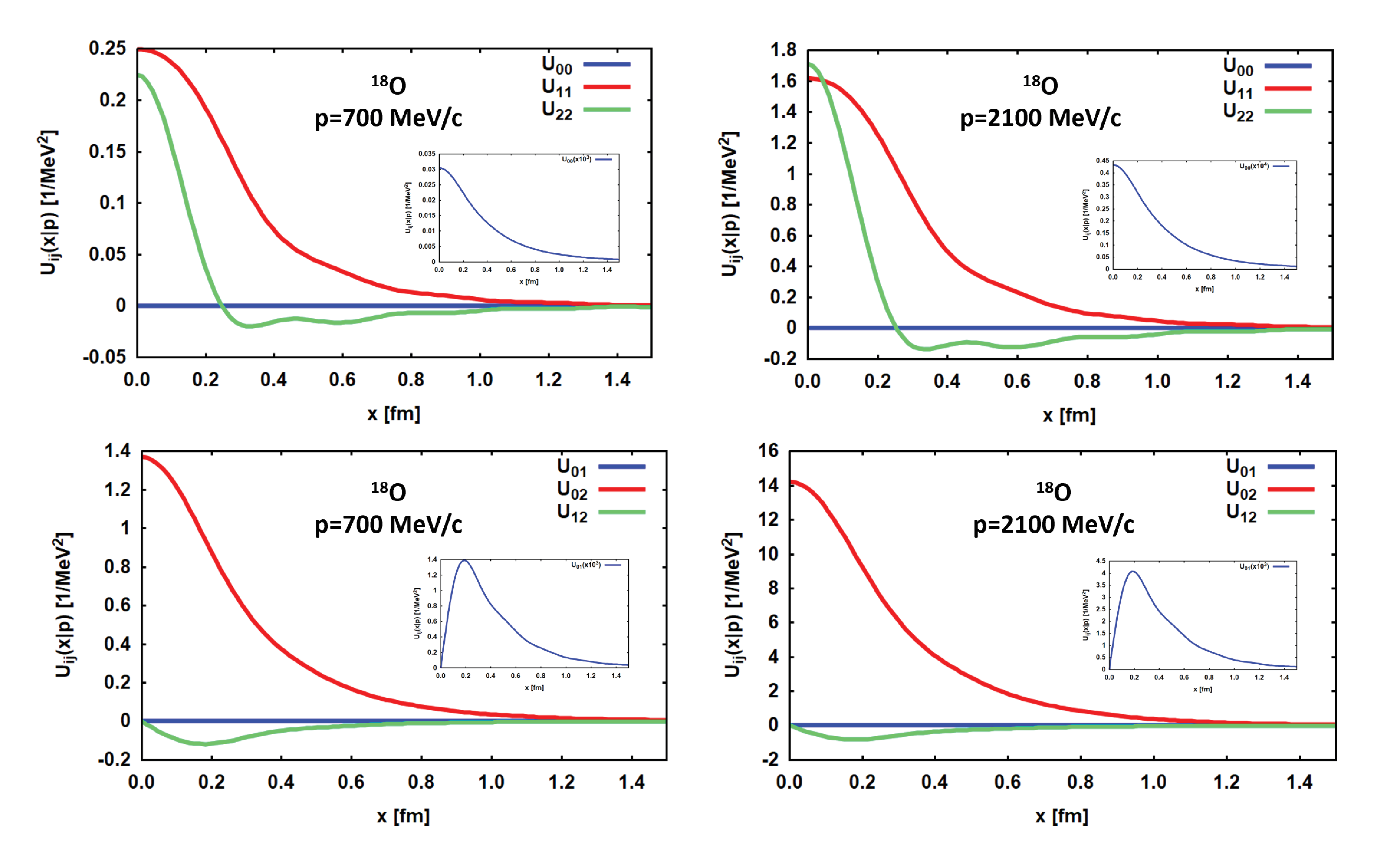}
\caption{Pion
 potentials in $^{18}$O for $p=700$~MeV/c (left column) and $p=2100$~MeV/c (right column) are shown as functions of the distance $x$ between the two nucleons participating in the DCE transition. Diagonal combinations of vertex operators as shown in the upper row. The~potentials for mixed operator combinations are displayed in the lower row. The~(scaled) potentials $U_{00}$ and $U_{01}$ are shown in the inserts. The~two momenta correspond to $p\sim \frac{1}{3}k_\alpha $ and $p\sim k_\alpha$, respectively, of~the DCE reaction induced by $^{18}$O$+{}^{40}$Ca at $T_{lab}=270$~MeV.}
\label{fig:Uij_18O}

\end{figure}

\begin{figure}[H]

\includegraphics[width = 13.5cm]{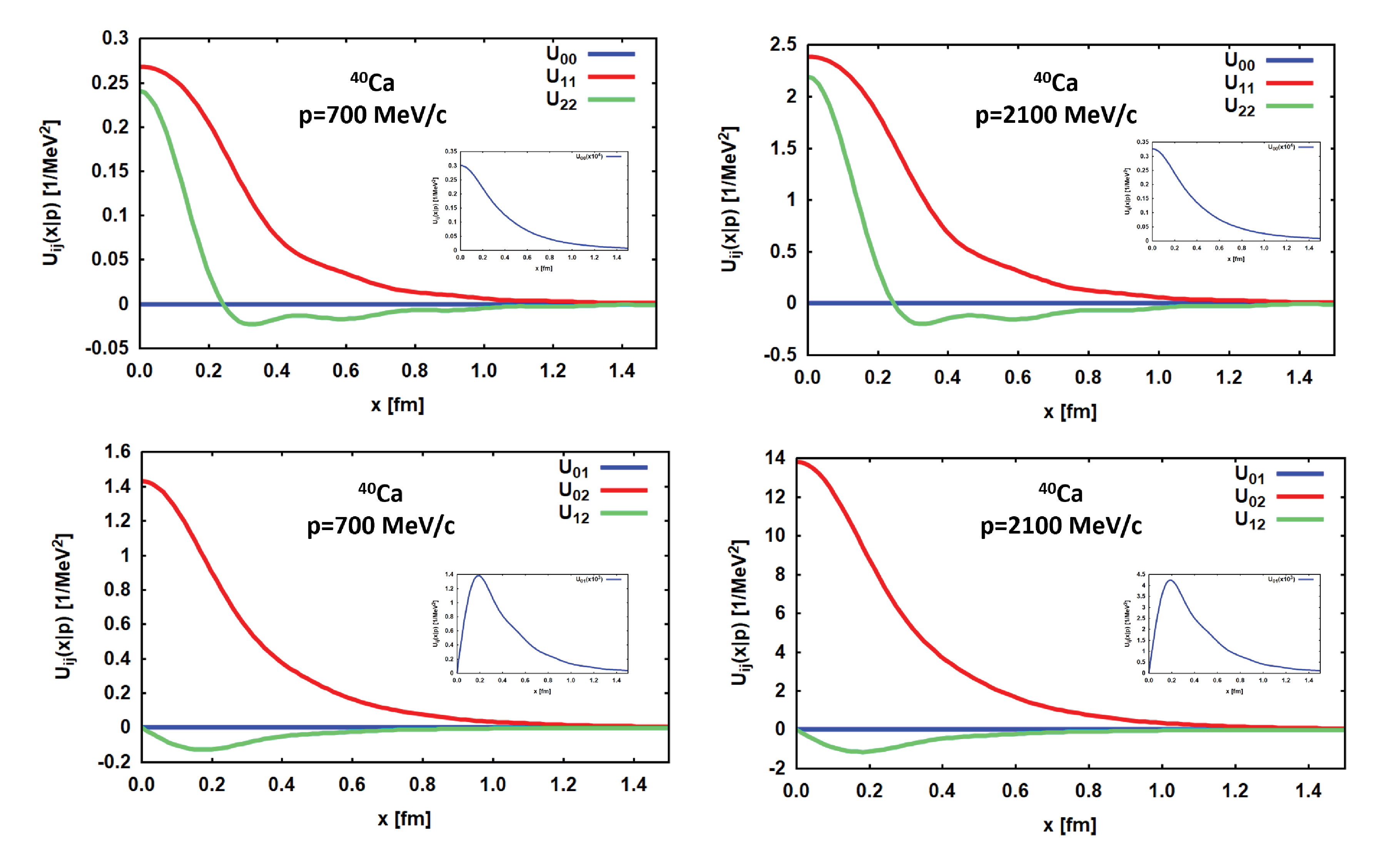}
\caption{Pion
 potentials in $^{40}$Ca for $p=700$~MeV/c (left column) and $p=2100$~MeV/c (right column) are shown as functions of the distance $x$ between the two nucleons participating in the DCE transition. Diagonal combinations of vertex operators as shown in the upper row, and the~potentials for mixed operator combinations are displayed in the lower row. The~(scaled) potentials $U_{00}$ and $U_{01}$ are shown in the inserts. The~two momenta correspond roughly to $p\sim \frac{1}{3}k_\alpha $ and $p\sim k_\alpha$, respectively, of~the DCE reaction induced by $^{18}$O$+{}^{40}$Ca at $T_{lab}=270$~MeV.}
\label{fig:Uij_40Ca}

\end{figure}

Typical results for the pion potentials $U_{ij}$ for the reaction  $^{18}$O$+{}^{40}$Ca at $T_{lab}=270$~MeV are shown in Figure~\ref{fig:Uij_18O} and in Figure~\ref{fig:Uij_40Ca}, respectively. As~discussed above, ISI and FSI favor momenta $\mathbf{p}_{1,2}$ which are centered around the on-shell momenta of the entrance and exit channels, $k_\alpha\sim k_\beta \sim 2100$~MeV/c. Accordingly, the~potentials are displayed at $p\sim k_\alpha$  and $p=\frac{1}{3}k_\alpha$. In~magnitude, the potentials increase with momentum, which seems to be especially pronounced for the P-wave parts $T_{1,2}$. However, as~a look to Equation~\eqref{eq:Uii} and Equation~\eqref{eq:Uij}, respectively, reveals, the enhancement is largely due to the explicit dependence of the P-wave potentials on powers of $p$. Compared to that dependence, the~S-wave form factors $U_{00}$ remain in small-to-moderate magnitude. For~$p\ll k_\alpha$, the enhancement effect decreases, and the S-wave potentials become relatively more important.
Comparing the oxygen and calcium  potentials, one observes a rather mild dependence on the nuclear system as~is expected for a short-range~phenomenon.

The s-channel $\pi^0$ exchange establishes in fact a rather tight two--nucleon correlation. Overall, the~range of the potentials rarely reaches 40\% of the range of pion exchange $r_\pi\sim 1/m_\pi\sim 1.4$~fm.  Hence, the~MDCE process is of a pronounced short-range character. The~correlated pair of SCE vertices acts as a virtual, polarized pion dipole source. Comparisons of the data of the DCE reaction induced by $^{18}$O$+{}^{40}$Ca at $T_{lab}=270$~MeV can be found elsewhere~\cite{Cappuzzello:2022ton}.

\unskip

\subsection{Transition Matrix~Elements}
In closure approximation and with the pion potential formalism, the TMEs are obtained in the condensed form
\be\label{eq:WAB}
\mathcal{W}_{AB}(\mathbf{p}_1,\mathbf{p}_2)=\lan \mathcal{I}^{(\pi)}_{2\mp 2}\ran
\sum_{i,j=0,1,2}\mathcal{M}^{(ij)}_{AB}(\mathbf{p}_1,\mathbf{p}_2).
\ee
Thus, the~transition $A\to B$ is described by a sum of nine partial TMEs
\be\label{eq:Mij}
\mathcal{M}^{(ij)}_{AB}(\mathbf{p}_1,\mathbf{p}_2)=
\lan B|e^{-i\mathbf{p}_2\cdot \mathbf{r}_2}W^{(ij)}_{AB}(\mathbf{x}|\mathbf{p}_1,\mathbf{p}_2)
e^{i\mathbf{p}_1\cdot \mathbf{r}_1}I^{(N)}_{2\pm 2}|A\ran
\ee
which are determined by the transition potentials $W^{(ij)}_{AB}$. They are defined and studied in detail in
Appendix \ref{app:PiPot}. There, it is also shown that the useful and successful approach is to express momentum and spin operators in the basis of spherical unit vectors. In~that basis, one finds that $W^{(ij)}_{AB}$ are dyadic tensor forms. The~x-dependence is given by Yukawa-type form factors of a rather short range of less than half of the range of a (static) pion-exchange potential. Hence,using contact interactions might be a meaningful approximation which, however, will not be considered further~here.

The two-body operator connecting, in Equation~\eqref{eq:Mij}, the initial and final states is in fact separable into one-body operators. That property is evident for the plane wave factor, considering that $\mathbf{x}=\mathbf{r}_1-\mathbf{r}_2$, and~also the potentials $W^{(ij)}_{AB}$ are given by products of one-body operators.  In~practical calculations, the~plane waves are expanded into partial waves in $\mathbf{r}_1$ and $\mathbf{r}_2$, and by the formalism introduced in Appendix \ref{app:PiPot} the potentials can also be treated accordingly. At~the end,  Equation \eqref{eq:Mij} reduces to a (finite) sum of a number of multipole components which are determined by the angular momentum and parity selection rules of the DCE transition $A(Z,N|J^{\pi_A}_A)\to B(Z\pm 2,N\mp 2|J^{\pi_B}_B)$.

With the bi-spherical harmonics
\be
\mathcal{Y}_{(\ell_1\ell_2)\ell m}(\widehat{\mathbf{x}},\widehat{\mathbf{y}})=
\sum_{m_1m_2}\left(\ell_1m_1\ell_2m_2|\ell m \right)
Y_{\ell_1m_1}(\widehat{\mathbf{x}})Y_{\ell_1m_1}(\widehat{\mathbf{y}}).
\ee
we find
\be\label{eq:MijMulti}
\mathcal{M}^{(ij)}_{AB}(\mathbf{p}_1,\mathbf{p}_2)=\sum_{\ell_1\ell_2,\ell m}(-)^{\ell+m}
\mathcal{Y}_{(\ell_1\ell_2)\ell -m}(\widehat{\mathbf{p}}_1,\widehat{\mathbf{p}}_2)
M^{(ij)}_{(\ell_1\ell_2)\ell m}(\mathbf{p}_1,\mathbf{p}_2)
\ee

The multipole TMEs are given by rank-2 isotensor two-body multipole operators
\be\label{eq:MijL1L2}
M^{(ij)}_{(\ell_1\ell_2)\ell m}(\mathbf{p}_1,\mathbf{p}_2)=
\lan B|\left[R_{\ell_2}(\mathbf{r}_2|p_2)\otimes R_{\ell_2}(\mathbf{r}_1|p_1) \right]_{\ell m}
W^{(ij)}_{AB}(\mathbf{x}|\mathbf{p}_1,\mathbf{p}_2)\mathcal{I}^{(N)}_{2\pm 2}|A\ran .
\ee

For $i,j=0,1$, the MDCE transition operators are of a spin--scalar character, and the matrix elements describe non-spinflip double-Fermi (FF) excitation. The~FF modes are described by
spin--scalar one-body operators which are given by  Riccati--Bessel functions $j_\ell (x)$:
\be\label{eq:Rlm}
R_{\ell m}(\mathbf{r}|p)=j_{\ell}(pr)i^\ell Y_{\ell m}(\hat{\mathbf{r}}).
\ee

For $i=j=2$, spin--vector transition operators are encountered which give rise to double excitations of Gamow--Teller (GG) modes, which include a spin--vector transition of natural and unnatural parity. If~$i=2$ but $j=0,1$ or $i=0,1$ and $j=2$, we encounter two-body operators of mixed spin--scalar/spin--vector structure, leading to mixed FG and GF excitation by combination of Fermi and Gamow--Teller~modes.

The GG and mixed FG/GF modes are described by spin--vector one-body operators. Their derivation and especially proper implementation into the theory requires a remarkable amount of angular momentum recoupling. The~spin--vector formalism for DCE reactions was studied in detail in~\cite{Lenske:2024dsc} and will not be considered further here.
As was shown also in~\cite{Lenske:2024dsc}, the GG operators support total spin transfers $S=0,1,2$, to~be combined with the total orbital angular momentum transfer $\mathbf{L}$ to total angular momentum transfer  $\mathbf{J}=\mathbf{L}+\mathbf{S}$. That leads to a rich spectrum of transitions, e.g.,~a DCE reaction with $J^\pi=0^+$ may proceed by $L=0,S=0$ and $L=2,S=2$ partial~contributions.

\subsection{Transition Matrix Elements in Collinear~Approximation}
For arbitrary values of $\mathbf{p}_1$ and $\mathbf{p}_2$, the evaluation and the practical handling of the TMEs are theoretically and numerically formidable tasks. The~efforts, however, are substantially reduced for collinear external momenta, i.e.,~$\mathbf{p}_1||\mathbf{p}_2$ and also $\mathbf{p}_1-\mathbf{p}_2=\mathbf{q}_{12}$ and $(\mathbf{p}_1+\mathbf{p}_2)/2=\mathbf{P}_{12}$ are collinear. Further simplifications are obtained by imposing, in addition, the stronger constraint $\mathbf{p}_1=\mathbf{p}_2=\mathbf{p}$, which implies $|\mathbf{q}_{12}|=0$ and $\mathbf{P}_{12}=\mathbf{p}$. Then, Equation~\eqref{eq:Mij} simplifies to
\be\label{eq:MijColl}
\widetilde{\mathcal{M}}^{(ij)}_{AB}(\mathbf{p})=
\lan B|e^{i\mathbf{p}\cdot \mathbf{x}}\widetilde{\mathcal{W}}^{(ij)}_{AB}(\mathbf{x}|\mathbf{p})\mathcal{I}^{(N)}_{2\pm 2}|A\ran =
\lan B|\widehat{\mathcal{W}}^{(ij)}_{AB}(\mathbf{x}|\mathbf{p})\mathcal{I}^{(N)}_{2\pm 2}|A\ran ,
\ee
where $\widehat{\mathcal{W}}^{(ij)}_{AB}$ contains the plane wave factor.
As an example, we consider the $L=0,S=0$ component of $A(Z,N|0^+) \to B(Z\pm 2,N\mp 2|0^+)$  double-Fermi transitions. Hence, only the spin--scalar S-wave ($i=j=0$), P-wave ($i=j=1$) and the mixed S/P-wave parts  ($i=0,j=1$) and  ($i=1,j=0$) are considered. In~Appendix \ref{app:Collinear}, the spin--scalar collinear transition potentials
$\widehat{\mathcal{W}}^{(ij)}_{AB}$ are derived, and their multipole structure is investigated.
For $0^+ \to 0^+$ transitions, the complexity of the potentials is reduced further. For~that case, explicit expression are found also in Appendix \ref{app:Collinear}.

Following~\cite{Lenske:2021jnr}, we assume that the states in the DCE daughter nucleus $B$ are obtained by acting with appropriate many-body operators on the ground state of the parent nucleus~$A$:
\be\label{eq:DCEState}
| B(Z \pm 2,N \mp 2),J_BM_B \ran  \simeq
\sum_{{C_1}{C_2}} {z_{{C_1}{C_2}}^{{J_B}{M_B}}}
\left[ \Omega _{{C_1}}^\dag  \otimes \Omega _{{C_2}}^\dag  \right]_{{J_B}{M_B}}
| A(Z,N),J_AM_A\ran +\ldots ,
\ee
where higher-order quasiparticle configurations may contribute but will not be reached in leading order by the DCE transition operators. The~same set of operators and the underlying basis of single article wave functions are used to express the transition potentials in second quantization. In~practice, nuclear ground-state properties are described with Hartree--Fock--Bogolyubov (HFB) theory, and  Quasiparticle Random Phase Approximation (QRPA) is used for  excited SCE-type states, see~\cite{Lenske:2018jav}.

Without going further into the details of the nuclear structure approach, the~essence of the approach is that the TMEs, Equation~\eqref{eq:MijColl}, are given by nuclear transition form factors
\be
\rho^{(C)}_{\lambda\mu}(p)\sim \lan A|[\Omega _{{C}},R_{\lambda \mu}(\mathbf{r}|p)]|A\ran ,
\ee

The TMEs are obtained by the scheme developed in Appendix \ref{app:Collinear}. As~a recipe, we have to replace in the expression derived in the appendix the operators $R_{\lambda \mu}$ by the Fourier--Bessel form factors $\rho^{(C)}_{\lambda \mu}(\mathbf{p})$ and finally perform the momentum~integrals.

\begin{figure}[H]

\includegraphics[width = 13cm]{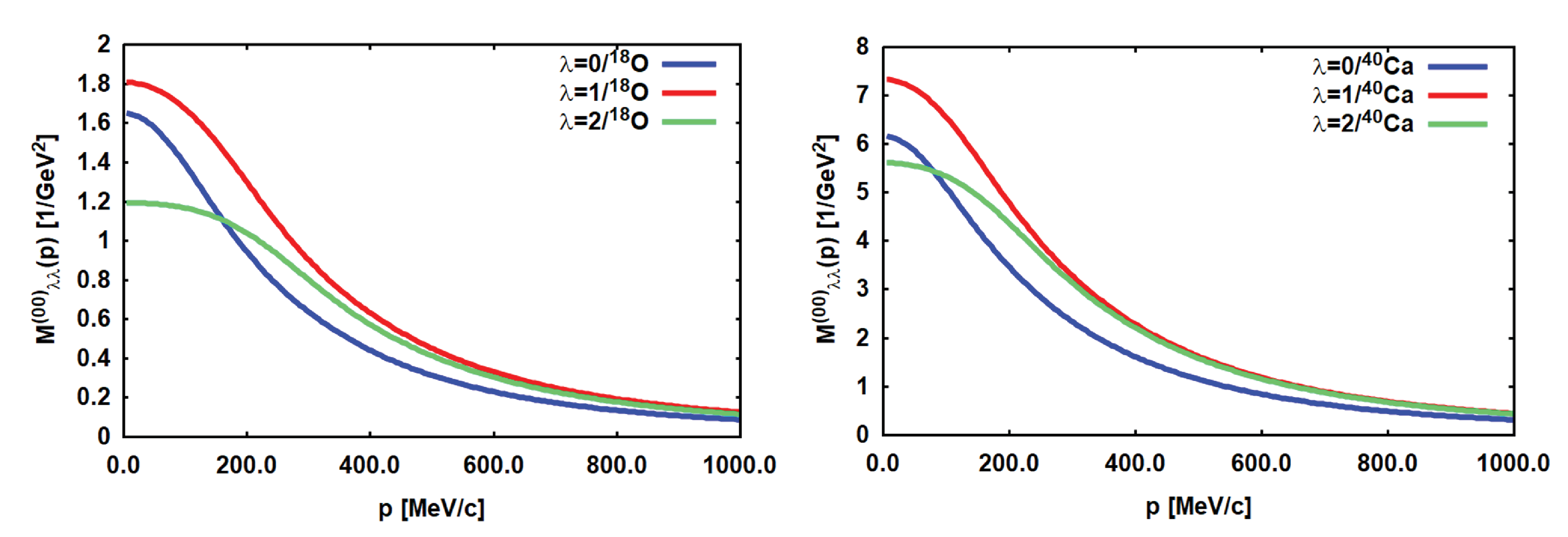}
\caption{DCE non-spinflip double-Fermi ($L=0$) transition matrix elements for $^{18}$O$\to ^{18}$Ne (left) and $^{40}$Ca$\to ^{40}$Ar (right), respectively, are shown in collinear approximation. The~real parts of the double S-wave TME $M^{(00)}_{(\lambda\lambda)L}$ for $\lambda=0,1,2$ are compared.
 See text for further discussion. }
\label{fig:M00}

\end{figure}

\begin{figure}[H]

\includegraphics[width = 13cm]{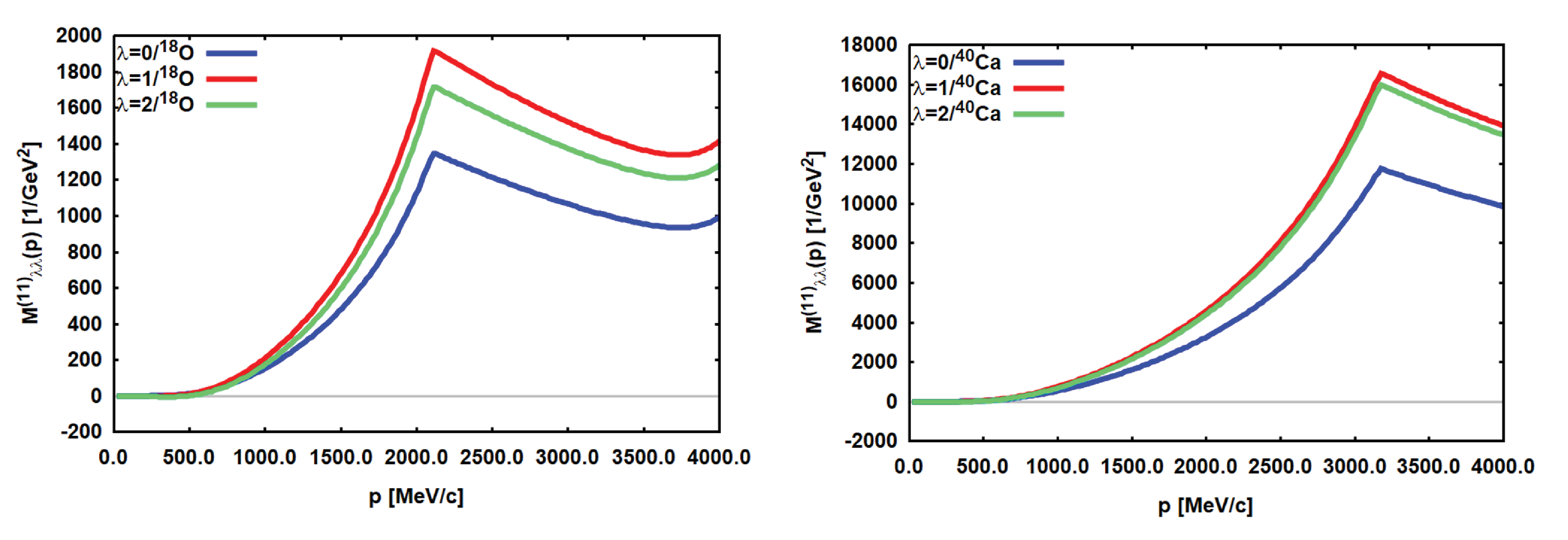}
\caption{DCE non-spinflip
double-Fermi ($L=0$) transition matrix elements for $^{18}$O$\to ^{18}$Ne (left) and $^{40}$Ca$\to ^{40}$Ar (right), respectively, are shown in collinear approximation. The~real parts of the double P-wave TME $M^{(11)}_{(\lambda\lambda)L}$ for $\lambda=0,1,2$ are compared.
 See text for further discussion. }
\label{fig:M11}

\end{figure}

\begin{figure}[H]

\includegraphics[width = 13cm]{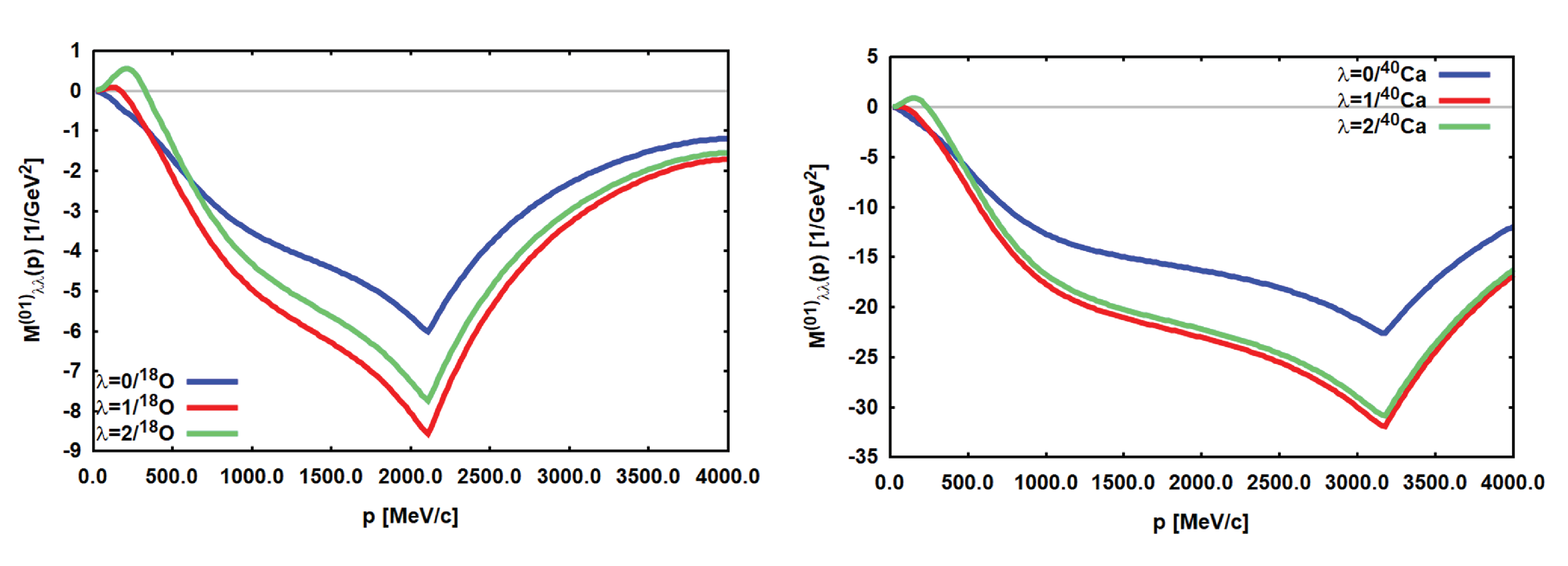}
\caption{DCE non-spinflip
double-Fermi ($L=0$) transition matrix elements for $^{18}$O$\to ^{18}$Ne (left) and $^{40}$Ca$\to ^{40}$Ar (right), respectively, are shown in collinear approximation. The~real parts of the mixed S/P-wave TME $M^{(01)}_{(\lambda\lambda)L}$ for $\lambda=0,1,2$ are compared.
 See text for further discussion. }
\label{fig:M01}

\end{figure}

In Figures
~\ref{fig:M00}--\ref{fig:M01}, partial TMEs, Equation~\eqref{eq:MijL1L2}, of~$0^+ \to 0^+$ transitions in $^{18}$O$\ \to{}^{18}$Ne and $^{40}$Ca$\ \to{}^{40}$Ar are shown. The~TMEs are relevant for the DCE reaction $^{40}$Ca$(^{18}$O$,^{18}$Ne$)^{40}$Ar studied in~\cite{Cappuzzello:2022ton,Cappuzzello:2015ixp}. State-independent average transition densities are used, which are averaged over the spectral distributions and normalized to the respective non-energy weighted multipole sum rule, corresponding to the \emph{unit strength form factors} introduced in~\cite{Lenske:2021jnr}. Hence, the~results are representative of monopole FF modes in $^{18}$Ne and in $^{40}$Ar, relative to the respective parent~nuclei.

\unskip

\unskip

Comparing the results, the~most outstanding feature are the differences between the S-wave and the P-wave TMEs. The~double S-wave TME, Figure~\ref{fig:M00}, contribute only at small momenta close to the threshold. The~TME involving P-wave amplitudes, \mbox{Figures~\ref{fig:M11} and \ref{fig:M01}}, increase strongly with momentum, exceeding the strength of the S-wave TME by large factors. The~P-wave enhancement is largely an effect of the additional polynomial momentum dependencies up to order $p^4$, see Appendix \ref{app:Collinear}. However, it has to be remembered that the shown results are the bare TMEs before ISI/FSI renormalization. After~renormalization, i.e.,~in a full distorted wave calculation, the high-momentum regions will especially be quenched in addition to the overall reduction by about two to three orders of magnitude, thus  considerably damping the apparent~enhancement.

For arbitrary total angular momentum $J^\pi$, various combinations of partial contributions of angular momentum $J^{\pi_1}_1$ and $J^{\pi_2}_2$ are allowed, constrained, however, by~parity, $\pi=\pi_1\pi_2$, and~otherwise limited only by the shell structure and other related properties of the nucleus under consideration. For~the $0^+$ case, this means that in principle, all pairs of transition densities of equal angular momentum $\ell_1=\ell_2=\lambda$ may contribute. For~the $J=L=0$ case, this property of the TME is illustrated in the figures by showing the partial TME,  Equation~\eqref{eq:MijL1L2}, for~$\lambda=0,1,2$. In~magnitude and shape, the partial TMEs are rather similar. Thus, we conclude that the MDCE operators support a large spectrum of multipolarities as is~typical for short-range~dynamics.

An eye-catching feature visible in all plots is the kinks. They appear at the momenta where the intermediate $\pi +C$ channels cross from below the on-shell boundary, which produces a pole in the propagator. That happens at $p\sim 2156$~MeV/c and $p=3225$~MeV/c for $\pi +^{18}$F and $\pi +^{40}$K, respectively. Another feature is the crossing of the on-shell boundary of the $\pi + N$ subsystems at the slightly smaller momenta $p\sim 2120$~MeV/c for A=18 and $p\sim 3180$~MeV/c for A=40. The~location of these thresholds depends however, on~the modeling of in-medium pion dynamics, which here is not considered, as~mentioned before. Above~these momenta, the~potentials develop imaginary parts of moderate strength which are not shown~here.

\section{Relation of Heavy Ion DCE Dynamics to Double Beta~Decay}\label{sec:DCE-DBD}
\unskip

\subsection{Leptonic and Hadronic DCE~Processes}
Besides hadronic DCE reactions, there is only one other process known to change nuclear charges by two units, namely, double beta decay by weak interactions. The similarities between hadronic DCE (HDCE) reactions and leptonic DBD (LDBD), elucidated recently in \cite{Lenske:2024var} are worth closer consideration. An overall striking similarity is already that both hadronic and leptonic DCE may proceed in two distinct versions: in HDCE, these are the DSCE and MDCE reaction mechanisms, and in LDBD, these are the $2\nu 2\beta$ and the $0\nu 2\beta$ processes. Moreover, due to obvious reasons, HDCE and LDBD utilize unavoidably the same kind of nuclear configurations, which implies that spectroscopic information gained in one type of DCE will be of high value for research on the other kind of DCE.

The multitudes of similarities will surely be realized also on the level of elementary processes. However, we have to keep in mind that in the sense of the \emph{standard model}, HDCE and LDBD occur on asymptotic low-energy scales. Hence, LDBD does not give direct access to electro-weak gauge boson physics, and HDCE physics is highly unlikely to probe directly  quark--gluon QCD dynamics. The hidden background scale is another connecting feature of the two DCE sectors. Nevertheless, the fundamental dynamics is of course reflected in operator structures,  coupling constants, form factors and other features of the involved interactions. In the DBD area, the connections to fundamental dynamics have been studied for decades in much detail, see \cite{Tomoda:1990rs,Ejiri:2019ezh}. In the HDCE sector investigations of comparable intensity are at an emerging level.

\begin{figure}
\includegraphics[width = 11cm]{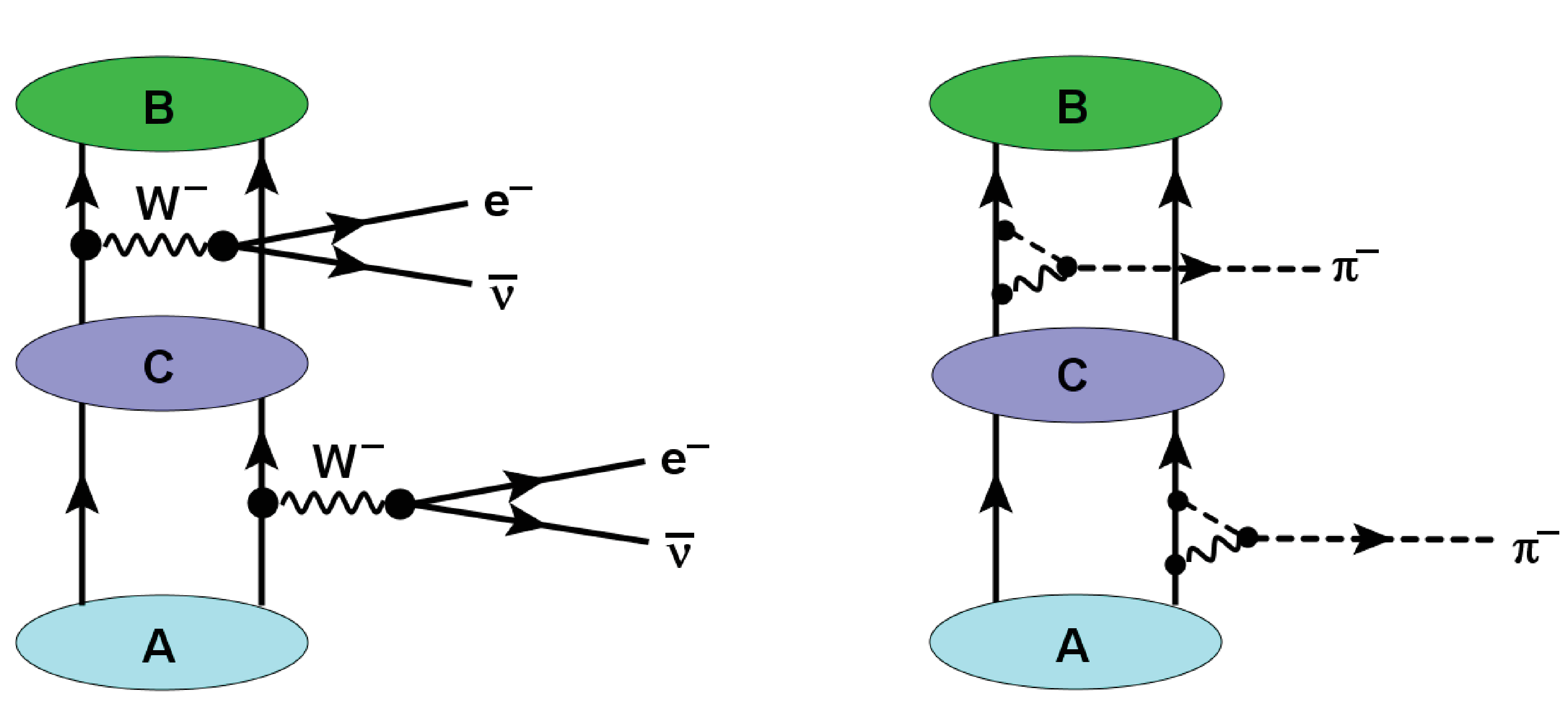}
\caption{Diagrams describing the $A(Z,N)\to B(Z+2,N-2)$ transition for two-neutrino DBD (left) and hadronic DSCE in one of the interacting nuclei participating in a DCE reaction (right). In the DSCE case, the wavy line indicates a virtual $\rho^-$ meson, which decays in to a reabsorbed $\pi^0$ and a virtual $\pi^-$ meson. See text and Ref. \cite{Lenske:2024var} for further discussion. }
\label{fig:DSCE_NME}
\end{figure}

In  Figure \ref{fig:DSCE_NME}, two-neutrino DBD and DSCE are compared for a set of selected diagrams, chosen for emphasizing the similarities. The emission of the two $e^-\bar{\nu}_e$ pairs is initiated by unobserved $W^-$ vector bosons, acting far off the their respective mass shells. For comparison, a DSCE subprocess is displayed, where  highly virtual $\rho^-$ vector mesons decay into neutral and charged pions, both far off their respective mass shells. The triangle diagram would appear in a field-theoretical description of NN scattering. The $\pi^0$ mesons are reabsorbed at the emission point within the same nucleus, while the two virtual $\pi^-$ mesons leave the interaction zone on their way to a nucleon in the reaction partner. Since a pair of neutrons is changed into a pair of protons, the emitting nucleus is left in a $p^2n^{-2}$ configuration. In the acceptor nucleus, the $\pi^-$ mesons initiate a complementary DCE process, in which a pair of protons is converted into a pair of neutrons and a $n^2p^{-2}$ state emerges. Two-neutrino DBD is a second-order process of two (uncorrelated) single beta decay events, characterized by almost point-like interactions. The strength is determined by the weak axial and vector coupling constants $g_{A,V}\sim \mathcal{O}(1)$ which describe already very accurately the decay properties. In DSCE, the pions are emitted by nucleon sources also in a point-like manner, where finite seize effects are parameterized into the pion--nucleon coupling constants $g_{\pi N}$.
The considerably larger value of $g_{\pi N}\sim \mathcal{O}(10)$ and of the other meson--nucleon couplings demands to solve the NN scattering problem in all orders with the full set of mesons and to use the NN T matrix in DSCE calculations.

\begin{figure}[H]

\includegraphics[width = 10cm]{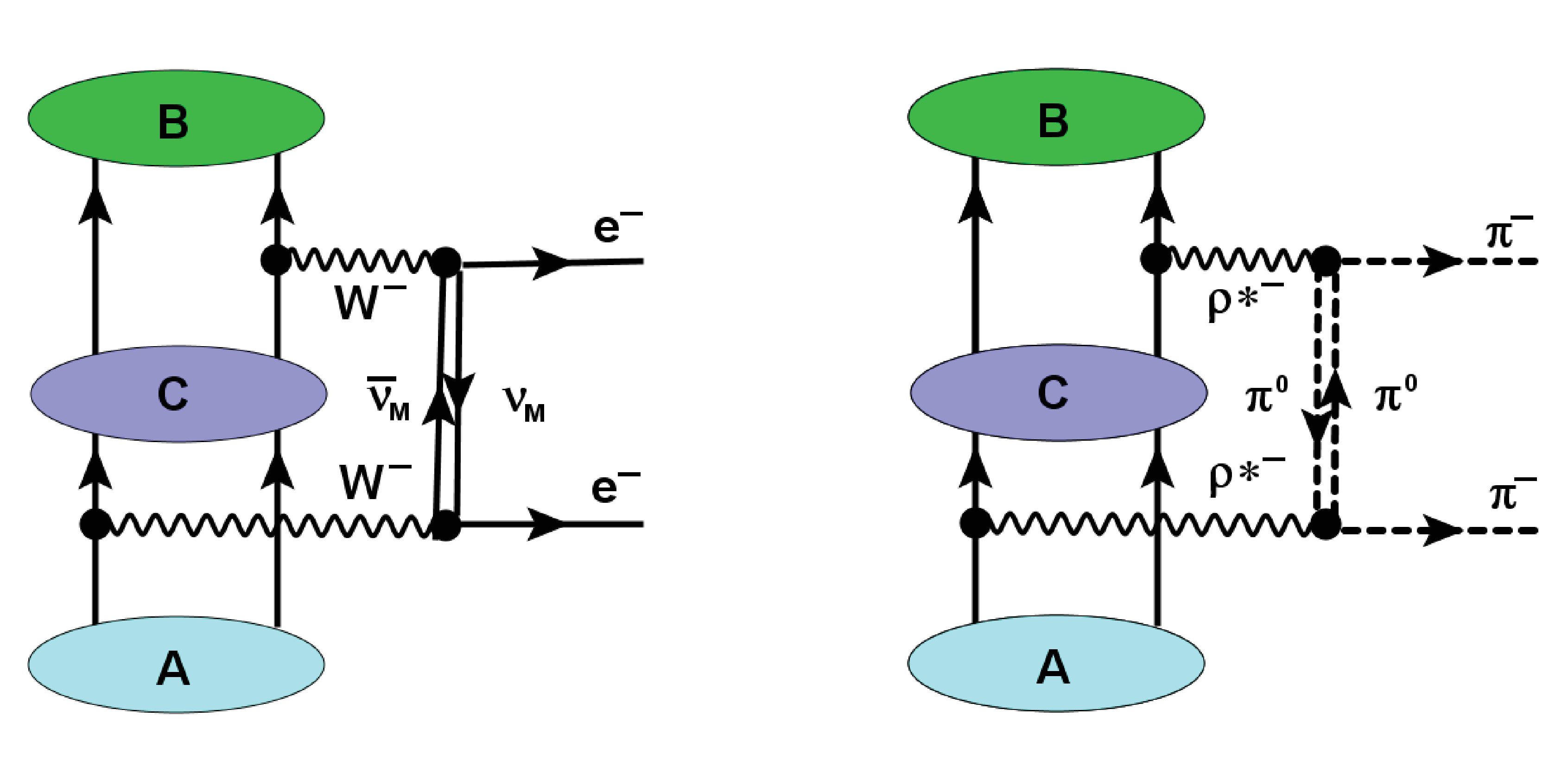}
\caption{Graphical illustration of a $A(Z,N)\to B(Z+2,N-2)$ nuclear DCE transition for neutrinoless Majorana DBD (left) and the hadronic MDCE modes in one of the interacting nuclei (right). See text and Ref. \cite{Lenske:2024var} for further discussion.}
\label{fig:MDCE_NME}

\end{figure}

The hidden connections between MDBD and MDCE are exemplified by the two diagrams shown in Figure \ref{fig:MDCE_NME}. The MDBD process starts again by a pair of virtual $W^-$ bosons, now materializing, however, in  Majorana neutrinos
$\nu_M =\bar{\nu}_M$ and a $e^-$ pair, which leaves the nucleus on the mass shell \cite{Tomoda:1990rs,Ejiri:2019ezh}. The MDCE event resembling the closest the Majorana decay is depicted by the diagram on the right side of Figure \ref{fig:MDCE_NME}. As before, virtual $\rho^-$ mesons and their subsequent decay into $\pi^0\pi^-$ are the initiators of the DCE process. The initial rho-mesons may be produced in s-channel pion--nucleon resonance formation--decay processes or result from t-channel pion--nucleon interactions.
However, different from the DSCE case, here the neutral pions are exchanged between the decay vertices, i.e., they take the role of the MDBD neutrinos. In the previous sections, the $\pi^0$ exchange led to the pion potentials, which established a short-range correlation between the two $\pi N$ SCE events. The charged pions leave the DCE vertex in highly virtual states. But different from  the DSCE scenario, they are emitted by a correlated source. As a result, they have imprinted the two--nucleon correlation of their origin and will transmit that information to the other nucleus. The result is a combination of $p^2n^{-2}$ in one nucleus accompanied by a $n^2p^{-2}$ in the other nucleus. Under reaction--theoretical aspects, we encounter a pair of virtual complementary  pion--nucleon DCE reactions as explored in the previous sections.

\vspace{-6pt}

\vspace{-16pt}

\subsection{Lepton MDCE in Heavy Ion DCE Reactions?}
\begin{figure}[H]

\includegraphics[width = 6.5cm]{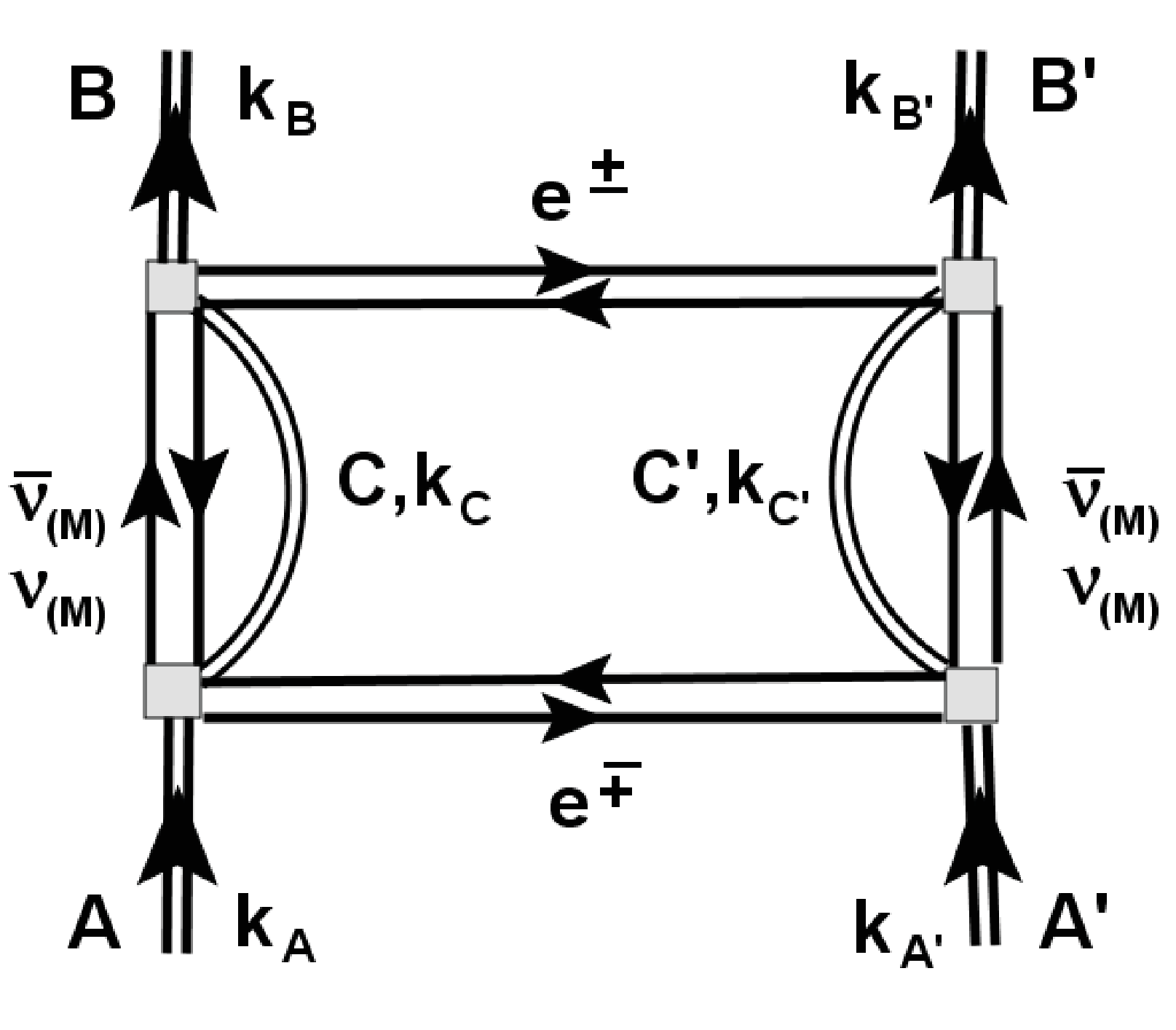}
\caption{Diagrammatical illustration of the leptonic MDCE process discussed in the text. The~charged pions, responsible for hadronic MDCE processes, are replaced by electrons and positrons exchanged in the t-channel, while the neutral pions, propagating in the s-channel, are replaced by neutrinos and antineutrinos, respectively, or~by Majorana neutrinos, if~they exist. See text for further~discussion. }
\label{fig:MDCE_Lepton}

\end{figure}

As an outlook to future work, we point finally to possible purely leptonic contributions in a heavy ion MDCE reaction. Leaving aside the much weaker interactions of an electro-weak process, in principle, the ions may also interact by the exchange of leptons. In~Figure~\ref{fig:MDCE_Lepton}, as an example, we show a diagram illustrating a DCE reaction by $e^\pm$ t-channel exchange between the ions. By~inverse beta-decay on a nucleon---without or with excitation of an $N^*$ resonance---the $e^\pm$ are converted into a neutrino or antineutrino, respectively, which propagates in the s channel, interacts with another nucleon, and is converted into a lepton of a charge which is complementary to the one of the incoming lepton. As~indicated in Figure~\ref{fig:MDCE_Lepton}, the $\nu / \bar{\nu}$ may be replaced by Majorana neutrinos $\nu_M / \bar{\nu}_M$. Obviously, such processes are of special interest for MDCE and MDBD physics because they contain as a subprocess the same dynamical structure as neutrinoless DBD. The~graph shows that this hitherto neglected MDCE process relies completely on electro-weak interactions, probing directly the Majorana hypothesis in an off-shell process embedded into the environment of a heavy ion DCE reaction. It is also of interest that the diagram is of the same topology as the hadronic counterpart in Figure~\ref{fig:MDCE}. The charged pions are replaced by electrons or positrons, respectively. The~neutrinos and antineutrinos take over the role of the neutral pions and propagate between the two SCE vertices, thus also establishing a short-range correlation. In~the scenario involving Majorana neutrinos, the~heavy ion DCE reaction corresponds to a double Majorana DBD process, occurring, however, off~the mass shell. SCE scattering of leptons on nuclei is an important issue in understanding the interactions of high--energy cosmic neutrinos with matter. In~that context, SCE reactions induced by charged leptons and neutrinos are studied extensively in theory and experiment~\cite{AlvarezRuso:1997mx,Alvarez-Ruso:2017oui,Alvarez-Ruso:2021dna,Martini:2010ex,Chanfray:2021tsp,Ankowski:2022thw,deGouvea:2022gut}.

Hence, dynamically leptonic MDCE would include a process which depends on the same kind of interactions as expected for neutrinoless Majorana DBD (MDBD). Of~special interest is that the vertices necessarily are determined by the same $e^\pm \leftrightarrow \nu_M\bar{\nu}_M$ conversion mechanisms as assumed for MDBD. In~particular, lepton MDCE will probe in both of the interacting nuclei directly the Pontecorvo–Maki–Nakagawa–Sakata (PMNS) mass matrix~\cite{Maki:1960ut,Maki:1962mu,Lu:2024xkf}, albeit under the conditions of a nuclear~reaction.

\vspace{-3pt}

As a caveat, the~interactions involved in leptonic MDCE will be much weaker than hadronic MDCE. Nevertheless, it is tempting and worthwhile to investigate further that kind of heavy ion DCE subprocess, if~not for heavy ion reactions but for the DCE reaction with light ion or pion beams, respectively. The~search for a signal might be tedious, but looking for interference signals of the weak lepton MDCE amplitude with the dominating hadronic MDCE amplitude could be a promising approach. Since lepton MDCE is determined by long-range interactions, signatures will most likely be best observable at extreme forward~angles.

\section{Summary}\label{sec:SumOut}
A generic feature of heavy ion DCE reactions is the versatility of reaction mechanisms by which the transition from the initial to the final channel can proceed. Occasionally, the~related ambiguities are considered a severe disadvantage of research with heavy ion beams. That point of view is much too pessimistic because in reality, it is of advantage to be able to investigate all facets of a physical system under the same, well-defined experimental conditions and describe the results consistently by the theoretical apparatus of nuclear many-body theory. The~theoretical task and  challenge is to overcome the traditional separation of nuclear reaction and nuclear structure physics. Heavy ion DCE physics demands a combined approach as indispensable for any research on quantum mechanical many-body systems. In~the NUMEN project, this decisive aspect is realized by the multimethod approach as discussed in~\cite{Cappuzzello:2022ton}.

In this work, we investigated the theory of the Majorana (MDCE) mechanism which is an especially interesting part of heavy ion DCE reactions. As~it was emphasized repeatedly, MDCE theory requires to go much beyond traditional concepts of nuclear reaction and structure theory. First of all, as~a hitherto never considered aspect, the~MDCE scenario relies on pion--nucleon dynamics which---from the beginning, most likely unexpected---suddenly involves subnuclear degrees of freedom as nucleon resonances into a low-energy nuclear process. Already, that aspect makes it worth the effort of investigating DCE~reactions.

The MDCE process relies on a hitherto unknown mechanism, namely, a  dynamically induced rank-2 isotensor interaction.  One of the central results was to introduce the MDCE closure approximation, which allowed to derive pion potentials and two-body nuclear matrix elements connecting directly the entrance and the DCE exit channels. The~pion potentials include combinations of spin and momentum scalar parts, spin--scalar longitudinal and spin--vector transversal momentum--vector components, all attached to a rank-2 isotensor operator. This rich operator structure allows widespread spectroscopic studies, allowing a detailed tomography of the nuclear wave functions. However, experimentally and theoretically such studies are highly demanding because they require to observe, analyze, and~interpreted energy--momentum distributions over large~ranges.

An especially appealing aspect of heavy ion DCE physics is the conceptional closeness to double beta decay research. That relationship was elucidated in some detail by considering the deeper levels underlying weak and strong DCE processes. They meet at the level of QCD and electro-weak physics. Clearly, neither DBD nor DCE reactions proceed at those fundamental levels. Rather, both types of process are determined by low-energy realizations of the two fundamental theories of the current standard model of physics. However, the~comparison of weak and strong DCE processes at the fundamental level is helpful to understand that nuclear DBD and nuclear DCE phenomena are finally nothing but two realizations of the same kind of fundamental processes. The~differences in dynamics and strengths of DBD and hadronic DCE are due to the breaking of the fundamental symmetries in our physical low-energy~environment.

As an interesting outlook to future work, the~closeness of low-energy DBD and DCE physics was elucidated further by pointing to another competing reaction mechanism in heavy ion DCE reactions. Nothing forbids  MDCE reactions from proceeding by the exchange of leptons. Leptonic MDCE proceeds by electro-weak dynamics but relies on diagrams of the same topology as investigated in this paper in detail for hadronic MDCE. It is left for future work to understand the dynamics and physics of lepton MDCE in detail and explore the competition of the two seemingly very different but~interfering types of weak and strong MDCE reaction~mechanisms.

As a closing remark, we emphasize again that in MDCE, reactions are not governed by NN interactions as is the case for DSCE reactions. MDCE reactions are determined by  pion--nucleon interactions, which provide the required isospin operator structures for an effective rank-2 isotensor interaction. In~MDCE reactions, the colliding ions dynamically generate their own and specific isotensor interactions. Charge and baryon number conservation and isospin symmetry require that a $\Delta Z=+2$ transition in one nucleus must be accompanied by a $\Delta Z=-2$ transition in the other nucleus. Obviously, all of the involved transitions are allowed and possible by strong nuclear interactions. Hence, hadronic DCE is not suppressed or even forbidden by violating fundamental laws of the standard model as required for $0\nu 2\beta$ decay. While MDBD is constrained trivially to appear on the mass shell, MDCE reactions take advantage of the presence of another nucleus which gives access to a broad spectrum of off-shell processes and new research opportunities. Hence, it depends on our theoretical and experimental skills to identify and prepare the proper conditions under which rare hadronic or even leptonic MDCE events will become~observable.

\vspace{6pt}

\authorcontributions{H.L.: Conceptualization, methodology, original draft preparation and funding acquisition,
J.B.: investigation and methodology, M.C.: investigation, administration, supervision, and funding
acquisition, D.G.: investigation and formal analysis.}

\funding{H. Lenske acknowledges financial support in part by DFG, grant Le439/16-2, and INFN/LNS
Catania.}

\dataavailability{Not applicable.}

\conflictsofinterest{The authors declare no conflicts of interest.}

\appendixtitles{yes} 
\appendixstart
\appendix
\section[\appendixname~\thesection]{Momentum Structure of Distorted Waves and Distortion~Amplitudes}\label{app:DW}

With ISI and FSI, the~momentum relation is changed because the diffractive and absorptive interactions of the optical potential $U_{opt}$ admix a certain amount of off-shellness into the theory as~seen immediately by expressing the distorted waves by their integral equation, e.g.,
\bea
|\chi^{(+)}_{\mathbf{k}}\ran&=&|\phi_{\mathbf{k}}\ran+
\int \frac{d^3k'}{(2\pi)^3}|\phi_{\mathbf{k}'}\ran \frac{1}{E_k-E_{k'}+i\eta}
\lan \phi_{\mathbf{k}'}|U_{opt}|\chi^{(+)}_{\mathbf{k}}\ran \label{eq:chi}\\
&=&|\phi_{\mathbf{k}}\ran+
\int \frac{d^3k'}{(2\pi)^3}|\phi_{\mathbf{k}'}\ran \frac{1}{E_k-E_{k'}+i\eta}
\lan \phi_{\mathbf{k}'}|U_{opt}|\phi_{\mathbf{k}}\ran \ldots.
\eea
$\phi_{\mathbf{k}}$ and $\phi_{\mathbf{k}'}$ are plane waves and $E_{k,k'}$ are the kinetic energies defined by the momenta $\mathbf{k},\mathbf{k}'$, $E_{k}\sim k^2$.
Thus, a~finite range optical potential will always lead to a wave functions with a momentum distribution centered at the physical asymptotic momentum $\mathbf{k}$ but with a finite width, which in leading order is determined by the 3D Fourier transform of the~potential.

As anticipated in Section~\ref{sec:ISIFSI}, the distorted wave may indeed by cast into the form
\be
\chi^{(+)}_{\mathbf{k}}(\mathbf{r})=
\lan \mathbf{r}|\chi^{(+)}_{\mathbf{k}}\ran=e^{i\mathbf{k}\cdot \mathbf{r}}\left(1-h_\mathbf{k}(\mathbf{r})\right)
\ee
and referring to Equation~\eqref{eq:chi}, we derive
\be
h_\mathbf{k}(\mathbf{r})=-e^{-i\mathbf{k}\cdot \mathbf{r}} \int \frac{d^3k'}{(2\pi)^3}e^{i\mathbf{k}'\cdot \mathbf{r}} \frac{1}{E_k-E_{k'}+i\eta}
\lan \phi_{\mathbf{k}'}|U_{opt}|\chi^{(+)}_{\mathbf{k}}\ran
\ee
where we changed to the coordinate representation
$\lan  \mathbf{r}|\phi_{\mathbf{k}}\ran=e^{i\mathbf{k}\cdot \mathbf{r}}$.

Now, we are at the position for gaining further insight into the meaning and physical content of the distortion coefficients:
\bea
D(\mathbf{p},\mathbf{k})&=&\frac{1}{(2\pi)^3}\lan \phi_\mathbf{p}|\chi^{(+)}_{\mathbf{k}}\ran=
\delta(\mathbf{k}-\mathbf{p})+\frac{1}{E_k-E_{p}+i\eta}\frac{1}{(2\pi)^3}
\lan \phi_{\mathbf{p}}|U_{opt}|\chi^{(+)}_{\mathbf{k}}\ran \\
&\approx&
\delta(\mathbf{k}-\mathbf{p})-i\pi\delta(E_k-E_p)U_{opt}(\mathbf{p},\mathbf{k})_{|\mathbf{p}|=|\mathbf{k}|}+
\frac{P}{E_k-E_{p}}U_{opt}(\mathbf{p},\mathbf{k})
 \ldots
\eea
where we  introduce $U_{opt}(\mathbf{k},\mathbf{k}')=\frac{1}{(2\pi)^3}\lan \phi_{\mathbf{k}}|U_{opt}|\phi_{\mathbf{k}'}\ran$ and the Cauchy decomposition of the energy denominator is used. The~distortion amplitudes defined in Equation~\eqref{eq:fafb} are identified as the Fourier transform
\bea\label{eq:fdist}
f(\mathbf{p},\mathbf{k})&=&\int \frac{d^3r}{(2\pi)^3}e^{-i(\mathbf{p}-\mathbf{k})\cdot \mathbf{r}}h_\mathbf{k}(\mathbf{r})\\
&=&-\frac{1}{E_k-E_{p}+i\eta}\frac{1}{(2\pi)^3}
\lan \phi_{\mathbf{p}}|U_{opt}|\chi^{(+)}_{\mathbf{k}}\ran .
\eea

Finally, we note that on the momentum shell $|\mathbf{p}|=|\mathbf{k}|$, defined by the pole part of the energy denominator, we retrieve the optical model elastic scattering amplitude, $T_{opt}(\mathbf{k},\mathbf{k}')\simeq \lan \phi_{\mathbf{k}'}|U_{opt}|\chi^{(+)}_{\mathbf{k}}\ran$ where $\mathbf{k}'\cdot\mathbf{k}=k^2\cos{\theta}$.

\section[\appendixname~\thesection]{Evaluation of the MDCE Box Diagram without ISI and FSI: Plane~Waves}\label{app:BoxPW}
The total available energy in the rest frame of the incident $A+A'$ system is defined by the sum  $P_\alpha= k_{A}+k_{A'}$ of the four-momenta $k_{A,A'}$ of the incoming ions, leading to the Lorentz-invariant Mandelstam energy $s_\alpha=P^2_\alpha(k_{A}+k_{A'})^2$ and by energy--momentum conservation $s_\alpha=P^2_\beta=(k_B+k_{B'})^2$. In~the laboratory frame with a beam of ions with rest mass $M_{A'}$ impinging with kinetic energy $T_{lab}$ on the target nuclei with rest mass $M_A$, the invariant energy is defined by $s_\alpha=(T_{lab}+M_{A'}+M_{A})^2-T_{lab}(T_{lab}+2M_{A'})$. In~the rest frame, the ions carry the four-momenta $k_{A,A'}=(E_{A,A'}^T,\mp \mathbf{k}_\alpha)^T$ with the invariant relative three-momentum and energies
\be
k^2_\alpha=\frac{1}{4s_\alpha\frac{}{}}(s_\alpha-(M_A+M_{A'})^2)(s_\alpha-(M_A-M_{A'})^2)\quad  ; \quad  E_{A,A'}=\sqrt{M^2_{A,A'}+k^2_\alpha}.
\ee

The outgoing ions $B,B'$ leave the interaction zone with four-momenta $k_{B,B'}=(E_{B,B'},\pm \mathbf{k}_\beta)^T$. In~the rest frame, they are given by
\be
k^2_\beta=\frac{1}{4s_\alpha\frac{}{}}(s_\alpha-(M_B+M_{B'})^2)(s_\alpha-(M_B-M_{B'})^2)\quad  ; \quad  E_{B,B'}=\sqrt{M^2_{B,B'}+k^2_\beta}.
\ee
The reaction proceeds by momentum transfers in the t-channel $q_{\alpha\beta}=k_{A}-k_{B}=-(k_{A'}-k_{B'})$ and in the u-channel  $p_{\alpha\beta}=k_{A}-k_{B'}=-(k_{A'}-k_{B})$, leading to the well-known invariants $t=q^2_{\alpha\beta}$ and $u=p^2_{\alpha\beta}$. Together with the invariant total energy, we retrieve the well-known Mandelstam relation
\be\label{eq:Mandelstam}
s+t+u=M^2_{A} + M^2_{A'} + M^2_{B} + M^2_{B'}
\ee

The charged pions are exchanged in the t channel with momenta
\be
p_1=k_1-k_{A}=k_2+k_{A'} \quad ; \quad p_2=k_1-k_{B}=k_2+k_{B'} .
\ee

Hence, we find
\be
p_2-p_1=k_A-k_B =k_{B'}-k_{A'}=q_{\alpha\beta} \quad ; \quad k_1-k_2=k_{A}+k_{A'}=k_{B}+k_{B'}=P_\alpha ,
\ee
and the t-channel and s-channel momenta are related by
\be
k_1+k_2-(p_1+p_2)=p_{\alpha\beta}.
\ee

Summing the squares of these relations, the~Mandelstam relation, Equation~\eqref{eq:Mandelstam}, is recovered but now expressed by the internal momenta of the box diagram. Since the four internal momenta of the box diagram are constrained by three invariants, we realize that the initially four independent momentum integrals have collapsed to a single momentum integration. A~meaningful choice is to use one of the t-channel momenta as the independent variable, e.g.,~$\mathbf{p}_1$.

Considering the reaction in the ion--ion rest frame, in~the plane wave (PW) limit, the momenta are fixed by Equation~\eqref{eq:K0}, from~which we find~immediately the following:
\begin{itemize}
\item $k_A=(E_A(k_\alpha),-\mathbf{k}_\alpha)^T$ and $k_B=(E_B(k_\beta),+\mathbf{k}_\beta)^T$;
\item $k_{A'}=(E_{A'}(k_\alpha),+\mathbf{k}_\alpha)^T$ and $k_{B'}=(E_{B'}(k_\beta),-\mathbf{k}_\beta)^T$;
\item $p_1=(0,\mathbf{p}_1)^T$ is a  purely space-like four momentum;
\item $p_2=(E_A-E_B,\mathbf{p}_2)^T$ includes formally the Q-value of the DCE reaction;
\item $\mathbf{p}_1=\mathbf{k}_\alpha$ and $\mathbf{p}_2=\mathbf{k}_\beta$ are fixed by the three momenta of the incoming and outgoing systems;
\item $k_1=p_1+k_A=(E_A(k_\alpha),\mathbf{0})^T$ is a purely time-like four vector;
\item $k_2=p_1-k_{A'}=(-E_{A'}(k_\alpha),\mathbf{0})^T$ is a purely time-like four-vector.
\end{itemize}
Thus, in~the PW limit, the intermediate s-channels have available the energies $k^2_1=E^2_A=M^2_A+k^2_\alpha$ and $k^2_2=E^2_{A'}=M^2_{A'}+k^2_\alpha$, showing that in the PW limit all momenta are~known.

Dynamically, the~intranuclear DCE transitions are given by sequential pion--nucleon SCE reactions, meaning that finally we have to resolve the intermediate configurations into their  pion--nucleon substructures. The~first step is to recognize that the invariant energy available for the $\pi^0+C$ configurations is defined by the purely time-like four-vector $k_1$. Thus, energetically $s_{\pi C}=k^2_1=E^2_A=k^2_\alpha + M^2_A$ defines the on-shell conditions. In~their rest frame, the pion and the SCE--excited nucleus $C$
are moving with the invariant three-momentum $k^2_\gamma=(s_{\pi C}-(M^*_C+m_\pi)^2)(s_{\pi C}-(M^*_C-m_\pi)^2)/(4s_{\pi C})$. Their four-momenta are $k_C=(E_c(k_\gamma),-\mathbf{k}_\gamma)^T$ and $k_\pi=(E_\pi(k_\gamma),\mathbf{k}_\gamma)^T$ and $k_C+k_\pi=k_1$. On~the pion--nucleon level, we have available in the average the energy $s_{\pi N}=k2_1/A^2=E^2_A/A^2$ from which we obtain, as before, the relative pion--nucleon momentum $\mathbf{k}$, $k^2=(s_{\pi N}-(m_N+m_\pi)^2)(s_{\pi N}-(m_N-m_\pi)^2)/(4s_{\pi N})$. As~a side result, we find the equivalent (fictitious) pion energy in the laboratory frame,
\be\label{eq:Tlab}
T_{lab}=\frac{1}{2m_N}\left(s_{\pi N}-(m_N+m_\pi)^2\right).
\ee
which attains positive values as long as $s_{\pi N}>(m_N+m_\pi)^2)$. However,
depending on the energy of the initial $A+A'$--system, we may encounter $s_{\pi N}<(m_N+m_\pi)^2$ and consequently $T_{lab}<0$, thus entering into the subthreshold region. Classically, that energy region is of course forbidden, but~not in quantum mechanics which, however, also inhibits to explore experimentally the below-threshold regions.  The~pion--nucleon interactions become virtual processes, governed by $\mathcal{T}_{\pi N}$ which is located outside of the physically accessible region. Hence, it is the task of theory to provide a description which allows to extend $\mathcal{T}_{\pi N}$ from the experimentally accessible regime into the subthreshold energy regions encountered for pion--nucleon interactions in nuclear reaction like a heavy ion DCE~reaction.

Obviously, there is no unique choice for (virtual) pion--nucleon kinematics under the conditions of a heavy ion reaction. The~DCE reaction in total is a highly dynamical process and as such corresponds to sampling over many different $\pi^0 + C$ configurations. The~\emph{mean energy approach} sketched above takes that into account in the~average.

\section[\appendixname~\thesection]{Evaluation of the MDCE Box Diagram with ISI and FSI: Distorted~Waves}\label{app:BoxDW}
A realistic description of heavy ion DCE reactions requires of course the inclusion of elastic ion--ion interactions. With~ISI/FSI, the interacting ions occupy regions of the configuration space beyond the on-shell point. In~the ion--ion rest frame, the off-shellness is defined by the distribution of the three-momenta $\mathbf{p}_{1,2}$ around the respective on-shell momenta $\mathbf{k}_{\alpha,\beta}$. The~width and shape in general of the distribution are controlled by the distortion amplitudes $f_{\alpha,\beta}$. According to Appendix \ref{app:DW}, Equation~\eqref{eq:fdist}, for~a known optical potential, modeling the elastic ion--ion self-energies, the~distributions are unambiguously known as determined by the half off-shell elastic scattering amplitudes of the incoming and outgoing~ions.

Hence, ISI and FSI dissolve the strict momentum relations of the PW limit. This process, however, is of a purely virtual character which introduces a dynamically generated uncertainty in the three-momenta but does not alter the conserved energy. Hence, in~the ion--ion rest frame, $A$ and $A'$ carry momenta $k_{A,A'}=(E_{A,A'}(\mathbf{p}_1),\mp \mathbf{p}_1)^T$, while the on-shell energies $k^2_{A,A'}=M^2_{A,A'}$ are retained.
Accordingly, the~exit channel is described by $k_{B,B'}=(E_{B,B'}(\mathbf{p}_2),\pm \mathbf{p}_2)^T$, $k^2_{B,B'}=M^2_{B,B'}$.

As a result, also the intermediate pion--nucleus, and~consequently the pion--nucleon, channels are affected because $k_{1,2}=(E_{A,A'}(\mathbf{p}_1),\mathbf{0})^T$ depend on the virtual momentum $\mathbf{p}_1$. Thus, $s_{\pi C}$ and $s_{\pi N}$ depend on $\mathbf{p}_1$. Since the MDCE reaction amplitude, Equation \eqref{eq:Mab_momentum}, is given finally by integrations over the virtual momenta $\mathbf{p}_{1,2}$, the~reaction $A+A'\to B+B'$ proceeds as a sampling over a distribution of off-shell nuclear transition form factors. The distribution, however, is centered at the on-shell form factor, which is directly related to the MDCE nuclear matrix~element.

\section[\appendixname~\thesection]{The Pion--Nucleon~T Matrix}\label{app:T012}
The form factors $T_k$, $k=0,1,2$ of~the pion--nucleon T matrix, Equation~\eqref{eq:TpiN}, are given by the partial wave amplitudes. The~
The $T_0$ component is given by formation of $\pi N$ S-wave $N^*$ resonances $S_{2I2J}$ of negative parity with
isospin $I=\frac{1}{2},\frac{3}{2}$  and total angular momentum $J^\pi=\frac{1}{2}^-$.
The form factors  $T_{1,2}$ of the longitudinal and transversal parts originate from $P_{2I2J}$ configurations of positive parity with isospin as before but $J^\pi=\frac{1}{2}^+,\frac{3}{2}^+$. The~most prominent P-wave resonances  are the Delta and the Roper resonances with spectroscopic notations $P_{33}(1232)$ and $P_{11}(1440)$, respectively.

At the energies considered here, MDCE reactions take place off the pion--nucleon mass shell. Hence, pion--nucleon scattering must be described by methods allowing to extrapolate $T_{\pi N}$ into off-shell energy regions. That goal is achieved by appropriately modeling the pion self-energies with the analytically given complex-valued form factors, where the parameters are adjusted to on-shell observables, which in our case are partial wave cross sections. In~order to obtain converged below-threshold results, in~practical calculations, resonances up to the mass region of about 2~GeV must be taken into~account.

The vertex form factors of the isovector pion--nucleon T matrix are obtained from the scattering amplitudes $U_{L2I2J}$ in the various pion--nucleon channels by proper isospin coupling.  The~pion--nucleon scattering amplitudes  are measured (and calculated) in the particle basis  $[\pi^\pm n]$, $[\pi^\pm p]$, $[\pi^0n]$, and~$[\pi^0p]$, respectively. The~pion--nucleon states are transformed to the isospin basis by Clebsch--Gordan coefficients. For example,
\be
|\pi^{-}p\ran = \sum_{T,T_3}( 1 -1\frac{1}{2}\frac{1}{2}|TT_3)|TT_3\ran
\ee
where $T=\frac{1}{2},\frac{3}{2}$ and in above case we have $ T_3=-\frac{1}{2}$. The~scattering amplitudes in the particle basis are set equal to the matrix elements which are obtained in the isospin basis for the isoscalar and the isovector interactions. The~pion--nucleon T matrix $\widehat{\mathcal{T}}=T_0+T_1\mathbf{T}_\pi\cdot \bm{\tau}$ is an isospin invariant operator and as also are the matrix elements.
The isoscalar and isovector operator form
of the T matrix is finally obtained by inversion from the scattering amplitudes in the particle basis. The~procedure is discussed~in~\cite{Feshbach:2003}.
The result is
\bea
T_0(k)&=&\frac{1}{3}F(k)\left(U_{S11}(k)-U_{S31}(k)\right) \\
T_1(k)&=&\frac{1}{3}F(k)\left(U_{P11}(k)+2U_{P13}(k)-U_{P31}(k)-2U_{P33}(k)\right) \\
T_2(k)&=&\frac{1}{3}F(k)\left(U_{P13}(k)-  U_{P11}(k)-U_{P33}(k)+U_{P31}(k)\right)
\eea
The factors of $\pm \frac{1}{3}$ and $\pm \frac{2}{3}$ are resulting from the isospin Clebsch--Gordan~coefficients.

The scattering amplitudes are normalized to units of $1/MeV$. With~the
the kinematical factor $F(k)=-4\pi\hbar^3/(2m_{\pi N})$ the T-matrix amplitudes are normalized to units of MeVfm$^3$. $m_{\pi N}$ is the pion--nucleon reduced mass.
$k=k(s_{\pi N}$ denotes the invariant relative pion--nucleon momentum which is determined by the invariant Mandelstam energy $s_{\pi N}$.

\section[\appendixname~\thesection]{Nuclear Matrix Elements and Pion~Potentials}\label{app:PiPot}

In closure approximation, the~TMEs are given in general by a superposition of nine~terms
\be
\mathcal{W}_{AB}(\mathbf{p}_1,\mathbf{p}_2)=\sum_{i,j=0,1,2}\lan \mathcal{I}^{(\pi)}_{2\mp 2} \ran
\lan B|e^{-i\mathbf{p}_2\cdot \mathbf{r}_3}W^{(ij)}_{AB}(\mathbf{x}|\mathbf{p}_1,\mathbf{p}_2)e^{i\mathbf{p}_1\cdot \mathbf{r}_1}
\mathcal{I}^{(N)}_{2\pm 2}(13)|A\ran
\ee
$\mathcal{I}^{(N)}_{2\pm 2}(13)=\left[\bm{\tau}_3\otimes \bm{\tau}_1\right]_{2\pm 2}$ is the nucleon rank-2 isotensor operator and $\lan \mathcal{I}^{(\pi)}_{2\pm 2}\ran$ denotes the pion counter apart. The~distance between the two SCE--vertices, i.e.,~the distance between the participating nucleons, is $\mathbf{x}=\mathbf{r}_1-\mathbf{r}_3$.

The three diagonal potentials are
\be\label{eq:W00}
W^{(00)}_{AB}(\mathbf{x}|\mathbf{p}_1,\mathbf{p}_2)=U_{00}(\mathbf{x}) ,
\ee
\be\label{eq:W11}
W^{(11)}_{AB}(\mathbf{x}|\mathbf{p}_1,\mathbf{p}_2)=U_{11}(\mathbf{x}|\mathbf{p}_1,\mathbf{p}_2) ,
\ee
\be\label{eq:W22}
W^{(22)}_{AB}(\mathbf{x}|\mathbf{p}_1,\mathbf{p}_2)=
\bm{\sigma}_3\cdot\overleftrightarrow{\mathbf{U}}_{22}(\mathbf{x}|\mathbf{p}_1,\mathbf{p}_2)\cdot\bm{\sigma}_1 .
\ee
As seen below, $U_{00}$ and $U_{11}$, respectively, are scalar forms giving rise to spin--scalar double excitations of Fermi character (FF). $U_{22}$ is a dyadic vector form featuring spin--vector double excitations of the Gamow--Teller type (GG).

The two non-diagonal terms inducing FF modes are
\be\label{eq:W01}
W^{(01)}_{AB}(\mathbf{x}|\mathbf{p}_1,\mathbf{p}_2)=U_{01}(\mathbf{x}|\mathbf{p}_1) ,
\ee
\be\label{eq:W10}
W^{(10)}_{AB}(\mathbf{x}|\mathbf{p}_1,\mathbf{p}_2)=U_{10}(\mathbf{x}|\mathbf{p}_2) .
\ee

The remaining four non--diagonal terms are of mixed spin character inducing FG  and GF modes, respectively:
\be\label{eq:W02}
W^{(02)}_{AB}(\mathbf{x}|\mathbf{p}_1,\mathbf{p}_2)=
\overrightarrow{\mathbf{U}}_{02}(\mathbf{x}|\mathbf{p}_1)\cdot \bm{\sigma}_1 ,
\ee
\be\label{eq:W20}
W^{(20)}_{AB}(\mathbf{x}|\mathbf{p}_1,\mathbf{p}_2)=
\bm{\sigma}_3\cdot \overleftarrow{\mathbf{U}}_{20}(\mathbf{x}|\mathbf{p}_2) ,
\ee
\be\label{eq:W12}
W^{(12)}_{AB}(\mathbf{x}|\mathbf{p}_1,\mathbf{p}_2)=
\overrightarrow{\mathbf{U}}_{12}(\mathbf{x}|\mathbf{p}_1,\mathbf{p}_2)\cdot \bm{\sigma}_1,
\ee
\be\label{eq:W20}
W^{(21)}_{AB}(\mathbf{x}|\mathbf{p}_1,\mathbf{p}_2)=
\bm{\sigma}_3\cdot \overleftarrow{\mathbf{U}}_{21}(\mathbf{x}|\mathbf{p}_1,\mathbf{p}_2) .
\ee

\subsection[\appendixname~\thesubsection]{The S-Wave~Potential}
\begin{figure}[H]

\includegraphics[width = 9.5cm]{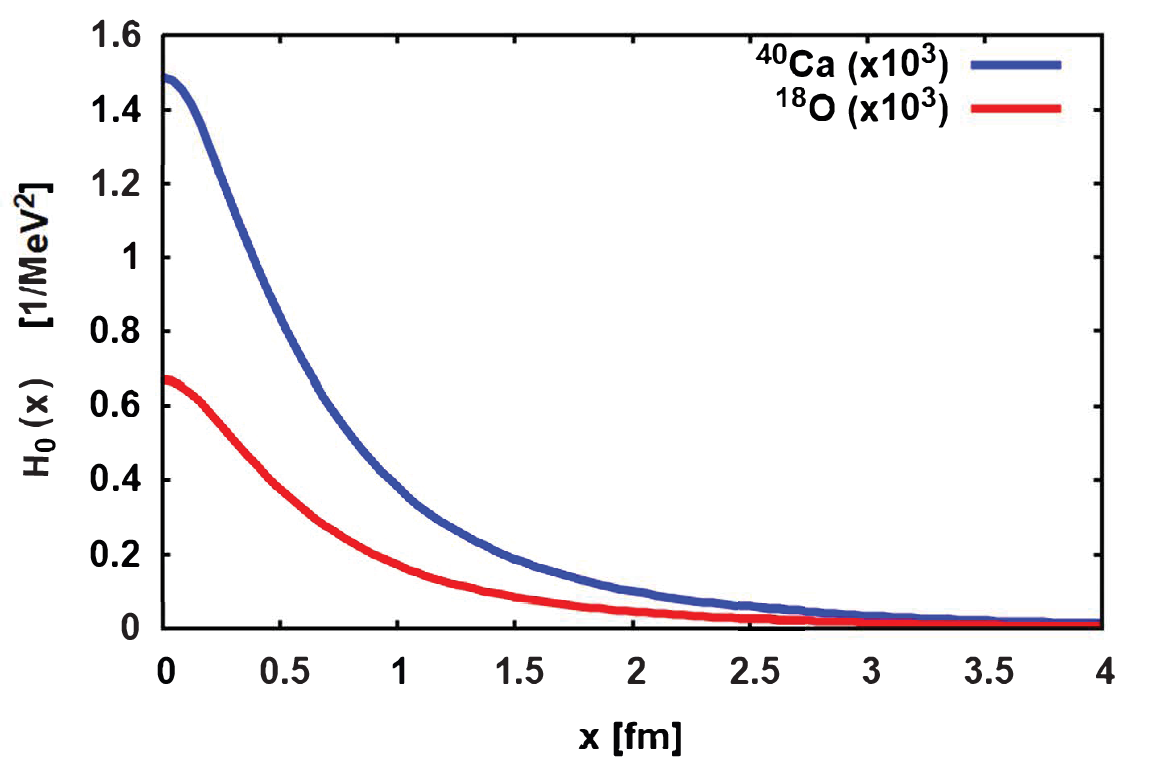}
\caption{The S-wave form factor $H_{0}(x)$ for $^{40}$Ca (blue) and $^{18}$O (red). The~full propagator, Equation~\eqref{eq:gc_red}, was used. See text for further~discussion. }
\label{fig:U00}

\end{figure}

The diagonal S-wave potential is easily evaluated:
\be \label{eq:U00}
U_{00}(\mathbf{x})= T_{0}(k_{\pi N})\int \frac{d^3k}{(2\pi)^3}g^{(+)}_\gamma(\mathbf{k}|\mathbf{p}_1)e^{i\mathbf{k}\cdot \mathbf{x}}
T_{0}(k_{\pi N})=H_0(x)
\ee
with the scalar monopole form factor
\be
H_0(x)=T^2_{0}(k_{\pi N})\frac{1}{2\pi^2}\int^{\infty}_0 dk k^2g^{(+)}_\gamma(\mathbf{k}|\mathbf{p}_1)j_0(kx),
\ee
where $j_\ell(kx)$ is the Riccati--Bessel function of order $\ell$. Numerically, the~form factor
resembles a regularized Yukawa potential as~seen in Figure~\ref{fig:U00}. The~magnitude of $H_0$ is defined $s_{\gamma,\gamma'}=E^2_{A,A'}(\mathbf{p}_1)$, see Equation~\eqref{eq:gc_red}.

\unskip

\subsection[\appendixname~\thesubsection]{The Diagonal P-Wave and the  Mixed S/P-Wave~Potentials}
The mixed S/P-wave potentials and the diagonal P-wave potentials are evaluated by expressing the momentum vectors in spherical coordinates, see~\cite{Lenske:2024dsc}. The~spherical basis is defined by the bi-orthogonal set of unit vectors
$\{\mathbf{e}_m,\mathbf{e}^*_m\}$, $m=0,\pm 1$ :
\bea
\mathbf{e}_{\pm 1}=\frac{\pm }{\sqrt{2}}\left(\mathbf{e}_x\pm i\mathbf{e}_y \right)\quad ; \quad \mathbf{e}_0=\mathbf{e}_z\\
\mathbf{e}^*_m\cdot \mathbf{e}_n=\mathbf{e}_m\cdot\mathbf{e}^*_n=\delta_{mn}.
\eea

The vector products are
\be
\mathbf{e}_i\times \mathbf{e}_j=\epsilon^{ijk}\mathbf{e}_k
\ee
where the Levi--Civita tensor is evaluated with $\{-1,0,+1\}=\{123\}$.
In that basis, space-like three-vectors $\mathbf{V}$ are given by:
\be
\mathbf{V}=\sum_{m=0,\pm 1}\mathbf{e}_mV^*_m=\sum_{m=0,\pm 1}V_m\mathbf{e}^*_m .
\ee
where
\be
V_m=\mathbf{V}\cdot \mathbf{e}_m \quad ; \quad V^*_m=\mathbf{e}^*_m\cdot \mathbf{V}
\ee

A particularly useful result is that the spherical components of $\mathbf{V}$ can be expressed in terms of spherical harmonics
\be
V_m=V\sqrt{\frac{4\pi}{3}}Y_{1m}(\hat{\mathbf{V}}),
\ee
where $V=|\mathbf{V}|$ and the rank--1 spherical harmonics $Y_{1m}(\hat{\mathbf{V}})=Y_{1m}(\theta_V,\varphi_V)$, $m=0\pm 1$, describes the orientation of $\mathbf{V}$ in~3D--space.

In the mixed S/P-wave potentials and in the diagonal P-wave potentials, we encounter monadic and dyadic structures of tensorial rank 1 and rank-2, respectively:
\be\label{eq:Hright}
\overrightarrow{H}(\mathbf{x})=\int \frac{d^3k}{(2\pi)^3}g^{(+)}_\gamma(\mathbf{k}|\mathbf{p}_1)e^{i\mathbf{k}\cdot \mathbf{x}}\mathbf{k}.
\ee
\be\label{eq:Hleft}
\overleftarrow{H}(\mathbf{x})=\int \frac{d^3k}{(2\pi)^3}\mathbf{k}g^{(+)}_\gamma(\mathbf{k}|\mathbf{p}_1)e^{i\mathbf{k}\cdot \mathbf{x}}.
\ee
\be\label{eq:Hleftright}
\overleftrightarrow{H}(\mathbf{x})=\int \frac{d^3k}{(2\pi)^3}\mathbf{k}g^{(+)}_\gamma(\mathbf{k}|\mathbf{p}_1)e^{i\mathbf{k}\cdot \mathbf{x}}\mathbf{k}.
\ee

In the spherical basis, the~tensors become
\be\label{eq:HrightS}
\overrightarrow{H}(\mathbf{x})=
\sum_{m=0,\pm 1} H^{(01)}_{1m}(\mathbf{x})\mathbf{e}^*_m,
\ee
\be\label{eq:HleftS}
\overleftarrow{H}(\mathbf{x})=
\sum_{m=0,\pm 1}\mathbf{e}_m (-)^{1+m}H^{(10)}_{1-m}(\mathbf{x}).
\ee

The rank-2 tensor is treated accordingly but leads to a more involved structure:
\be\label{eq:HleftrightS}
\overleftrightarrow{H}(\mathbf{x})=
\sum_{\lambda=0,2;\mu}(-)^{\lambda+\mu}H^{(11)}_{\lambda-\mu}(\mathbf{x})\left[\mathbf{e}\otimes \mathbf{e} \right]_{\lambda\mu}
\ee
with the dyadic products
\be
\left[\mathbf{e}\otimes \mathbf{e} \right]_{\lambda\mu}=
\sum_{m,n=0,\pm 1}\left(1m1n|\lambda \mu \right)\mathbf{e}_m \mathbf{e}_n .
\ee

The multipole form factors are
\be
H^{(\ell_1\ell_2)}_{\lambda\mu}(\mathbf{x})= H^{(\ell_1\ell_2)}_{\lambda}(x)i^{\lambda}Y_{\lambda\mu}(\hat{\mathbf{x}})
\ee
with the radial form factor
\be
H^{(\ell_1\ell_2)}_{\lambda}(x)=C_{\lambda}(\ell_1\ell_2)\frac{1}{2\pi^2}\int^\infty_0 dk k^{2+\ell_1+\ell_2}g^{(+)}_\gamma(\mathbf{k}|\mathbf{p}_1)j_\lambda(kx).
\ee

The parity coefficient
\be
C_{\lambda}(\ell_1\ell_2)=\left(\frac{4\pi}{3}\right)^{\frac{\ell1+\ell_2}{2}}\sqrt{\frac{(2\ell_1+1)(2\ell_2+1)}{4\pi(2\lambda+1)}}\left(\ell_10\ell_20|\lambda 0 \right)\frac{1}{2}\left(1+(-)^{\ell_1+\ell_2+\lambda} \right)
\ee
vanishes if $\ell_1+\ell_2+\lambda$ is an odd number. A~collection of Clebsch--Gordan coefficients is found in Table~\ref{tab:CGC}. Other values of relevance for the present purpose are obtained readily by means of the symmetry relations
\be \left(\ell_1,-m_1,\ell_2,-m_2|\lambda,-\mu \right)=(-)^{\ell_1+\ell_2-\lambda}\left(\ell_1,m_1,\ell_2,m_2|\lambda,\mu \right)\nonumber
\ee
and also
\be
\left(\ell_2,m_2,\ell_1,m_1|\lambda,\mu \right)=(-)^{\ell_1+\ell_2-\lambda}\left(\ell_1,m_1,\ell_2,m_2|\lambda,\mu \right)\nonumber .
\ee

\begin{table}[H]
 \caption{Short table
 of Clebsch--Gordon~coefficients. }\label{tab:CGC}
  \centering
\begin{tabularx}{\textwidth}{CCCC}
\toprule
  $\boldsymbol{\ell_1,m_1}$ & $\boldsymbol{\ell_2,m_2}$ & $\boldsymbol{\lambda,\mu}$ & $\boldsymbol{\left(\ell_1m_1\ell_2m_2|\lambda\mu \right)}$ \\\midrule
  0, 0
 & 0, 0 & 0, 0 & $1$ \\\midrule
  1, 0 & 1, 0 & 0, 0 & $-\frac{1}{\sqrt{3}}$ \\\midrule
  1, 1 & 1, $-$1 & 0, 0 & $\frac{1}{\sqrt{3}}$ \\\midrule
   1, 0 & 1, 0 & 2, 0 & $\sqrt{\frac{2}{3}}$ \\\midrule
  1, 1 & 1, $-$1 & 1, 0 & $\frac{1}{\sqrt{2}}$ \\\midrule
  1, 1 & 1, $-$1 & 2, 0 & $\frac{1}{\sqrt{6}}$ \\\midrule
  1, 1 & 1, 0 & 2, 1 & $\frac{1}{\sqrt{2}}$ \\\midrule
  1, 1 & 1, 1 & 2, 2 & $1$ \\
  \bottomrule
\end{tabularx}

\end{table}

The diagonal P-wave transition potentials are given by:
\be\label{eq:U11}
U_{11}(\mathbf{x}|\mathbf{p}_1,\mathbf{p}_2)=\frac{1}{m^4_\pi}T_1(k_{\pi N})\mathbf{p}_2\cdot \overleftrightarrow{H}(\mathbf{x})\cdot \mathbf{p}_1 T_1(k_{\pi N}),
\ee
and
\be\label{eq:U22}
\overleftrightarrow{\mathbf{U}}_{22}(\mathbf{x}|\mathbf{p}_1,\mathbf{p}_2)=-\frac{1}{m^4_\pi}T_2(k_{\pi N})\bm{\sigma}_3\cdot\left(\mathbf{p}_2\times \overleftrightarrow{H}(\mathbf{x})\times \mathbf{p}_1\right)\cdot\bm{\sigma}_1 T_2(k_{\pi N}).
\ee

The mixed S/P-wave transition potentials are
\be\label{eq:U01}
U_{01}(\mathbf{x}|\mathbf{p}_1)=\frac{1}{m^2_\pi}T_0(k_{\pi N})\overrightarrow{H}(\mathbf{x})\cdot \mathbf{p}_1 T_1(k_{\pi N}),
\ee
\be\label{eq:U10}
U_{10}(\mathbf{x}|\mathbf{p}_2)=\frac{1}{m^2_\pi}T_1(k_{\pi N})\mathbf{p}_2\cdot \overleftarrow{H}(\mathbf{x}) T_0(k_{\pi N}),
\ee
\be\label{eq:U02}
\overrightarrow{\mathbf{U}}_{02}(\mathbf{x}|\mathbf{p}_1)=-\frac{1}{m^2_\pi}T_0(k_{\pi N})\overrightarrow{H}(\mathbf{x})\times \mathbf{p}_1\cdot \bm{\sigma}_1 T_2(k_{\pi N}),
\ee
\be\label{eq:U20}
\overleftarrow{\mathbf{U}}_{20}(\mathbf{x}|\mathbf{p}_1)=\frac{1}{m^2_\pi}T_2(k_{\pi N})\bm{\sigma}_3\cdot\mathbf{p}_2\times \overleftarrow{H}(\mathbf{x})\cdot \mathbf{p}_1 T_0(k_{\pi N}),
\ee
\be\label{eq:U12}
\overrightarrow{\mathbf{U}}_{12}(\mathbf{x}|\mathbf{p}_1,\mathbf{p}_2)=-\frac{1}{m^4_\pi}T_1(k_{\pi N})\mathbf{p}_2\cdot\overleftrightarrow{H}(\mathbf{x})\times \mathbf{p}_1\cdot \bm{\sigma}_1 T_2(k_{\pi N}),
\ee
\be\label{eq:U21}
\overleftarrow{\mathbf{U}}_{21}(\mathbf{x}|\mathbf{p}_1,\mathbf{p}_2)=\frac{1}{m^4_\pi}T_2(k_{\pi N})\bm{\sigma}_3\cdot\mathbf{p}_2\times \overleftrightarrow{H}(\mathbf{x})\cdot \mathbf{p}_1 T_1(k_{\pi N}).
\ee

\section[\appendixname~\thesection]{Spin--Scalar Transition Potentials in Collinear~Approximation}\label{app:Collinear}
In the collinear limit, it is mathematically advantageous to include the plane wave factors into the potentials. The~modified spin--scalar transition potentials are introduced in Equation~\eqref{eq:MijColl}. They are given by
\be
\widehat{\mathcal{W}}^{(ij)}_{AB}(\mathbf{x}|\mathbf{p},\mathbf{p})=
e^{i\mathbf{p}\cdot \mathbf{x}}\widetilde{\mathcal{W}}^{(ij)}_{AB}(\mathbf{x}|\mathbf{p},\mathbf{p})
=\sum_{i=0,1,j\geq i}\mathcal{W}_{ij}(\mathbf{\mathbf{r}_1,\mathbf{r}_2|p}),
\ee
where $\mathbf{x}=\mathbf{r}_1-\mathbf{r}_2$.
By a proper change of integration variables $\mathbf{k}\to \mathbf{q}=\mathbf{k}+\mathbf{p}$, we find the transition potentials
\be
\mathcal{W}_{ij}(\mathbf{x}|\mathbf{p})=T_{i}(k_{\pi N})\int \frac{d^3q}{(2\pi)^3}G_{ij}(\mathbf{q},\mathbf{p})e^{i\mathbf{q}\cdot \mathbf{x}}T_{j}(k_{\pi N}),
\ee
with the integration kernels
\bea
G_{00}(\mathbf{q},\mathbf{p})&=&g^{(+)}_\gamma(|\mathbf{q}-\mathbf{p}|)\\
G_{01}(\mathbf{q},\mathbf{p})&=&\frac{1}{m^2_\pi}\mathbf{p}\cdot(\mathbf{q}-\mathbf{p})g^{(+)}_\gamma(|\mathbf{q}-\mathbf{p}|)\\
G_{11}(\mathbf{q},\mathbf{p})&=&\frac{1}{m^4_\pi}\left(\mathbf{p}\cdot(\mathbf{q}-\mathbf{p}) \right)^2
g^{(+)}_\gamma(|\mathbf{q}-\mathbf{p}|).
\eea

We define $t=\cos{\vartheta_{pq}}$, express powers of t by sums of Legendre polynomials $P_\ell(t)$, use the addition theorem of spherical harmonics, and~find
\bea
\mathbf{p}\cdot(\mathbf{q}-\mathbf{p})&=&pqt-p^2=pqP_1(t)-p^2P_0(t)\\
&=&
4\pi\left(pq\sum_mY^*_{1m}(\widehat{\mathbf{p}})Y_{1m}(\widehat{\mathbf{q}})-
p^2Y^*_{00}(\widehat{\mathbf{p}})Y_{00}(\widehat{\mathbf{q}})\right)
\eea

The two-body potential obtained from the double P-wave amplitude contains the polynomial
\bea
\left(\mathbf{p}\cdot(\mathbf{q}-\mathbf{p})\right)^2&=&p^2q^2t^2+p^4-2p^3qt\\
&=&\frac{2}{3}p^2q^2P_2(t)-2p^3qP_1(t)+(\frac{1}{3}p^2q^2+p^4)P_0(t)\nonumber
\eea
which is converted to
\bea
&&\left(\mathbf{p}\cdot(\mathbf{q}-\mathbf{p})\right)^2=4\pi\\
&&\times\left(\frac{2}{3}p^2q^2\sum_mY^*_{2m}(\widehat{\mathbf{p}})Y_{2m}(\widehat{\mathbf{q}})
-2p^3q\sum_mY^*_{1m}(\widehat{\mathbf{p}})Y_{1m}(\widehat{\mathbf{q}})
+(\frac{1}{3}p^2q^2+p^4)Y^*_{00}(\widehat{\mathbf{p}})Y_{00}(\widehat{\mathbf{q}})\nonumber
\right)
\eea

The propagator is expanded accordingly in spherical harmonics:
\be\label{eq:gMultiPol}
g^{(+)}_\gamma(|\mathbf{q}-\mathbf{p}|)=
4\pi\sum_{\lambda \mu}g_\lambda(p,q)Y^*_{\lambda \mu}(\widehat{\mathbf{p}})
Y_{\lambda \mu}(\widehat{\mathbf{q}}),
\ee

Since the kernel of the double S-wave potential is identical to the propagator, we already find the multipole expansion of $G_{0}(\mathbf{q},\mathbf{p}$.
The multipoles of the kernels involving P-wave amplitudes are found by combining Equation~\eqref{eq:gMultiPol} with the corresponding polynomial pre-factors. Formally, the~final results are of a similar structure:
\be
G_{ij}(\mathbf{q},\mathbf{p})=4\pi\sum_{\lambda\mu}Y^*_{\lambda\mu}(\widehat{\mathbf{p}})Y_{\lambda\mu}(\widehat{\mathbf{q}})
g^{(ij)}_{\lambda}(q,p).
\ee

However, the~multipole form factors are of a more complex form. The~mixed S/P-wave form factor is
\be
g^{(01)}_{\lambda}(q,p)=\frac{4\pi}{m^2_\pi} \sum_{\ell=0}^{\lambda+1} (\left(pqA^2_{\ell 1 \lambda} -p^2\delta_{\ell \lambda}
A^2_{\ell 0 \lambda}\right)g_\ell(p,q)
\ee
and in the double P-wave case, we find
\bea
g^{(11)}_{\lambda}(q,p)=\frac{4\pi}{m^4_\pi}
\sum_{\ell=0}^{\lambda+2} \left(
\frac{2}{3}p^2q^2A^2_{\ell 2 \lambda}-2p^3qA^2_{\ell 1 \lambda} +(\frac{1}{3}p^2q^2+p^4)\delta_{\ell \lambda}
A^2_{\ell 0 \lambda}\right)g_\ell(p,q).
\eea

The coefficient
\be
A_{\ell_1\ell_2 \ell}=\sqrt{\frac{(2\ell_1+1)(2\ell_2+1)}{4\pi(2\ell_+1)}}(\ell_10\ell_20|\ell 0)\frac{1}{2}
\left(1+(-)^{\ell_1+\ell_2+\ell} \right)
\ee
defines the parity selection rule that $\ell_1+\ell_2+\ell$ must be an even~number.

Finally, the~plane wave is expanded into partial waves in the coordinates $\mathbf{r}=\mathbf{r}_{1,2}$. Separating spherical harmonics depending on the momenta, we find
\be
e^{i\mathbf{q}\cdot \mathbf{x}}=(4\pi)^2\sum_{\ell_1\ell_1\ell m}A_{\ell_1\ell_2\ell}(-)^{\ell+m}Y_{\ell -m}(\widehat{\mathbf{q}})\left[R_{\ell_2}(\mathbf{r}_2|q)\otimes R_{\ell_2}(\mathbf{r}_1|q) \right]_{\ell m}.
\ee

The angular integration can be performed in closed form and we find:
\bea
&&\mathcal{W}_{ij}(\mathbf{r}_1,\mathbf{r}_2|\mathbf{p})=T_{i}(k_{\pi N})T_{j}(k_{\pi N})\\
&&\times 8\sum_{\ell_1\ell_1 L M}A_{\ell_1\ell_2 L}(-)^{L}Y^*_{L M}(\widehat{\mathbf{p}})\int^\infty_0 dq q^2g^{(ij)}_{L}(q,p)
\left[R_{\ell_2}(\mathbf{r}_2|q)\otimes R_{\ell_2}(\mathbf{r}_1|q) \right]_{L M} \nonumber.
\eea

For $L=0$, the two-body potentials simplify to the operator
\be
\mathcal{W}^{(0)}_{ij}(\mathbf{r}_1,\mathbf{r}_2|\mathbf{p})=T_{i}(k_{\pi N})T_{j}(k_{\pi N})
\sum_{\ell}(2\ell+1)\frac{2}{\pi}\int^\infty_0 dq q^2g^{(ij)}_{0}(q,p)
R_{\ell}(\mathbf{r}_2|q) R_{\ell}(\mathbf{r}_1|q) .
\ee

The multipole kernels are
\bea
g^{(00)}(p,q)&=&g_0(p,q)\\
g^{(01)}_{0}(q,p)&=&\frac{1}{m^2_\pi}\left(3pqg_1(p,q)-p^2g_0(p,q)\right),\\
g^{(11)}_{0}(q,p)&=&
\frac{1}{m^4_\pi}\left(\frac{10}{3}p^2q^2g_2(p,q)-6p^3qg_1(p,q)+(\frac{1}{3}p^2q^2+p^4)g_0\frac{}{}(p,q)\right).
\eea

The spin--vector potentials $\mathcal{W}_{02}$, $\mathcal{W}_{12}$, and~$\mathcal{W}_{22}$ can be evaluated by the same techniques, but~additional work is needed for the proper treatment of the vector products, for~which the formalism of Appendix \ref{app:PiPot} will be found~useful.


\begin{adjustwidth}{-\extralength}{0cm}
\reftitle{References}


\PublishersNote{}
\end{adjustwidth}
\end{document}